\documentclass[aps,prd, notitlepage, onecolumn,superscriptaddress,floatfix,letterpaper,nofootinbib, longbibliography]{revtex4-1}

\usepackage{amsmath, amssymb, graphics, setspace, multirow}
\usepackage{amsmath, amsthm, amssymb,slashed, pifont}
\usepackage{pifont}

\usepackage[usenames, dvipsnames]{color}
\usepackage[svgnames]{xcolor}
\usepackage[colorlinks,citecolor=RoyalBlue, urlcolor=RoyalBlue, linkcolor=RoyalBlue ]{hyperref} 

\usepackage[normalem]{ulem}

\newcommand{\cblue}[1]{\textcolor{blue}{#1}}

\usepackage{booktabs}

\definecolor{mygray}{gray}{0.6}

\usepackage{upgreek}
\usepackage{bbm}

\newenvironment{myfont}[2][]{\csname#2\endcsname[#1]}{}

\usepackage{slashed}
\usepackage[makeroom]{cancel}
\usepackage[normalem]{ulem}
\usepackage{soul}
\newcommand{\stkout}[1]{\ifmmode\text{\sout{\ensuremath{#1}}}\else\sout{#1}\fi}

\usepackage{sseq}
\usepackage[all,cmtip]{xy}
\usepackage{tikz-cd}
\usepackage{tikz}
\usetikzlibrary{fit,matrix}
\usetikzlibrary{decorations.markings}
\usetikzlibrary{tikzmark,decorations.pathreplacing,positioning}

\usepackage{amsfonts}

\newcommand{\bea}{\begin{eqnarray}}
\newcommand{\eea}{\end{eqnarray}}
\def\be{\begin{equation}}
\def\ee{\end{equation}}

\newcommand{\e}{\hspace{1pt}\mathrm{e}}

\newcommand{\ii}{\hspace{1pt}\mathrm{i}\hspace{1pt}}
\newcommand{\jj}{\hspace{1pt}\mathrm{j}\hspace{1pt}}
\newcommand{\kk}{\hspace{1pt}\mathrm{k}\hspace{1pt}}

\definecolor{red}{rgb}{1,0,0}
\definecolor{blue}{rgb}{0,0,1}
\definecolor{dblue}{rgb}{0,0,0.4}
\definecolor{green}{rgb}{0,1,0}
\definecolor{black}{rgb}{0,0,0}
\definecolor{white}{rgb}{1,1,1}

\definecolor{brn}{rgb}{.8,.4,.0}
\definecolor{redo}{rgb}{1,.5,.0}
\definecolor{ddgrn}{rgb}{0,0.4,0}
\definecolor{dgrn}{rgb}{0,0.55,0}
\definecolor{dbl}{rgb}{0,0,0.5}

\newcommand{\white}[1]{\color{white}{#1}}

\usepackage[bbgreekl]{mathbbol}
\usepackage{amscd}

\newcommand{\Z}{\mathbb{Z}}
\newcommand{\C}{\mathbb{C}}
\newcommand{\R}{\mathbb{R}}

\newcommand{\bH}{\mathbb{H}}

\newcommand{\dd}{\hspace{1pt}\mathrm{d}}

\newcommand{\Eq}[1]{Eq.~(\ref{#1})}

\newcommand{\bpm}{\begin{pmatrix}}
\newcommand{\epm}{\end{pmatrix}}
\newcommand{\bmm}{\begin{matrix}}
\newcommand{\emm}{\end{matrix}}

\def\Z{{\mathbb{Z}}}

\def\R{{\mathbb{R}}}
\def\C{{\mathbb{C}}}

\def\bm{{\mathbf{m}}}

\def \Z{\mathbb{Z}}

\newcommand {\emptycomment}[1]{}

\newcommand{\U}{{\rm U}}

\newcommand{\rO}{{\rm O}}

\usepackage{centernot}
\newcommand{\nn}{{\nonumber}}

\newcommand{\App}[1]{App.~\ref{#1}} 

\usepackage{enumitem} 
\usepackage{mathtools,varwidth}

\usepackage{datetime}

\begin{document}

\title{C-R-T Fractionalization in the First Quantized Hamiltonian Theory} 

\author{Yang-Yang Li}
\email{yang-yang.li@stonybrook.edu}
\affiliation{Department of Physics and Astronomy, State University of New York at Stony Brook, NY 11794, USA}

\author{Zheyan Wan}
\email{wanzheyan@bimsa.cn}
\affiliation{Beijing Institute of Mathematical Sciences and Applications, Beijing 101408, China}

\author{Juven Wang}
\email{jw@lims.ac.uk}
\homepage{http://sns.ias.edu/~juven/}
\affiliation{London Institute for Mathematical Sciences, Royal Institution, W1S 4BS, UK}
\affiliation{Center of Mathematical Sciences and Applications, Harvard University, MA 02138, USA}

\author{Shing-Tung Yau}
\email{styau@tsinghua.edu.cn}

\affiliation{Yau Mathematical Sciences Center, Tsinghua University, Beijing 100084, China}
\affiliation{Beijing Institute of Mathematical Sciences and Applications, Beijing 101408, China}
\author{Yi-Zhuang You}
\email{yzyou@physics.ucsd.edu}
\affiliation{Department of Physics, University of California San Diego, CA 92093, USA}

\begin{abstract} 
Symmetry analysis is a cornerstone of modern physics, with charge- and spacetime-orientation-reversal (CRT) symmetry being a subject of particular interest. Recent research has revealed that the CRT symmetry for fermions exhibits a fractionalization distinct from the $\mathbb{Z}_2^{\mathcal{C}} \times \mathbb{Z}_2^{\mathcal{R}} \times \mathbb{Z}_2^{\mathcal{T}}$ symmetry for scalar bosons. In fact, the CRT symmetry for fermions can be extended by internal symmetries such as fermion parity $\mathbb{Z}_2^F$, chiral symmetry $\mathbb{Z}_2^\chi$, and continuous symmetries, thereby forming a group extension of the aforementioned $\mathbb{Z}_2$ direct product, and suffices to rule out bilinear mass terms. In the conventional framework, a Majorana fermion is defined by a single Dirac fermion with trivial charge conjugation. However, this definition encounters a fundamental challenge when the spacetime dimension $d+1=5,6,7\bmod8$, where the real dimension of Majorana fermion (dim$_{\mathbb{R}}\chi_{\mathcal{C}\ell(d,0)}$) aligns with the real dimension of Dirac fermion (dim$_{\mathbb{R}}\psi_{\mathcal{C}\ell(d)}$), rather than being half as in other dimensions. This peculiarity necessitates the introduction of a symplectic Majorana fermion, defined by a pair of Dirac fermions with trivial charge conjugation, to account for the discrepancy. To include these two types of Majorana fermions, we embed the Majorana theory in $n_{\mathbb{R}}$ and define the Majorana fermion field as a representation of the real Clifford algebra which exhibits an 8-fold Bott periodicity. Within the Hamiltonian formalism, we identify the 8-fold CRT-internal symmetry groups across general spatial dimensions. In the case of Dirac fermions, the fermion field is defined as a representation of the complex Clifford algebra, which has a 2-fold Bott periodicity. Interestingly, we discover that the CRT-internal symmetry groups exhibit an 8-fold periodicity that is distinct from that of the complex Clifford algebra. In certain dimensions where distinct mass terms can span a mass manifold, the CRT-internal symmetries can act non-trivially upon this mass manifold. Employing domain wall reduction method, we are able to elucidate the relationships between symmetries across different dimensions.
\end{abstract}

\maketitle

\tableofcontents

\section{Introduction and Summary}

In modern physics, the analysis of symmetry has emerged as a cornerstone and has profoundly influenced the paradigm of theoretical research, giving rise to a plethora of intriguing topics, including symmetry-protected topological (SPT) phases~\cite{RyuSPT0803.2786, Kitaevperiod0901.2686,  AIP0905.2029,Wen1111.6341, CWang1401.1142, 1406.0307CTHsiehMorimotoRyu, Senthil1405.4015, PrakashJW2011.12320, TurzilloYou2012.04621} and Wigner-Dyson-Altland-Zirnbauer symmetry classification~\cite{Wigner_1951, Dyson1962, AltlandZirnbauer9602137}. Among these symmetries, charge conjugation ($\mathcal{C}$), mirror reflection ($\mathcal{R}$), and time-reversal ($\mathcal{T}$) symmetries~\cite{Luders1954, SchwingerPhysRev.82.914, Pauli1955, LUDERS19571, StreaterWightman1989, shimizu1985, Wang2207.14813, Dolan2009.12557, Socolovsky0404038, VafaWittenPRL1984xg, Gaiotto2017yupZoharTTT1703.00501, Wan2018zql1812.11968} stand out as particularly well-known. Charge conjugation operates as a unitary transformation, interconverting excitations and anti-excitations. For instance, in the context of a Dirac fermion, this is encapsulated by
\begin{equation}
    \mathcal{C}\psi\mathcal{C}^{-1}=\psi^*,\quad \mathcal{C}\psi^*\mathcal{C}^{-1}=\psi,
\end{equation}
where $\psi^*$ denotes the ``anti-Dirac fermion" within the complex conjugate space.
This transformation is an internal symmetry, manifesting itself only through an active transformation on a particle or field. In contrast, mirror reflection $\mathcal{R}$ inverts the spatial coordinate and time-reversion $\mathcal{T}$ reverses the direction of time. These transformations are passive with respect to the spacetime coordinates $(t,x)=(t,x_1,...,x_d)$:
\bea
    \mathcal{R}_i(t,x)\mathcal{R}_i^{-1}&=&(t,x_1,...,-x_i,...,x_d),\nn\\
    \mathcal{T}(t,x)\mathcal{T}^{-1}&=&(-t,x).
\eea
By combining mirror reflections across all spatial directions, we arrive at parity $\mathcal{P}$, a concept extensively discussed in the literature. However, when the spatial dimension $d\bmod 2=0$, parity $\mathcal{P}$ is subsumed within the spatial rotation symmetry group $SO(d)\subset SO^+(d,1)$, and thus is not considered an independent discrete symmetry separate from Lorentz symmetries. This insight prompts us to prioritize the consideration of CRT symmetries~\cite{Witten1508.04715, Freed1604.06527} over CPT in the general context of spacetime dimensions beyond the conventional $3+1d$ framework. For instance, the CRT theorem~\cite{10.21468/SciPostPhys.8.4.062} in a general dimension asserts that the combined CRT symmetry is unbreakable -- a proposition not universally valid for CPT. Moreover, the canonical conditions for CRT symmetries are satisfied across all dimensions, which is a crucial prerequisite for domain wall reduction. In contrast, CPT symmetries fail to exhibit canonical properties in $d=3\bmod 4$. These considerations have motivated our focus on CRT over CPT in the ensuing discussion.

For a scalar boson $\phi$, the symmetries $\mathcal{C}$, $\mathcal{R}$, and $\mathcal{T}$ generate a direct product symmetry group $G_{\phi}=\mathbb{Z}_2^{\mathcal{C}}\times\mathbb{Z}_2^{R}\times\mathbb{Z}_2^{\mathcal{T}}$, where each symmetry independently generates a $\mathbb{Z}_2$ group. However, However, theoretical advancements have unveiled that fermion excitations can engender a more intricate symmetry structure~\cite{Affleck1988nt1989QuantumSpinChainsHaldaneGap, RyuSPT0803.2786, Kitaevperiod0901.2686, AIP0905.2029, Wen1111.6341, CWang1401.1142, 1406.0307CTHsiehMorimotoRyu, Metlitski20141406.3032, 1510.05663Metlitski, 1711.11587GPW, 10.21468/SciPostPhys.8.4.062, Kitaev2001chain0010440, FidkowskifSPT1, FidkowskifSPT2, 2020PhRvL.124w6804B, DelmastroGaiottoGomis2101.02218, Senthil1405.4015, Wen1610.03911} such as $\mathbb{Z}_4^{\mathcal{T}F}$ and non-Abelian $\mathbb{D}_8^{\mathcal{CR},\mathcal{C}}$ groups~\cite{2109.15320}. The deviation from the direct product structure is referred to as symmetry fractionalization, which is associated with quantum anomalies and typically manifests at the boundaries of a system. In many-body quantum systems, CRT fractionalization exhibits even richer structural nuances~\cite{Gu1308.2488, PrakashJW2011.12320, PrakashJW2011.13921, TurzilloYou2012.04621}. This process of symmetry fractionalization can also be comprehended through group extensions~\cite{2109.15320, 2312.17126Wan:2023nqe, Wang2017locWWW1705.06728, WWZ1912.13504, Gaiotto2014kfa1412.5148, WanWWZ1904.00994, Wan2018djlW2.1812.11955}. According to Ref.~\cite{2109.15320}, symmetry fractionalization implies that the matter field is not in the linear representation of the original symmetry group $G$, but in the projective representation of $G$ and in the linear representation of the extended group $\tilde{G}$. The extended group $\tilde{G}$ is characterized by the short exact sequence $1\to N\to \tilde{G}\to G\to 1$, where $N$ represents certain internal symmetries. In Ref.~\cite{2312.17126Wan:2023nqe}, $N$ is chosen to be the fermion parity $\mathbb{Z}_2^F$ which flips the sign of the fermion. However, the additional internal symmetries of the Dirac fermion render the extended group $\tilde{G}$ non-unique, suggesting a more intuitive approach is to encompass all internal symmetries within $N$.

In the conventional framework, a Majorana fermion is defined by a single Dirac fermion with trivial charge conjugation~\cite{Peskin1995, Weinberg1995, Zee2003, Srednicki2007, Polchinski:1998rr, Stone:2020vva2009.00518, Dirac1928, Lancaster:2014pza}. This definition is based on the Majorana fermion's real dimension being half that of the Dirac fermion: (dim$_{\mathbb{R}}\chi_{\mathcal{C}\ell(d,0)}=1/2$dim$_{\mathbb{R}}\psi_{\mathcal{C}\ell(d)}$), a relationship derived from the representation theory of Clifford algebras (see Secs.~\ref{sec:Maj_field} and \ref{sec:Dirac_field}). However, this definition falters when the spacetime dimension satisfies $d+1=5,6,7\bmod8$, leading to dim$_{\mathbb{R}}\psi_{\mathcal{C}\ell(d,0)}=$dim$_{\mathbb{R}}\psi_{\mathcal{C}\ell(d)}$. To address this, we define a symplectic Majorana fermion~\cite{Stone:2020vva2009.00518, deligne2000quantum} as two Dirac fermions with trivial charge conjugation, as discussed in Appendices~\ref{app:Maj_2}-\ref{app:Maj_3}).
The original quaternion representation of symplectic Majorana fermions is not always convenient for analysis, which is why it was not included in Ref.~\cite{2312.17126Wan:2023nqe}. To simplify, we embed Majorana fermions in $n_{\mathbb{R}}$ where $n$ is the real dimension of the representation. This allows us to select gamma matrices that construct the Majorana equation as a real equation, with the trivial charge conjugation constraint simplifying to $\psi=\psi^*$, yielding a real fermion field. Therefore in this article, we systematically analyze the properties of Majorana fermions in the embedded $n_{\mathbb{R}}$ space, defining them uniformly as a real Grassmannian field, acting as an irreducible representation of 8-fold real Clifford algebra.

To elucidate the Clifford algebra structure within the embedded $\mathbb{R}(n)$ space, we scrutinize the internal symmetry of the Clifford algebra generated by the corresponding gamma matrices. For instance, the Sp$(1)$ internal symmetry of the quaternion algebra is identifiable within the embedded $\mathbb{R}(n)$ space. These internal symmetries play a pivotal role as they can rotate the mass terms, culminating in the formation of a mass manifold. Notably, the mass manifold exhibits an especially intricate structure in spacetime dimensions $d+1=5,6,7\bmod 8$ with Sp$(1)$ symmetries,  a feature not addressed in Ref.~\cite{2312.17126Wan:2023nqe}.

Our results are multifaceted and encompass a range of scenarios:

\begin{itemize}
    \item We verify the results in Ref.~\cite{2312.17126Wan:2023nqe} using Hamiltonian theory rather than Lagrangian theory in \cite{2312.17126Wan:2023nqe}.
    \item We define the Majorana fermion as a real Grassmannian field, acting as an irreducible representation of 8-fold real Clifford algebra to include the symplectic Majorana fermion in spacetime dimension $d+1=5,6,7\bmod 8$ and conventional Majorana fermion.
    \item We determine the invariant group for Majorana and Dirac fermion, including internal symmetries and CRT symmetries. The results are listed in Tabs.~\ref{tab:CRT-internal_Maj},\ref{tab:CRT-internal_Maj_Weyl},\ref{tab:CRT-internal_Dirac}, and \ref{tab:CRT-internal_Weyl}.
    \item We analyze the mass manifold generated by multiple mass terms. The symmetries can act nontrivially on the manifold, and the symmetry reduction under domain wall is independent of the choice of mass term. The main results are collected in Secs.~\ref{sec:Maj_mass},\ref{sec:Maj_mass&sym},\ref{sec:Dirac_mass}, and \ref{sec:Dirac_mass&sym}.
    \item CRT-internal symmetries are sufficient to rule out all bilinear mass terms. We systematically assign these symmetries in all dimensions for both Majorana and Dirac fermions and prove that we can use these symmetries to induce a gapless regime. Notably, the gapless regime is caused by symmetry constraint instead of anomaly, so we can still use 4- or more-fermion interaction to symmetrically deform to a symmetric gapped phase.
\end{itemize}

In this work, we commence with an examination of the real Majorana fermion, as detailed in Sec.~\ref{sec:Maj}. We define the Majorana fermion field as a real Grassmannian field, acting as an irreducible representation of 8-fold real Clifford algebra. When the spatial dimension $d=1,5\bmod 8$, massless Majorana fermion splits into two isomorphic Cartan subalgebras, and the Majorana field in each subalgebra is defined as Majorana-Weyl fermion. We define the corresponding symmetries for both massless Majorana and Majorana-Weyl fermions, encompassing Lorentz symmetry, RT symmetry, and internal symmetries that are intrinsic to the Clifford algebra $\mathcal{C}\ell(d,0)$. The results of these symmetries are compiled in Tabs.~\ref{tab:CRT-internal_Maj}-\ref{tab:CRT-internal_Maj_Weyl}, where we observe an 8-fold periodicity in the symmetry group. To further investigate the richer structures, we extend the mass terms from the representation space of $\mathcal{C}\ell(d,0)$ to that of $\mathcal{C}\ell(d,1)$, as outlined in Tab.~\ref{tab:explicit-mass-extension}. This extension allows distinct mass terms to form mass manifolds, on which the aforementioned symmetries can act non-trivially, as demonstrated in Tabs.~\ref{tab:CRT-mass_d=3}-\ref{tab:CRT-mass_d=5}. By employing domain wall reduction methods, we establish relations between symmetries in different dimensions, which are detailed in Tabs.~\ref{tab:sym_reduction&mass}-\ref{tab:sym_operator_reduction}.

Similarly, we also delve into the properties of the complex Dirac fermion, as discussed in Sec.~\ref{sec:Dirac}. We define the Dirac fermion field as a complex Grassmannian field, acting as an irreducible representation of 2-fold complex Clifford algebra. In odd spatial dimensions, massless Dirac fermion splits into two isomorphic Cartan subalgebras, and the Dirac field in each subalgebra is defined as Weyl fermion. We therefore define the corresponding symmetries for both massless Dirac and Weyl fermions, including Lorentz symmetry, CRT symmetry, and internal symmetries that are intrinsic to the Clifford algebra $\mathcal{C}\ell(d)$. Notably, the symmetries exhibit an 8-fold periodicity rather than a 2-fold one. These results are systematically compiled in Tabs.~\ref{tab:CRT-internal_Dirac}-\ref{tab:CRT-internal_Weyl}. To further explore the richer structures, we extend the mass terms from the representation space of $\mathcal{C}\ell(d)$ to that of $\mathcal{C}\ell(d+1)$, as detailed in Tabs.~\ref{fig:mass-extension-Dirac}-\ref{tab:mass-manifold_Dirac}. Distinct mass terms can form mass manifolds, on which the symmetries above can act non-trivially, as demonstrated in Tabs.~\ref{tab:CRT-mass_d=1_Dirac}-\ref{tab:CRT-mass_d=7_Dirac}. By employing domain wall reduction methods, we establish relations between symmetries in different dimensions, which are detailed in Tabs.~\ref{tab:sym_reduction&mass_D}-\ref{tab:sym_operator_reduction_D}.

\section{Majorana Fermion}\label{sec:Maj}

In the well-known theory of Dirac equation in 3+1$d$ spacetime, we derive the Hamiltonian for a fermion by applying the square root towards the square of the relativistic energy $E^2=\sum_i k_i^2+m^2$:

\begin{equation}
    H=\frac{1}{2}\int \dd^3x \psi^\dagger\left(\sum_{i=1}^3 \alpha_i\ii\partial_i+\beta m\right)\psi
\end{equation}
where $\alpha_i$ and $\beta$ are 4$\times$4 gamma matrices, satisfying

\begin{equation}
    \{\alpha_i,\alpha_j\}=2\delta_{ij},\quad \{\alpha_i,\beta\}=0,\quad \beta^2=1
\end{equation}
to ensure the relativistic energy spectrum, and $\psi$ is a Dirac spinor with 4 components (or flavors). To include the conventional Majorana fermion ($d+1=0,1,2,3,4\bmod 8$) and the symplectic Majorana fermion ($d+1=5,6,7\bmod 8$), we uniformly define the Majorana fermion in the embedded $n_{\mathbb{R}}$ space as a real Grassmannian field acting as an irreducible representation of real Clifford algebra (see discussion in Sec.~\ref{sec:Maj_field_model}). In Sec.~\ref{sec:Maj_sym}, we'll introduce CRT and internal symmetries intrinsic to our minimal Majorana fermion model. The mass terms also form a non-trivial manifold in some dimensions as we'll give a discussion in Sec.~\ref{sec:Maj_mass}. The existence of mass terms will break some internal symmetries, and we'll show that the CRT and internal symmetries together are enough to rule out all possible mass bilinear terms in Sec.~\ref{sec:Maj_mass&sym}. Finally in Sec.~\ref{sec:Maj_CPT_dom_wall}, we'll use the domain wall reduction method to give the relation between symmetry groups in different dimensions.

\subsection{Field Theory Models}\label{sec:Maj_field_model}

We first start with the field theory model of the real fermion $\chi$ whose complex conjugation $\chi^*$ is itself. To give a general field theory of the real Majorana fermion in a generalized dimension, we duplicate it to form a spinor $\chi$ with multiple components (or flavors), and the Hamiltonian is the ``square root" of $\sum_i k_i^2+m_i^2$ analogous to the case in 3+1$d$ spacetime.

\subsubsection{Hamiltonian}\label{sec:Maj_Ham}

A free \textbf{Majorana fermion} theory in \((d+1)\)-dimensional \textit{spacetime} with \(n\)-dimensional \textit{mass manifold} is described by
the field theory Hamiltonian

\begin{equation}
    H=\frac{1}{2}\int \dd^d x\chi^\mathsf{T} h \chi,
\end{equation}
where $h$ is the square root of $\sum_i k_i^2+m_i^2$ with $d+n$ anticommuting matrices $\alpha_i$ and $\beta_i$:

\begin{equation}\label{eq:Hamiltonian}
    h=\sum _{i=1}^d\alpha _i\ii\partial _i+\sum _{i=1}^n\beta _i m_i.
\end{equation}

In general, the matrix $h$ must satisfy

\begin{equation}\label{eq:h1}
h=-h^*=-h^\mathsf{T}=h^{\dagger },
\end{equation}
for the following reasons:

\begin{itemize}
    \item \textit{Hermiticity} of the Hamiltonian: \(H=H^{\dagger }\). Since

    \begin{equation}
    H^{\dagger }=\frac{1}{2}\int \dd^d x\chi ^{\dagger }h^{\dagger } \chi ^{\dagger \mathsf{T}}=\frac{1}{2}\int \dd^dx \chi^\mathsf{T}h^{\dagger } \chi ,
    \end{equation}
    we must have \(h=h^{\dagger }\). Here, we have used the fact that \(\chi\) is a real field, s.t. \(\chi ^{\dagger \mathsf{T}}=\chi ^*=\chi\).

    \item Compatible with the \textit{fermion} statistics: \(\chi _i\chi _j=-\chi _j\chi _i\), s.t.

    \begin{equation}
    \frac{1}{2}\int \dd^dx \chi^\mathsf{T}h \chi =\frac{1}{2}\int \dd^dx \sum_{i,j}h_{ij}\chi_i\chi_j=-\frac{1}{2}\int \dd^dx \sum_{i,j}h_{ij}\chi_j\chi_i= -\frac{1}{2}\int \dd^dx \chi^\mathsf{T}h^\mathsf{T} \chi ,
    \end{equation}
    therefore, we must have \(h=-h^\mathsf{T}\).
\end{itemize}

Combining \(h=h^{\dagger }\) and \(h=-h^\mathsf{T}\), we conclude that \(h\) must satisfy \Eq{eq:h1}.

Now given the specific form of $h$ in \Eq{eq:Hamiltonian}, in which momentum operators \(\ii\partial _i\) and real scalar mass terms \(m_i\) satisfy

\begin{equation}
\ii\partial _i=-\left(\ii\partial _i\right)^\mathsf{T}=-\left(\ii\partial _i\right){}^*=\left(\ii\partial _i\right){}^{\dagger },\quad m_i=m_i^\mathsf{T}=m_i^*=m_i^{\dagger
},
\end{equation}
we must require

\begin{equation}
\alpha _i=\alpha _i^*=\alpha _i^\mathsf{T}=\alpha _i^{\dagger },\quad \beta _i=-\beta _i^*=-\beta _i^\mathsf{T}=\beta _i^{\dagger },
\end{equation}
in order for the condition \Eq{eq:h1} to hold for \(h\) in \Eq{eq:Hamiltonian}.

In conclusion, the Hamiltonian \(H\) is fully specified by a \textbf{real Clifford algebra} \(\mathcal{C}\ell (d,n)\)~\cite{Spin-geometry, Murayama2007NotesOC, 2008arXiv0805.0311G}, that defines \(\alpha _i\) and \(\ii \beta _i\)
as its \textit{Hermitian generators}, satisfying:

\begin{itemize}
    \item \textit{Anticommutation} relations

    \begin{equation}
    \left\{\alpha _i,\alpha _j\right\}=2\delta _{i\, j},\quad \left\{\ii \beta _i,\ii \beta _j\right\}=-2\delta _{i\, j},\quad \left\{\alpha _i,\ii \beta _j\right\}=0,
    \end{equation}

    \item \textit{Real} conditions

    \begin{equation}\label{eq:real}
    \alpha _i=\alpha _i^*=\alpha _i^\mathsf{T}=\alpha _i^{\dagger }\in \mathbb{R}(\cdot ),\quad \beta _i=-\beta _i^*=-\beta _i^\mathsf{T}=\beta _i^{\dagger }\in
    \mathbb{I}(\cdot ).
    \end{equation}
\end{itemize}

\subsubsection{Real Clifford Algebra}\label{sec:Maj_Clif}

After defining the free Majorana fermion field theory specified by real Clifford algebra $\mathcal{C}\ell(d,n)$, it's conducive for us to further analyze its algebraic structure.

Real Clifford algebras have the following recursive relations:

\bea
\mathcal{C}\ell (p,q)&\cong& \mathcal{C}\ell (q,p-2)\otimes _{\mathbb{R}}\mathbb{R}(2),\nn\\
\mathcal{C}\ell (p,q)&\cong& \mathcal{C}\ell (q-2,p)\otimes _{\mathbb{R}}\mathbb{H}(1),\nn\\
\mathcal{C}\ell
(p,q)&\cong& \mathcal{C}\ell (p-1,q-1)\otimes _{\mathbb{R}}\mathbb{R}(2),\nn\\
\eea
where $\mathbb{R}(n)$, $\mathbb{C}(n)$, and $\mathbb{H}(n)$ denotes the real, complex, and quaternion algebra characterized by $n\times n$ matrices whose components are real number, complex number, and quaternion as their linear representation. The tensor product of division algebras over the real field is given by

\begin{equation}
\begin{array}{c|ccc}
 \otimes _{\mathbb{R}} & \mathbb{R} & \mathbb{C} & \mathbb{H} \\
\hline
 \mathbb{R} & \mathbb{R} & \mathbb{C} & \mathbb{H} \\
 \mathbb{C} & \mathbb{C} & \mathbb{C}\oplus \mathbb{C} & \mathbb{C}(2) \\
 \mathbb{H} & \mathbb{H} & \mathbb{C}(2) & \mathbb{R}(4) \\
\end{array}
.
\end{equation}
These relations above suffices to derive the 8-fold Bott periodicity for the real Clifford algebra:

\bea
    \mathcal{C}\ell (p+8,q)\cong \mathcal{C}\ell (q,p+6)\otimes_{\mathbb{R}}\mathbb{R}(2)\cong \mathcal{C}\ell(p+4,q)\otimes_{\mathbb{R}}\mathbb{H}(2)\cong \mathcal{C}\ell(q,p+2)\otimes_{\mathbb{R}}\mathbb{H}(4)\cong \mathcal{C}\ell(p,q)\otimes_{\mathbb{R}}\mathbb{R}(16),\nn\\
    \mathcal{C}\ell (p,q+8)\cong \mathcal{C}\ell (q+6,p)\otimes_{\mathbb{R}}\mathbb{H}(1)\cong \mathcal{C}\ell(p,q+4)\otimes_{\mathbb{R}}\mathbb{H}(2)\cong \mathcal{C}\ell(q+2,p)\otimes_{\mathbb{R}}\mathbb{R}(8)\cong \mathcal{C}\ell(p,q)\otimes_{\mathbb{R}}\mathbb{R}(16).\label{eq:Bott}
\eea
This 8-fold periodicity allows us to list the real Clifford algebras $\mathcal{C}\ell(d,n)$ in a finite periodic table in Tab.~\ref{tab:Cl(d,n)}

\begin{table}[h]
    \centering
    \begin{doublespace}
    \noindent\(\begin{array}{c|cccccccc}
     \, d\,\backslash \, n & 0 & 1 & 2 & 3 & 4 & 5 & 6 & 7 \\
    \hline
     0 & \mathbb{R}(1) & \mathbb{C}(1) & \mathbb{H}(1) & 2\mathbb{H}(1) & \mathbb{H}(2) & \mathbb{C}(4) & \mathbb{R}(8) & 2\mathbb{R}(8) \\
     1 & 2\mathbb{R}(1) & \mathbb{R}(2) & \mathbb{C}(2) & \mathbb{H}(2) & 2\mathbb{H}(2) & \mathbb{H}(4) & \mathbb{C}(8) & \mathbb{R}(16) \\
     2 & \mathbb{R}(2) & 2\mathbb{R}(2) & \mathbb{R}(4) & \mathbb{C}(4) & \mathbb{H}(4) & 2\mathbb{H}(4) & \mathbb{H}(8) & \mathbb{C}(16) \\
     3 & \mathbb{C}(2) & \mathbb{R}(4) & 2\mathbb{R}(4) & \mathbb{R}(8) & \mathbb{C}(8) & \mathbb{H}(8) & 2\mathbb{H}(8) & \mathbb{H}(16) \\
     4 & \mathbb{H}(2) & \mathbb{C}(4) & \mathbb{R}(8) & 2\mathbb{R}(8) & \mathbb{R}(16) & \mathbb{C}(16) & \mathbb{H}(16) & 2\mathbb{H}(16) \\
     5 & 2\mathbb{H}(2) & \mathbb{H}(4) & \mathbb{C}(8) & \mathbb{R}(16) & 2\mathbb{R}(16) & \mathbb{R}(32) & \mathbb{C}(32) & \mathbb{H}(32) \\
     6 & \mathbb{H}(4) & 2\mathbb{H}(4) & \mathbb{H}(8) & \mathbb{C}(16) & \mathbb{R}(32) & 2\mathbb{R}(32) & \mathbb{R}(64) & \mathbb{C}(64) \\
     7 & \mathbb{C}(8) & \mathbb{H}(8) & 2\mathbb{H}(8) & \mathbb{H}(16) & \mathbb{C}(32) & \mathbb{R}(64) & 2\mathbb{R}(64) & \mathbb{R}(128) \\
    \end{array}\)
    \end{doublespace}
    \caption{8-fold periodic table for real Clifford algebra $\mathcal{C}\ell(d,n)$. \(2\mathbb{R}(1)\) is a short-hand notation for \(\mathbb{R}(1)\oplus \mathbb{R}(1)\), and so on.}
    \label{tab:Cl(d,n)}
\end{table}

For further analytical derivation, it's intuitive to choose a specific representation for these matrices $\alpha_i$ and $\beta_i$ in the real Clifford algebra. To be concrete, we may choose the following explicit representations in massless ($n$=0) and massive ($n$=1) cases that we're most interested in:

\begin{itemize}
    \item \(\mathcal{C}\ell (d,0)\) - \textbf{massless} Majorana fermions (boundary)

    The Majorana fermion on the boundary (or domain-wall) is described by massless Majorana fermion Hamiltonian, specified by matrices $\alpha_i$ in real Clifford algebra $\mathcal{C}\ell(d,0)$. The explicit representation for these matrices is given in Tab.~\ref{tab:Cl(d,0)}.

    \begin{table}[h]
        \centering
        \begin{doublespace}
        \noindent\(\begin{array}{c|ccccccc|cc}
         \, d & \alpha _1 & \alpha _2 & \alpha _3 & \alpha _4 & \alpha _5 & \alpha _6 & \alpha _7 & \chi  & \dim _{\mathbb{R}}\chi  \\
        \hline
         0 & \text{} & \text{} & \text{} & \text{} & \text{} & \text{} & \text{} & 1_{\mathbb{R}} & 1 \\
         1 & \sigma ^3 & \text{} & \text{} & \text{} & \text{} & \text{} & \text{} & 1_{\mathbb{R}}^+\oplus 1_{\mathbb{R}}^- & 2 \\
         2 & \sigma ^1 & \sigma ^3 & \text{} & \text{} & \text{} & \text{} & \text{} & 2_{\mathbb{R}} & 2 \\
         3 & \sigma ^{10} & \sigma ^{22} & \sigma ^{30} & \text{} & \text{} & \text{} & \text{} & 2_{\mathbb{C}} & 4 \\
         4 & \sigma ^{100} & \sigma ^{212} & \sigma ^{220} & \sigma ^{300} & \text{} & \text{} & \text{} & 2_{\mathbb{H}} & 8 \\
         5 & \sigma ^{3100} & \sigma ^{3212} & \sigma ^{3220} & \sigma ^{3232} & \sigma ^{3300} & \text{} & \text{} & 2_{\mathbb{H}}^+\oplus 2_{\mathbb{H}}^-
        & 16 \\
         6 & \sigma ^{1000} & \sigma ^{3100} & \sigma ^{3212} & \sigma ^{3220} & \sigma ^{3232} & \sigma ^{3300} & \text{} & 4_{\mathbb{H}} & 16 \\
         7 & \sigma ^{1000} & \sigma ^{2002} & \sigma ^{3100} & \sigma ^{3212} & \sigma ^{3220} & \sigma ^{3232} & \sigma ^{3300} & 8_{\mathbb{C}} & 16 \\
        \end{array}\)
        \end{doublespace}
        \caption{Explicit representation for $\mathcal{C}\ell(d,0)$ describing massless Majorana fermions on the boundary. dim$_{\mathbb{R}}\,\chi$ is the smallest flavor number of Majorana fermions $\chi_i$ we need to write down a Hamiltonian in Eq.~(\ref{eq:Hamiltonian}), which can be calculated through the Clifford algebra structure.}
        \label{tab:Cl(d,0)}
    \end{table}

    \item \(\mathcal{C}\ell (d,1)\) - \textbf{massive} Majorana fermions (bulk)

    The Majorana fermion in the bulk is described by massive Majorana fermion Hamiltonian, specified by matrices $\alpha_i$ and $\beta_1$ in real Clifford algebra $\mathcal{C}\ell(d,1)$. The explicit representation for these matrices is given in Tab.~\ref{tab:Cl(d,1)}.

    \begin{table}[h]
        \centering
        \begin{doublespace}
        \noindent\(\begin{array}{c|cccccccc|cc}
         \, d & \alpha _1 & \alpha _2 & \alpha _3 & \alpha _4 & \alpha _5 & \alpha _6 & \alpha _7 & \beta _1 & \chi  & \dim _{\mathbb{R}}\chi  \\
        \hline
         0 & \text{} & \text{} & \text{} & \text{} & \text{} & \text{} & \text{} & \sigma ^2 & 1_{\mathbb{C}} & 2 \\
         1 & \sigma ^3 & \text{} & \text{} & \text{} & \text{} & \text{} & \text{} & \sigma ^2 & 2_{\mathbb{R}} & 2 \\
         2 & \sigma ^{31} & \sigma ^{33} & \text{} & \text{} & \text{} & \text{} & \text{} & \sigma ^{32} & 2_{\mathbb{R}}^+\oplus 2_{\mathbb{R}}^- & 4 \\
         3 & \sigma ^{10} & \sigma ^{22} & \sigma ^{30} & \text{} & \text{} & \text{} & \text{} & \sigma ^{21} & 4_{\mathbb{R}} & 4 \\
         4 & \sigma ^{100} & \sigma ^{212} & \sigma ^{220} & \sigma ^{300} & \text{} & \text{} & \text{} & \sigma ^{211} & 4_{\mathbb{C}} & 8 \\
         5 & \sigma ^{3100} & \sigma ^{3212} & \sigma ^{3220} & \sigma ^{3232} & \sigma ^{3300} & \text{} & \text{} & \sigma ^{1002} & 4_{\mathbb{H}} & 16
        \\
         6 & \sigma ^{31000} & \sigma ^{33100} & \sigma ^{33212} & \sigma ^{33220} & \sigma ^{33232} & \sigma ^{33300} & \text{} & \sigma ^{32000} & 4_{\mathbb{H}}^+\oplus
        4_{\mathbb{H}}^- & 32 \\
         7 & \sigma ^{10000} & \sigma ^{20020} & \sigma ^{31000} & \sigma ^{32120} & \sigma ^{32200} & \sigma ^{32320} & \sigma ^{33000} & \sigma ^{20212}
        & 8_{\mathbb{H}} & 32 \\
        \end{array}\)
        \end{doublespace}
        \caption{Explicit representation for $\mathcal{C}\ell(d,1)$ describing massive Majorana fermions in the bulk. dim$_{\mathbb{R}}\,\chi$ is the smallest flavor number of Majorana fermions $\chi_i$ we need to write down a Hamiltonian in Eq.~(\ref{eq:Hamiltonian}), which can be calculated through the Clifford algebra structure.}
        \label{tab:Cl(d,1)}
    \end{table}

\end{itemize}

\subsubsection{Majorana Field}\label{sec:Maj_field}

In this work, \(\chi\) is defined as a \textit{real Grassmannian} (Majorana) field, acting as an \textbf{irreducible representation} of the Clifford algebra \(\mathcal{C}\ell
(d,n)\). Furthermore, we notice that when (and only when) \(d-n=1,5 \bmod 8\), the Clifford algebra \(\mathcal{C}\ell (d,n)\) \textit{splits}
into two isomorphic Cartan subalgebras, 

\begin{equation}
\mathcal{C}\ell (d,n)=\mathcal{C}\ell (d,n)^+\oplus \mathcal{C}\ell (d,n)^-.
\end{equation}

The subalgebras \(\mathcal{C}\ell (d,n)^{\pm }\) are split by the following projection operator defined by

\begin{equation}
\eta \text{:=}\ii^n\prod _{i=1}^d\alpha _i\prod _{j=1}^n\beta _j=\pm 1,
\end{equation}
where \(\eta\) is the \textit{pseudo scalar} in \(\mathcal{C}\ell (d,n)\). In this case, the Majorana field \(\chi\) can be further projected into
the representation space of each subalgebra \(\mathcal{C}\ell (d,n)^{\pm }\),
\bea
\chi ^{\pm }\text{:=}\frac{1\pm \eta }{2}\chi,
\eea
defined as the \textbf{Majorana-Weyl} fermion, which only occurs at \(d-n=1,5 \bmod 8\).

The component number of the Majorana fermion $\chi$ is calculated by the \textbf{real representation dimension} \(\dim _{\mathbb{R}}\chi\), chosen to be the (minimal) dimension of a real vector space in which the representation \(\chi\) can be faithfully embedded. The dimension is counted from the corresponding vector space of $\chi$ analogous to the Clifford algebra structure, following the rules:

\begin{equation}
\dim _{\mathbb{R}}2^k{}_{\mathbb{R}}=2^k,\quad \dim _{\mathbb{R}}2^k{}_{\mathbb{C}}=2^{k+1},\quad \dim _{\mathbb{R}}2^k{}_{\mathbb{H}}=2^{k+2},
\end{equation}
where $n_{\mathbb{R}}$, $n_{\mathbb{C}}$, and $n_{\mathbb{H}}$ denotes the real, complex, and quaternion vector spaces characterized by $n$ component vectors filled with real number, complex number, and quaternion.

The 8-fold \textbf{Bott periodicity} of the real Clifford algebra \(\mathcal{C}\ell (d,n)\) in both \(d\) and \(n\) (given in Eq.~(\ref{eq:Bott})) directly implies the 8-fold periodicity in the corresponding vector space:

\begin{equation}
\chi _{\mathcal{C}\ell (d+8,n)}=\chi _{\mathcal{C}\ell (d,n)}\otimes 16_{\mathbb{R}},\quad \chi _{\mathcal{C}\ell (d,n+8)}=\chi _{\mathcal{C}\ell (d,n)}\otimes 16_{\mathbb{R}},
\end{equation}

Due to the Bott periodicity, it will be sufficient to enumerate real Clifford algebras and their corresponding vector spaces for \(d=0,\ldots,7\) in the massless Majorana fermion case and Majorana-Weyl fermion case:

\begin{itemize}
    \item \textbf{Majorana fermion} (massless case, i.e. \(n=0\))

    The Majorana fermion on the boundary (or domain-wall) is represented in a vector space, where different component corresponds to different Majorana flavor (copy). The structure of the vector space corresponds to the Clifford algebra $\mathcal{C}\ell (d,0)$, and is listed in Tab.~\ref{tab:chi_Cl(d,0)}.
    
    \item \textbf{Majorana-Weyl fermion} (massless case, i.e. \(n=0\))

    The Majorana-Weyl fermion on the boundary (or domain-wall) is also represented in a vector space. The structure of the vector space corresponds to the Clifford subalgebra $\mathcal{C}\ell (d,0)^+$, and is listed in Tab.~\ref{tab:chi_Cl(d,0)+}.

\begin{table}[htbp]
\begin{minipage}{.5\columnwidth}
  \centering
  \begin{equation*}
    \begin{array}{c|ccc}
     d & \mathcal{C}\ell (d,0) & \chi  & \dim _{\mathbb{R}}\chi  \\
    \hline
     0 & \mathbb{R}(1) & 1_{\mathbb{R}} & 1 \\
     1 & \mathbb{R}(1)\oplus \mathbb{R}(1) & 1_{\mathbb{R}}^+\oplus 1_{\mathbb{R}}^- & 2 \\
     2 & \mathbb{R}(2) & 2_{\mathbb{R}} & 2 \\
     3 & \mathbb{C}(2) & 2_{\mathbb{C}} & 4 \\
     4 & \mathbb{H}(2) & 2_{\mathbb{H}} & 8 \\
     5 & \mathbb{H}(2)\oplus \mathbb{H}(2) & 2_{\mathbb{H}}^+\oplus 2_{\mathbb{H}}^- & 16 \\
     6 & \mathbb{H}(4) & 4_{\mathbb{H}} & 16 \\
     7 & \mathbb{C}(8) & 8_{\mathbb{C}} & 16 \\
    \end{array}
    \end{equation*}
  \caption{Clifford algebra $\mathcal{C}\ell(d,0)$, vector space of massless Majorana fermion $\chi$ and its real representation dimension dim$_{\mathbb{R}}\,\chi$.}
  \label{tab:chi_Cl(d,0)}
\end{minipage}
\hfill
\begin{minipage}{.4\columnwidth}
  \centering
\begin{equation*}
\begin{aligned}
    &\begin{array}{c|ccc}
     d & \mathcal{C}\ell (d,0)^+ & \chi ^+ & \dim _{\mathbb{R}}\chi ^+ \\
    \hline
     1 & \mathbb{R}(1)^+ & 1_{\mathbb{R}}^+ & 1 \\
     5 & \mathbb{H}(2)^+ & 2_{\mathbb{H}}^+ & 8 \\
    \end{array}\\
    &\text{}\\
    &\text{}
\end{aligned}
    \end{equation*}
  \caption{Clifford subalgebra $\mathcal{C}\ell(d,0)^+$, vector space of massless Majorana-Weyl fermion $\chi^+$ and its real representation dimension dim$_{\mathbb{R}}\,\chi^+$.}
  \label{tab:chi_Cl(d,0)+}
\end{minipage}
\end{table}
    
\end{itemize}

\subsection{Symmetries}\label{sec:Maj_sym}

In the previous sections, we've listed the explicit representation of the massless (boundary) Majorana fermions in Tab.~\ref{tab:Cl(d,0)}, and we can clearly observe that the $\alpha$ matrices cannot generate the complete $\mathbb{R}(\text{dim}_{\mathbb{R}}\chi_{\mathcal{C}\ell(d,0)})$ algebra in some dimensions. For example, for $d=1,5\bmod8$, $\alpha$ matrices span a diagonal matrix space with chiral symmetry operator $(-)^{\chi}=\sigma^3$ (or $\sigma^{3000}$ for $d=5$). These symmetries are called internal symmetries for the Clifford algebra, as we'll define more strictly later. Apart from internal symmetries, we'll also include the most well-known symmetries for physicists: Lorentz symmetry and CRT symmetries.

\subsubsection{Lorentz Symmetry}\label{sec:Lorentz}

Given the Clifford algebra \(\mathcal{C}\ell (d,n)\), the \textbf{Lorentz group} is generated by

\begin{itemize}
    \item \textbf{Boost} generators:
\bea
\alpha _i\ (\text{for } i=1,\ldots ,d),
\eea
with the property that \(\alpha _i=\alpha _i^*=\alpha _i^\mathsf{T}=\alpha _i^{\dagger }\),

    \item \textbf{Rotation} generators:
\bea
\Sigma _{i\, j}=\frac{\ii}{2}\left[\alpha _i,\alpha _j\right]\ (\text{for } i,j=1,\ldots ,d),
\eea
with the property that \(\Sigma _{i\, j}=-\Sigma _{i\, j}^*=-\Sigma _{i\, j}^\mathsf{T}=\Sigma _{i\, j}^{\dagger }\).

\end{itemize}

This can be checked in the action formulation, where \(S=\int \dd x^0L\) with the Lagrangian \(L\) in two possible forms depending on the \textit{metric
signature}:

\begin{itemize}
    \item \textbf{Minkowski spacetime}: \(g_{\mu \mu }=(-,+,\cdots )\)
\bea\label{eq:Minkowski}
L&=&\frac{1}{2}\int \dd^dx \chi^\mathsf{T}\ii\partial _0\chi -H\nn\\
&=&\frac{1}{2}\int \dd^dx \chi^\mathsf{T}\left(\ii\partial _0-\sum _{i=1}^d\alpha _i\ii\partial _i-\sum _{i=1}^n\beta _i m_i\right)\chi.
\eea
Lorentz symmetry \(\Lambda (\zeta ,\theta )\) (parametrized by the rabidity \(\zeta _i\) and the rotation angle \(\theta _{i\, j}\)) is implemented
as

\bea\label{eq:Lorentz-Minkowski}
&&\Lambda (\zeta ,\theta )\left(
\begin{array}{c}
 \partial _0 \\
 \partial _1 \\
 \partial _2 \\
 \vdots  \\
\end{array}
\right)\Lambda (\zeta ,\theta )^{-1}=\exp \left(
\begin{array}{cccc}
 0 & \zeta _1 & \zeta _2 & \cdots  \\
 \zeta _1 & 0 & -\theta _{1\, 2} & \cdots  \\
 \zeta _2 & \theta _{1\, 2} & 0 & \cdots  \\
 \vdots  & \vdots  & \vdots  & \ddots \\
\end{array}
\right)\left(
\begin{array}{c}
 \partial _0 \\
 \partial _1 \\
 \partial _2 \\
 \vdots  \\
\end{array}
\right),\nn\\
&&\Lambda (\zeta ,\theta )\chi  \Lambda (\zeta ,\theta )^{-1}=\exp \left(\frac{1}{2}\left(\sum _i\zeta _i\alpha _i+\ii\sum _{i<j}\theta _{i\,
j}\Sigma _{i\, j}\right)\right)\chi .
\eea
such that 

\(\partial _{\mu }\) transforms as the \textit{vector} representation of \(\text{SO}(d,1)\),

\(\chi\) transforms as the \textit{spinor} representation of \(\text{Spin}(d,1)\).

The detailed verification of the invariance of Lagrangian in \Eq{eq:Minkowski} under Lorentz boost and rotation in \Eq{eq:Lorentz-Minkowski} is discussed in \App{app:verif_Minkowski}.

\item \textbf{Euclidian spacetime}: \(g_{\mu \mu }=(+,+,\cdots )\)
\bea\label{eq:Euclidean}
L&=&\frac{1}{2}\int \dd^dx \chi^\mathsf{T}\partial _0\chi +H\nn\\
&=&\frac{1}{2}\int \dd^dx \chi^\mathsf{T}\left(\partial _0+\sum _{i=1}^d\alpha _i\ii\partial _i+\sum _{i=1}^n\beta _i m_i\right)\chi .
\eea
Lorentz symmetry \(\Lambda (\zeta ,\theta )\) (parametrized by the rabidity \(\zeta _i\) and the rotation angle \(\theta _{i\, j}\)) is implemented
as

\bea\label{eq:Lorentz-Euclidean}
&&\Lambda (\zeta ,\theta )\left(
\begin{array}{c}
 \partial _0 \\
 \partial _1 \\
 \partial _2 \\
 \vdots  \\
\end{array}
\right)\Lambda (\zeta ,\theta )^{-1}=\exp \left(
\begin{array}{cccc}
 0 & -\zeta _1 & -\zeta _2 & \cdots  \\
 \zeta _1 & 0 & -\theta _{1\, 2} & \cdots  \\
 \zeta _2 & \theta _{1\, 2} & 0 & \cdots  \\
 \vdots  & \vdots  & \vdots  & \ddots \\
\end{array}
\right)\left(
\begin{array}{c}
 \partial _0 \\
 \partial _1 \\
 \partial _2 \\
 \vdots  \\
\end{array}
\right),\nn\\
&&\Lambda (\zeta ,\theta )\chi  \Lambda (\zeta ,\theta )^{-1}=\exp \left(\frac{1}{2}\left(-\ii\sum _i\zeta _i\alpha _i+\ii\sum _{i<j}\theta _{i\,
j}\Sigma _{i\, j}\right)\right)\chi .
\eea

such that 

\(\partial _{\mu }\) transforms as the \textit{vector} representation of \(\text{SO}(d+1)\),

\(\chi\) transforms as the \textit{spinor} representation of \(\text{Spin}(d+1)\).

Again, the detailed verification of the invariance of Lagrangian in \Eq{eq:Euclidean} under Lorentz boost and rotation in \Eq{eq:Lorentz-Euclidean} is discussed in \App{app:verif_Euclidean}.

\end{itemize}

\subsubsection{Internal Symmetry}

The \textbf{internal symmetry} of the Majorana fermion corresponds to the \textit{invariant group} \(G(\mathcal{C}\ell (d,n))\) of the Clifford algebra
\(\mathcal{C}\ell (d,n)\), defined by the \textit{short exact sequence}:
\bea\label{eq:invariant-group}
1\to G(\mathcal{C}\ell (d,n))\to \rO\left(\dim _{\mathbb{R}}\chi \right)\to\text{Aut}(\mathcal{C}\ell (d,n))\to1,
\eea
where

\begin{itemize}
    \item \(\rO\left(\dim _{\mathbb{R}}\chi \right)\) - the \textbf{maximal} \textbf{orthogonal group} of \(\chi\) preserving its anticommutation relation \(\{\chi
    ,\chi^\mathsf{T}\}=\mathbf{1}\).

    \item \(\text{Aut}(\mathcal{C}\ell (d,n))\) - the \textbf{automorphism group} of \(\mathcal{C}\ell (d,n)\). Each automorphism is induced by a \textit{group
    conjugation} for \(g\in \rO\left(\dim _{\mathbb{R}}\chi \right)\) and \(h\in \mathcal{C}\ell (d,n)\),
    
    \begin{equation}
    h\rightarrow g^{-1}h g.
    \end{equation}
\end{itemize}

The short exact sequence \Eq{eq:invariant-group} indicates that \(G(\mathcal{C}\ell (d,n))\) is the \textbf{normal subgroup} of \(\rO\left(\dim _{\mathbb{R}}\chi \right)\)
that leaves \(\mathcal{C}\ell (d,n)\) invariant. The internal symmetries of the massless Majorana fermion and Majorana-Weyl fermion are demonstrated in the following:

\begin{itemize}
    \item \textbf{Majorana fermion} (massless case, i.e. \(n=0\))

    The internal symmetry of massless Majorana fermion is listed in Tab.~\ref{tab:internal_Maj}, which includes:
    
    (1) fermion parity $\mathbb{Z}_2^F$ which flips the sign of the Majorana fermion.

    (2) fermion chiral symmetry $\mathbb{Z}_2^\chi$ which add an additional minus sign to one of the Majorana-Weyl subspace in spacetime dimension $d+1=2,6\bmod 8$.

    (3) continuous symmetry generated by corresponding Lie algebra. The continuous symmetry only exists for spacetime dimension $d+1=0,4,5,6,7\bmod 8$.

    \begin{table}[h]
        \centering
        \begin{doublespace}
        \noindent\(\begin{array}{c|cc|ccc}
         \, d & \mathcal{C}\ell (d,0) & G(\mathcal{C}\ell (d,0)) & \mathbb{Z}_2^F & \mathbb{Z}_2^{\chi } & \text{Lie algebra} \\
        \hline
         0 & \mathbb{R}(1) & \mathbb{Z}_2 & -1 & \text{} & \text{} \\
         1 & \mathbb{R}(1)\oplus \mathbb{R}(1) & \mathbb{Z}_2\times \mathbb{Z}_2 & -\sigma ^0 & \sigma ^3 & \text{} \\
         2 & \mathbb{R}(2) & \mathbb{Z}_2 & -\sigma ^0 & \text{} & \text{} \\
         3 & \mathbb{C}(2) & \U(1) & -\sigma ^{00} & \text{} & \sigma ^{02} \\
         4 & \mathbb{H}(2) & \text{Sp}(1) & -\sigma ^{000} & \text{} & (\sigma ^{002},\sigma ^{021},\sigma ^{023}) \\
         5 & \mathbb{H}(2)\oplus \mathbb{H}(2) & \text{Sp}(1)\times \text{Sp}(1) & -\sigma ^{0000} & \sigma ^{3000} & (\sigma ^{0002},\sigma ^{0021},\sigma
        ^{0023}) \\
         6 & \mathbb{H}(4) & \text{Sp}(1) & -\sigma ^{0000} & \text{} & (\sigma ^{0002},\sigma ^{0021},\sigma ^{0023}) \\
         7 & \mathbb{C}(8) & \U(1) & -\sigma ^{0000} & \text{} & \sigma ^{0002} \\
        \end{array}\)
        \end{doublespace}
        \caption{The generators of the internal symmetries including fermion parity $\mathbb{Z}_2^F$, fermion chiral symmetry $\mathbb{Z}_2^\chi$, and continuous symmetry generated by Lie algebra for massless Majorana fermion.}
        \label{tab:internal_Maj}
    \end{table}

    \item \textbf{Majorana-Weyl fermion} (massless case, i.e. \(n=0\))

    The internal symmetry of massless Majorana-Weyl fermion is listed in Tab.~\ref{tab:internal_Maj-Weyl}, which includes:
    
    (1) fermion parity $\mathbb{Z}_2^F$ which flips the sign of the Majorana fermion.

    (2) continuous symmetry generated by corresponding Lie algebra. The continuous symmetry only exists for spacetime dimension $d+1=6\bmod 8$.

    \begin{table}[h]
        \centering
        \begin{doublespace}
        \noindent\(\begin{array}{c|cc|cc}
         \, d & \mathcal{C}\ell (d,0)^+ & G\left(\mathcal{C}\ell (d,0)^+\right) & \mathbb{Z}_2^F & \text{Lie algebra} \\
        \hline
         1 & \mathbb{R}(1)^+ & \mathbb{Z}_2 & -1 & \text{} \\
         5 & \mathbb{H}(2)^+ & \text{Sp}(1) & -\sigma ^{000} & (\sigma ^{002},\sigma ^{021},\sigma ^{023}) \\
        \end{array}\)
        \end{doublespace}
        \caption{The generators of the internal symmetries including fermion parity $\mathbb{Z}_2^F$ and continuous symmetry generated by Lie algebra for massless Majorana-Weyl fermion.}
        \label{tab:internal_Maj-Weyl}
    \end{table}
        
\end{itemize}

\subsubsection{CRT Symmetry}

For Majorana fermions, the charge conjugation \(\mathcal{C}\) is undefined since Majorana fermions are free of charges. Therefore in the following discussion, we'll only focus on the reflection \(\mathcal{R}_i\) and the time reversal \(\mathcal{T}\)
symmetry.

To define reflection $\mathcal{R}_i$, it's convenient for us first to define parity $\mathcal{P}$ and then use it to define reflections $\mathcal{R}_i$ in different directions:

\textbf{Parity} \(\mathcal{P}\): a \textit{unitary} symmetry, acting as 
\bea
\mathcal{P} \partial _i \mathcal{P}^{-1}&=&-\partial _i\ (\text{for } i=1,\ldots ,d),\nn\\
\mathcal{P} \chi  \mathcal{P}^{-1}&=&M_{\mathcal{P}}\chi .
\eea
\(M_{\mathcal{P}}\in \rO\left(\dim _{\mathbb{R}}\chi \right)\) must be in the \textit{maximal orthogonal group} of \(\chi\).

The massless Hamiltonian in \Eq{eq:Hamiltonian} changes under the parity transformation $\mathcal{P}$ as:

\bea
    \frac{1}{2}\int \dd^d x\chi^{\mathsf{T}}h\chi&\to&-\frac{1}{2}\int\dd^dx(M_{\mathcal{P}}\chi)^{\mathsf{T}}h(M_{\mathcal{P}}\chi)\nn\\
    &=&-\frac{1}{2}\int\dd^dx\chi^{\mathsf{T}}(M_{\mathcal{P}}^{\mathsf{T}}hM_{\mathcal{P}})\chi.
\eea

To keep the Hamiltonian invariant, $h$ should transform under $M_{\mathcal{P}}$ as

\begin{equation}
    M_{\mathcal{P}}^{\mathsf{T}}hM_{\mathcal{P}}=-h,
\end{equation}
which indicates that $M_{\mathcal{P}}$ should act on the Clifford
algebra as

\begin{equation}\label{eq:parity1}
M_{\mathcal{P}}^\mathsf{T}\alpha _iM_{\mathcal{P}}=-\alpha _i\ (\text{for } i=1,\ldots ,d).
\end{equation}

\textbf{Reflection} \(\mathcal{R}_i\): a \textit{unitary} symmetry, acting as 
\bea
\mathcal{R}_i\partial _j\mathcal{R}_i^{-1}&=&
\left\{
\begin{array}{ll}
 -\partial _i & j=i, \\
 \partial _j & j\neq i, \\
\end{array}
 \right.
\nn\\
\mathcal{R}_i\chi  \mathcal{R}_i^{-1}&=&\alpha _iM_{\mathcal{P}}\chi .
\eea
Given \Eq{eq:parity1}, one can easily prove that \(\alpha _iM_{\mathcal{P}}\) is also an orthogonal operator and acts on the Clifford algebra as expected:

\bea\label{eq:parity2}
\left(\alpha _iM_{\mathcal{P}}\right)^\mathsf{T}\alpha _j \left(\alpha _iM_{\mathcal{P}}\right)&=&M_{\mathcal{P}}^\mathsf{T}\alpha _i^\mathsf{T}\alpha
_j\alpha _iM_{\mathcal{P}}\nn\\
&=&M_{\mathcal{P}}^\mathsf{T}\left(
\left\{ 
\begin{array}{ll}
 \alpha _i & j=i \\
 -\alpha _j & j\ne i \\
\end{array}
\right.
\right)M_{\mathcal{P}}\nn\\
&=&
\left\{
\begin{array}{ll}
 -\alpha _i & j=i \\
 \alpha _j & j\ne i \\
\end{array}
\right.
.
\eea

Under this construction, we always have \(\forall i:\mathcal{R}_i^2=(-)^F\mathcal{P}^2\), where $F$ denotes the fermion number and is even for boson ($\partial_\mu$).

Similarly, time-reversal $\mathcal{T}$ can be defined as:

\textbf{Time reversal} \(\mathcal{T}\): an \textit{antiunitary} symmetry, acting as
\bea
\mathcal{T} \ii \mathcal{T}^{-1}&=&-\ii,\nn\\
\mathcal{T} \chi  \mathcal{T}^{-1}&=&\mathcal{K} M_{\mathcal{T}}\chi .
\eea
$\mathcal{K}$ is the complex conjugation operator. \(M_{\mathcal{T}}\in \rO\left(\dim _{\mathbb{R}}\chi \right)\) must be in the \textit{maximal orthogonal group} of \(\chi\).

The massless Hamiltonian in \Eq{eq:Hamiltonian} transforms under the time-reversion $\mathcal{T}$ as:

\bea
    \frac{1}{2}\int \dd^d x\chi^{\mathsf{T}}h\chi&\to&\frac{1}{2}\int\dd^dx(M_{\mathcal{T}}\chi)^{\mathsf{T}}h^*(M_{\mathcal{T}}\chi)\nn\\
    &=&-\frac{1}{2}\int\dd^dx\chi^{\mathsf{T}}(M_{\mathcal{T}}^{\mathsf{T}}hM_{\mathcal{T}})\chi.
\eea

To keep the Hamiltonian invariant, $h$ should transform under $M_{\mathcal{P}}$ as

\begin{equation}
    M_{\mathcal{T}}^{\mathsf{T}}hM_{\mathcal{T}}=-h,
\end{equation}
which indicates that $M_{\mathcal{T}}$ should act on the Clifford
algebra as

\begin{equation}
M_{\mathcal{T}}^\mathsf{T}\alpha _iM_{\mathcal{T}}=-\alpha _i\ (\text{for } i=1,\ldots ,d).
\end{equation}

To find specific representations for matrices \(M_{\mathcal{P}}\), \(M_{\mathcal{T}}\), we notice that the choices are ambiguous up to internal symmetry transformations \(g_{\mathcal{P}},g_{\mathcal{T}}\in
G(\mathcal{C}\ell (d,0))\),

\begin{equation}
M_{\mathcal{P}}\rightarrow g_{\mathcal{P}}M_{\mathcal{P}},\quad M_{\mathcal{T}}\rightarrow g_{\mathcal{T}}M_{\mathcal{T}}.
\end{equation}

To give further constraints on the explicit representation, it's intuitive to assume \textbf{canonical CRT} conditions~\cite{10.21468/SciPostPhys.8.4.062}. For Majorana fermions, \(\mathcal{C}\) is trivial, the canonical conditions are given by:

\begin{equation}
    (\mathcal{R}_i\mathcal{T})^2=1,\quad \mathcal{T}(\mathcal{R}_i\mathcal{T})=(-)^F (\mathcal{R}_i\mathcal{T})\mathcal{T}\ (\text{for } i=1,\ldots ,d).
\end{equation}
To realize these conditions, one convenient \textit{choice} is

\begin{equation}
M_{\mathcal{P}}=M_{\mathcal{T}}.
\end{equation}
Under this choice, we always have \(\mathcal{P}^2=\mathcal{T}^2\). In conclusion, for Majorana fermions, we can consistently assume the following:
\bea
&\mathcal{P}^2=\mathcal{T}^2=(-)^F\mathcal{R}_i^2,\quad (\mathcal{R}_i \mathcal{T})^2=1,\nn\\
&\mathcal{T}(\mathcal{R}_i\mathcal{T})=(-)^F (\mathcal{R}_i\mathcal{T})\mathcal{T}\ (\text{for } i=1,\ldots ,d).
\eea
The CRT and internal symmetries are summarized in Tabs.~\ref{tab:CRT-internal_Maj} and ~\ref{tab:CRT-internal_Maj_Weyl}, where we've chosen a specific direction for reflection $\mathcal{R}_1$. Other reflections can be generated through rotation in the Lorentz symmetry group, which we've not included in $G_{\text{CRTinternal}}$ for brevity.

\begin{itemize}
    \item \textbf{Majorana fermion} (massless case, i.e. \(n=0\))

    The internal symmetries $\mathbb{Z}_2^F$, $\mathbb{Z}_2^\chi$, and continuous symmetry generated by Lie algebra, along with RT symmetries generate the invariant group $G_{\text{CRTinternal}}$ for Majorana fermions, which is independent of explicit representation basis. [See Tab.~\ref{tab:CRT-internal_Maj}]

    \begin{table}[h]
    \centering
    \begin{doublespace}
    \noindent\(\begin{array}{c|cc|cccccc}
        d         & \mathcal{C}\ell(d,0)     & G_{\text{CRTinternal}}    & \mathbb{Z}_2^F   & \mathbb{Z}_2^\chi  & \text{Lie Algebra}  & \mathbb{Z}_2^{\mathcal{P}}   & \mathbb{Z}_2^{\mathcal{R}_1} & \mathbb{Z}_2^{\mathcal{T}}               \\ \hline
        0         & \mathbb{R}(1)                    & \mathbb{Z}_2^F\times\mathbb{Z}_2^{\mathcal{T}}                                   & -1             &      \text{}         &    \text{}                                    &   \text{}               &   \text{}               & \mathcal{K}                        \\
        1         & \mathbb{R}(1)\oplus\mathbb{R}(1) & \mathbb{D}_8^{\mathcal{T},\chi}\times\mathbb{Z}_2^{\mathcal{R}_1\mathcal{T}\chi} & -\sigma^{0}    & \sigma^3       &  \text{}                                    & \mathrm{i}\sigma^2      & \sigma^1                & \mathcal{K}\mathrm{i}\sigma^2      \\
        2         & \mathbb{R}(2)                    & \mathbb{D}_8^{\mathcal{T},\mathcal{R}_1}                                         & -\sigma^{0}    &       \text{}         &   \text{}                                      & \mathrm{i}\sigma^2      & \sigma^3                & \mathcal{K}\mathrm{i}\sigma^2      \\
        3         & \mathbb{C}(2)                    & \text{Pin}_-^{\mathcal{T}}(2)\times_{\mathbb{Z}_2^F}\mathbb{Z}_4^{\mathcal{JR}_1\mathcal{T}}      & -\sigma^{00}   &       \text{}         & \sigma^{02}                                 & \mathrm{i}\sigma^{23}   & \sigma^{33}             & \mathcal{K}\mathrm{i}\sigma^{23}   \\
        4         & \mathbb{H}(2)                    & \text{Spin}(3)\times_{\mathbb{Z}_2^F}\mathbb{D}_8^{\mathcal{J}_1\mathcal{R}_1,\mathcal{J}_1\mathcal{T}}    & -\sigma^{000}  &       \text{}         & (\sigma^{002},\sigma^{021},\sigma^{023})    & \ii\sigma^{230}      & \sigma^{330}  & \mathcal{K}\ii\sigma^{230}            \\
        5         & \mathbb{H}(2)\oplus\mathbb{H}(2) & \text{Spin}(4)\rtimes_{\mathbb{Z}_2^F}\mathbb{Z}_4^{\mathcal{T}}\times\mathbb{Z}_2^{\mathcal{R}_1\mathcal{T}\chi}   & -\sigma^{0000} & \sigma^{3000}   & (\sigma^{0002},\sigma^{0021},\sigma^{0023})  & \mathrm{i}\sigma^{2000} & \sigma^{1100} & \mathcal{K}\mathrm{i}\sigma^{2000} \\
        6         & \mathbb{H}(4)                    & \text{Spin}(3)\times_{\mathbb{Z}_2^F}\mathbb{D}_8^{\mathcal{T},\mathcal{R}_1}    & -\sigma^{0000} &       \text{}         & (\sigma^{0002},\sigma^{0021},\sigma^{0023}) & \mathrm{i}\sigma^{2000} & \sigma^{3000}           & \mathcal{K}\mathrm{i}\sigma^{2000} \\
        7         & \mathbb{C}(8)                    & \text{Pin}_+^{\mathcal{T}}(2)\times\mathbb{Z}_4^{\mathcal{JR}_1\mathcal{T}}      & -\sigma^{0000} &       \text{}         & \sigma^{0002}                               & \sigma^{2023}   & \mathrm{i}\sigma^{3023} & \mathcal{K}\sigma^{2023}           \\
        8         & \mathbb{R}(16)                   & \mathbb{D}_8^{\mathcal{R}_1,\mathcal{T}}                                         & -\sigma^{0000} &       \text{}         &   \text{}                                   & \sigma^{2023}   & \mathrm{i}\sigma^{3023} & \mathcal{K}\sigma^{2023}           \\
    \end{array}\)
    \end{doublespace}
    \caption{The invariant group of massless Majorana fermion in different dimensions, including fermion parity $\mathbb{Z}_2^F$, fermion chiral symmetry $\mathbb{Z}_2^\chi$, continuous symmetry generated by Lie algebra ($\mathcal{J}_i$), reflection $\mathbb{Z}_2^{\mathcal{R}_1}$, and time-reversal symmetry $\mathbb{Z}_2^{\mathcal{T}}$.}
    \label{tab:CRT-internal_Maj}
    \end{table}

    \item \textbf{Majorana-Weyl fermion} (massless case, i.e. \(n=0\))

    The internal symmetries $\mathbb{Z}_2^F$ and continuous symmetry generated by Lie algebra, along with combined $\mathcal{PT}$ symmetry generate the invariant group $G_{\text{CRTinternal}}$ for Majorana-Weyl fermions, which is independent of explicit representation basis. [See Tab.~\ref{tab:CRT-internal_Maj_Weyl}]

    \begin{table}[h]
        \centering
        \begin{doublespace}
        \noindent\(\begin{array}{c|cc|ccc}
         \, d & \mathcal{C}\ell (d,0)^+ & G_{\text{CRT}\text{internal}} & \mathbb{Z}_2^F & \text{Lie algebra} & \mathbb{Z}_2^{\mathcal{R}_1\mathcal{T}}
        \\
        \hline
         1 & \mathbb{R}(1)^+ & \mathbb{Z}_2^F\times \mathbb{Z}_2^{\mathcal{R}_1\mathcal{T}} & -1 & \text{} & \mathcal{K} \\
         5 & \mathbb{H}(2)^+ & \text{Spin}(3)\times \mathbb{Z}_2^{\mathcal{R}_1 \mathcal{T}} & -\sigma ^{000} & (\sigma ^{002},\sigma ^{021},\sigma ^{023})
        & \mathcal{K}\sigma^{100} \\
        \end{array}\)
        \end{doublespace}
        \caption{The invariant group of massless Majorana-Weyl fermion in different dimensions, including fermion parity $\mathbb{Z}_2^F$ and continuous symmetry generated by Lie algebra ($\mathcal{J}_i$), and combined $\mathbb{Z}_2^{\mathcal{R}_1\mathcal{T}}$ symmetry.}
        \label{tab:CRT-internal_Maj_Weyl}
    \end{table}

\end{itemize}

\subsubsection{Clifford Algebra Extension}

Though the invariant group including internal symmetries is independent of the choice of our explicit representation, the choice of $M_{\mathcal{P}}$ is still ambiguous. In fact, $\mathcal{P}^2$ is still not fixed in some dimensions.

Notably, the definition of $M_{\mathcal{P}}$ requires the anticommutation condition with all $\alpha_i$ matrices. Therefore, the choice of $M_{\mathcal{P}}$ is the same as extending an extra matrix to the Clifford algebra.

Given a \textit{massless} Majorana fermion theory specified by \(\mathcal{C}\ell (d,0)\), the \textbf{Cliffor algebra extension} concerns the ability to add
extra anticommuting terms to the theory without enlarging the representation dimension of \(\chi\). The extension is given by two possible sequences:
\bea
&\mathcal{C}\ell (d,0)\rightarrow \mathcal{C}\ell (d+1,0)^{(+)}\rightarrow \ldots .\rightarrow \mathcal{C}\ell (d+n,0)^{(+)},\nn\\
&\mathcal{C}\ell (d,0)\rightarrow \mathcal{C}\ell (d,1)^{(+)}\rightarrow \ldots .\rightarrow \mathcal{C}\ell (d,n)^{(+)},
\eea
where the last sequence is also called mass extension as we'll discuss in the next section.

\begin{table}[h]
\begin{center}
\begin{tikzpicture}
\matrix[matrix of math nodes,inner sep=1pt,row sep=1em,column sep=1em] (M)
{
 \, d\backslash \, n & 0 & 1 & 2 & 3 & 4 & 5 & 6 & 7 & \ \ {\white\backslash} 8 {\white\backslash}\ \ \\
 0 & \mathbb{R}(1) & \mathbb{C}(1) & \mathbb{H}(1) & 2\mathbb{H}(1) & \mathbb{H}(2) & \mathbb{C}(4) & \mathbb{R}(8) & 2\mathbb{R}(8) & \mathbb{R}(16) \\
 1 & 2\mathbb{R}(1) & \mathbb{R}(2) & \mathbb{C}(2) & \mathbb{H}(2) & 2\mathbb{H}(2) & \mathbb{H}(4) & \mathbb{C}(8) & \mathbb{R}(16) & 2\mathbb{R}(16) \\
 2 & \mathbb{R}(2) & 2\mathbb{R}(2) & \mathbb{R}(4) & \mathbb{C}(4) & \mathbb{H}(4) & 2\mathbb{H}(4) & \mathbb{H}(8) & \mathbb{C}(16) & \mathbb{R}(32) \\
 3 & \mathbb{C}(2) & \mathbb{R}(4) & 2\mathbb{R}(4) & \mathbb{R}(8) & \mathbb{C}(8) & \mathbb{H}(8) & 2\mathbb{H}(8) & \mathbb{H}(16) & \mathbb{C}(32) \\
 4 & \mathbb{H}(2) & \mathbb{C}(4) & \mathbb{R}(8) & 2\mathbb{R}(8) & \mathbb{R}(16) & \mathbb{C}(16) & \mathbb{H}(16) & 2\mathbb{H}(16) & \mathbb{H}(32) \\
 5 & 2\mathbb{H}(2) & \mathbb{H}(4) & \mathbb{C}(8) & \mathbb{R}(16) & 2\mathbb{R}(16) & \mathbb{R}(32) & \mathbb{C}(32) & \mathbb{H}(32) & 2\mathbb{H}(32) \\
 6 & \mathbb{H}(4) & 2\mathbb{H}(4) & \mathbb{H}(8) & \mathbb{C}(16) & \mathbb{R}(32) & 2\mathbb{R}(32) & \mathbb{R}(64) & \mathbb{C}(64) & \mathbb{H}(64) \\
 7 & \mathbb{C}(8) & \mathbb{H}(8) & 2\mathbb{H}(8) & \mathbb{H}(16) & \mathbb{C}(32) & \mathbb{R}(64) & 2\mathbb{R}(64) & \mathbb{R}(128) & \mathbb{C}(128) \\
 \,\,{\white\backslash}8{\white\backslash}\,\, & \mathbb{R}(16) & \mathbb{C}(16) & \mathbb{H}(16) & 2\mathbb{H}(16) & \mathbb{H}(32) & \mathbb{C}(64) & \mathbb{R}(128) & 2\mathbb{R}(128) & \mathbb{R}(256) \\
}
;

\draw (M-1-1.south west) -- (M-1-10.south east);
\draw (M-1-1.north east) -- (M-10-1.south east);

\draw[->,color=green] (M-3-2.east) -- (M-3-3.west);
\draw[->,color=green] (M-5-2.east) -- (M-5-3.west);
\draw[->,color=green] (M-6-2.east) -- (M-6-3.west);
\draw[->,color=green] (M-7-2.east) -- (M-7-3.west);
\draw[->,color=green] (M-6-3.east) -- (M-6-4.west);
\draw[->,color=green] (M-7-3.east) -- (M-7-4.west);
\draw[->,color=green] (M-7-4.east) -- (M-7-5.west);
\draw[->,color=green] (M-2-5.east) -- (M-2-6.west);
\draw[->,color=green] (M-2-6.east) -- (M-2-7.west);
\draw[->,color=green] (M-2-7.east) -- (M-2-8.west);
\draw[->,color=green] (M-2-9.east) -- (M-2-10.west);
\draw[->,color=green] (M-3-6.east) -- (M-3-7.west);
\draw[->,color=green] (M-3-7.east) -- (M-3-8.west);
\draw[->,color=green] (M-3-8.east) -- (M-3-9.west);
\draw[->,color=green] (M-4-3.east) -- (M-4-4.west);
\draw[->,color=green] (M-4-7.east) -- (M-4-8.west);
\draw[->,color=green] (M-4-8.east) -- (M-4-9.west);
\draw[->,color=green] (M-4-9.east) -- (M-4-10.west);
\draw[->,color=green] (M-5-4.east) -- (M-5-5.west);
\draw[->,color=green] (M-5-8.east) -- (M-5-9.west);
\draw[->,color=green] (M-5-9.east) -- (M-5-10.west);
\draw[->,color=green] (M-6-5.east) -- (M-6-6.west);
\draw[->,color=green] (M-6-9.east) -- (M-6-10.west);
\draw[->,color=green] (M-7-6.east) -- (M-7-7.west);
\draw[->,color=green] (M-8-3.east) -- (M-8-4.west);
\draw[->,color=green] (M-8-4.east) -- (M-8-5.west);
\draw[->,color=green] (M-8-5.east) -- (M-8-6.west);
\draw[->,color=green] (M-8-7.east) -- (M-8-8.west);
\draw[->,color=green] (M-9-4.east) -- (M-9-5.west);
\draw[->,color=green] (M-9-5.east) -- (M-9-6.west);
\draw[->,color=green] (M-9-6.east) -- (M-9-7.west);
\draw[->,color=green] (M-9-8.east) -- (M-9-9.west);
\draw[->,color=green] (M-10-5.east) -- (M-10-6.west);
\draw[->,color=green] (M-10-6.east) -- (M-10-7.west);
\draw[->,color=green] (M-10-7.east) -- (M-10-8.west);
\draw[->,color=green] (M-10-9.east) -- (M-10-10.west);

\draw[->,color=red] (M-4-2.east) -- (M-4-3.west);
\draw[->,color=red] (M-8-2.east) -- (M-8-3.west);
\draw[->,color=red] (M-2-4.east) -- (M-2-5.west);
\draw[->,color=red] (M-2-8.east) -- (M-2-9.west);
\draw[->,color=red] (M-5-3.east) -- (M-5-4.west);
\draw[->,color=red] (M-6-4.east) -- (M-6-5.west);
\draw[->,color=red] (M-7-5.east) -- (M-7-6.west);
\draw[->,color=red] (M-3-5.east) -- (M-3-6.west);
\draw[->,color=red] (M-3-9.east) -- (M-3-10.west);
\draw[->,color=red] (M-4-6.east) -- (M-4-7.west);
\draw[->,color=red] (M-5-7.east) -- (M-5-8.west);
\draw[->,color=red] (M-6-8.east) -- (M-6-9.west);
\draw[->,color=red] (M-7-9.east) -- (M-7-10.west);
\draw[->,color=red] (M-8-6.east) -- (M-8-7.west);
\draw[->,color=red] (M-9-3.east) -- (M-9-4.west);
\draw[->,color=red] (M-9-7.east) -- (M-9-8.west);
\draw[->,color=red] (M-10-4.east) -- (M-10-5.west);
\draw[->,color=red] (M-10-8.east) -- (M-10-9.west);

\draw[->,color=green] (M-3-2.south) -- (M-4-2.north);
\draw[->,color=green] (M-7-2.south) -- (M-8-2.north);
\draw[->,color=green] (M-8-2.south) -- (M-9-2.north);
\draw[->,color=green] (M-9-2.south) -- (M-10-2.north);
\draw[->,color=green] (M-4-3.south) -- (M-5-3.north);
\draw[->,color=green] (M-8-3.south) -- (M-9-3.north);
\draw[->,color=green] (M-9-3.south) -- (M-10-3.north);
\draw[->,color=green] (M-2-3.south) -- (M-3-3.north);
\draw[->,color=green] (M-5-4.south) -- (M-6-4.north);
\draw[->,color=green] (M-9-4.south) -- (M-10-4.north);
\draw[->,color=green] (M-2-4.south) -- (M-3-4.north);
\draw[->,color=green] (M-3-4.south) -- (M-4-4.north);
\draw[->,color=green] (M-6-5.south) -- (M-7-5.north);
\draw[->,color=green] (M-2-5.south) -- (M-3-5.north);
\draw[->,color=green] (M-3-5.south) -- (M-4-5.north);
\draw[->,color=green] (M-4-5.south) -- (M-5-5.north);
\draw[->,color=green] (M-7-6.south) -- (M-8-6.north);
\draw[->,color=green] (M-3-6.south) -- (M-4-6.north);
\draw[->,color=green] (M-4-6.south) -- (M-5-6.north);
\draw[->,color=green] (M-5-6.south) -- (M-6-6.north);
\draw[->,color=green] (M-8-7.south) -- (M-9-7.north);
\draw[->,color=green] (M-4-7.south) -- (M-5-7.north);
\draw[->,color=green] (M-5-7.south) -- (M-6-7.north);
\draw[->,color=green] (M-6-7.south) -- (M-7-7.north);
\draw[->,color=green] (M-9-8.south) -- (M-10-8.north);
\draw[->,color=green] (M-5-8.south) -- (M-6-8.north);
\draw[->,color=green] (M-6-8.south) -- (M-7-8.north);
\draw[->,color=green] (M-7-8.south) -- (M-8-8.north);
\draw[->,color=green] (M-2-9.south) -- (M-3-9.north);
\draw[->,color=green] (M-6-9.south) -- (M-7-9.north);
\draw[->,color=green] (M-7-9.south) -- (M-8-9.north);
\draw[->,color=green] (M-8-9.south) -- (M-9-9.north);
\draw[->,color=green] (M-3-10.south) -- (M-4-10.north);
\draw[->,color=green] (M-7-10.south) -- (M-8-10.north);
\draw[->,color=green] (M-8-10.south) -- (M-9-10.north);
\draw[->,color=green] (M-9-10.south) -- (M-10-10.north);

\draw[->,color=red] (M-2-2.south) -- (M-3-2.north);
\draw[->,color=red] (M-6-2.south) -- (M-7-2.north);
\draw[->,color=red] (M-3-3.south) -- (M-4-3.north);
\draw[->,color=red] (M-7-3.south) -- (M-8-3.north);
\draw[->,color=red] (M-4-4.south) -- (M-5-4.north);
\draw[->,color=red] (M-8-4.south) -- (M-9-4.north);
\draw[->,color=red] (M-5-5.south) -- (M-6-5.north);
\draw[->,color=red] (M-9-5.south) -- (M-10-5.north);
\draw[->,color=red] (M-6-6.south) -- (M-7-6.north);
\draw[->,color=red] (M-2-6.south) -- (M-3-6.north);
\draw[->,color=red] (M-7-7.south) -- (M-8-7.north);
\draw[->,color=red] (M-3-7.south) -- (M-4-7.north);
\draw[->,color=red] (M-8-8.south) -- (M-9-8.north);
\draw[->,color=red] (M-4-8.south) -- (M-5-8.north);
\draw[->,color=red] (M-9-9.south) -- (M-10-9.north);
\draw[->,color=red] (M-5-9.south) -- (M-6-9.north);
\draw[->,color=red] (M-2-10.south) -- (M-3-10.north);
\draw[->,color=red] (M-6-10.south) -- (M-7-10.north);

\end{tikzpicture}
\end{center}
\caption{Clifford algebra extension. 
The green arrow means regular extension. The red arrow means chiral extension. The extension sequence ends if we meet a red arrow and acquire a chiral term.
}
\label{tab:Clifford-algebra-extension}
\end{table}

As we've shown in Tab.~\ref{tab:Clifford-algebra-extension}, there are two possible extension in each direction:

\textbf{Regular extension}: if

\begin{equation}
\dim _{\mathbb{R}}\chi_{\mathcal{C}\ell (d,n)}=\dim _{\mathbb{R}}\chi_{\mathcal{C}\ell (d,n+1)},\quad \text{or} \quad \dim _{\mathbb{R}}\chi_{\mathcal{C}\ell (d,n)}=\dim _{\mathbb{R}}\chi_{\mathcal{C}\ell (d+1,n)},
\end{equation}
an anticommuting term can be added directly.

\textbf{Chiral extension}: if \(\mathcal{C}\ell (d,n+1)\cong \mathcal{C}\ell (d,n+1)^+\oplus \mathcal{C}\ell (d,n+1)^-\) (or \(\mathcal{C}\ell (d+1,n)\cong \mathcal{C}\ell (d+1,n)^+\oplus \mathcal{C}\ell (d+1,n)^-\)) splits and

\bea
&\dim _{\mathbb{R}}\chi_{\mathcal{C}\ell (d,n)}=\dim _{\mathbb{R}}\chi^{\pm }_{\mathcal{C}\ell (d,n+1)^{\pm }}=\frac{1}{2}\dim _{\mathbb{R}}\chi_{\mathcal{C}\ell
(d,n+1)},\nn \\
\text{or}\ &\dim _{\mathbb{R}}\chi_{\mathcal{C}\ell (d,n)}=\dim _{\mathbb{R}}\chi^{\pm }_{\mathcal{C}\ell (d+1,n)^{\pm }}=\frac{1}{2}\dim _{\mathbb{R}}\chi_{\mathcal{C}\ell
(d+1,n)},
\eea
an anticommuting term can be added by promoting the \textbf{Majorana} fermion to a \textbf{Majorana-Weyl} fermion in one of the chiral subalgebras (say \(\mathcal{C}\ell
(d,n+1)^+\) or \(\mathcal{C}\ell (d+1,n)^+\)). No further term can be added for a Majorana-Weyl fermion, so the chiral extension is always the \textit{end} of a extension sequence.

\begin{table}[h]
    \centering
    \begin{doublespace}
    \noindent\(\begin{array}{c|cccccccc|cccc|cccc}
     \, d & \alpha _1 & \alpha _2 & \alpha _3 & \alpha _4 & \alpha _5 & \alpha _6 & \alpha _7 & \alpha _8 & M_{\mathcal{P}}^+  & M_{\mathcal{P}}^+  & M_{\mathcal{P}}^+  & M_{\mathcal{P}}^+  & M_{\mathcal{P}}^-  & M_{\mathcal{P}}^-  & M_{\mathcal{P}}^- & M_{\mathcal{P}}^-  \\
    \hline
     1 & \sigma ^3 & \text{} & \text{} & \text{} & \text{} & \text{} & \text{} & \text{} & \sigma ^1 & \text{} & \text{} & \text{} & \ii\sigma ^2 & \text{} & \text{} & \text{} \\
     2 & \sigma ^1 & \sigma ^3 & \text{} & \text{} & \text{} & \text{} & \text{} & \text{} & \text{} & \text{} & \text{} & \text{} & \ii\sigma ^2 & \text{} & \text{} & \text{} \\
     3 & \sigma ^{10} & \sigma ^{22} & \sigma ^{30} & \text{} & \text{} & \text{} & \text{} & \text{} & \text{} & \text{} & \text{} & \text{} & \ii\sigma ^{21} & \ii\sigma ^{23} & \text{} & \text{} \\
     4 & \sigma ^{100} & \sigma ^{212} & \sigma ^{220} & \sigma ^{300} & \text{} & \text{} & \text{} & \text{} & \sigma ^{232} & \text{} & \text{}
    & \text{} & \ii\sigma ^{211} & \ii\sigma ^{213} & \ii\sigma ^{230}
    & \text{} \\
     5 & \sigma ^{3100} & \sigma ^{3212} & \sigma ^{3220} & \sigma ^{3232} & \sigma ^{3300} & \text{} & \text{} & \text{} & \sigma ^{1000} & \sigma ^{2002} & \sigma
    ^{2021} & \sigma ^{2023} & \ii\sigma ^{1002} & \ii\sigma ^{1021} & \ii\sigma
    ^{1023} & \ii\sigma ^{2000} \\
     6 & \sigma ^{1000} & \sigma ^{3100} & \sigma ^{3212} & \sigma ^{3220} & \sigma ^{3232} & \sigma ^{3300} & \text{} & \text{} & \sigma ^{2002} & \sigma ^{2021} & \sigma ^{2023}
    & \text{} & \ii\sigma ^{2000} & \text{} & \text{}
    & \text{} \\
     7 & \sigma ^{1000} & \sigma ^{2002} & \sigma ^{3100} & \sigma ^{3212} & \sigma ^{3220} & \sigma ^{3232} & \sigma ^{3300} & \text{} & \sigma ^{2021} & \sigma ^{2023} & \text{}
    & \text{} & \text{} & \text{} & \text{}
    & \text{} \\
     8 & \sigma ^{1000} & \sigma ^{2002} & \sigma ^{2021} & \sigma ^{3100} & \sigma ^{3212} & \sigma ^{3220} & \sigma ^{3232} & \sigma ^{3300} & \sigma ^{2023} & \text{} & \text{} & \text{}& \text{} & \text{} & \text{} & \text{} \\
    \end{array}\)
    \end{doublespace}
    \caption{Explicit representation of the matrix $M_{\mathcal{P}}$. The explicit Clifford algebra extensions correspond to the results in Tab.~\ref{tab:Clifford-algebra-extension}.}
    \label{tab:explicit-mass-extension}
\end{table}

Extension $\mathcal{C}\ell(d,0)$ to $\mathcal{C}\ell(d+1,0)$ corresponds to the choice $M_{\mathcal{P}}=\alpha_{d+1}$, which means $\mathcal{P}^2=1$, while extension $\mathcal{C}\ell(d,0)$ to $\mathcal{C}\ell(d,1)$ corresponds to the choice $M_{\mathcal{P}}=\ii\beta_{1}$, which means $\mathcal{P}^2=(-)^F$. In this sense, one can simply check Tab.~\ref{tab:Clifford-algebra-extension} to see the choice of $M_{\mathcal{P}}$ we can make. If only the first extension is allowed, then $\mathcal{P}^2=1$ cannot be modified to $(-)^F$ by the internal symmetry. Similarly, if only the second extension (mass extension) is allowed, then $\mathcal{P}^2=(-)^F$ cannot be modified to $1$ by the internal symmetry. If both extension is allowed, then the choice of $M_{\mathcal{P}}$ is still ambiguous, and we can modify $\mathcal{P}^2=1$ and $(-)^F$ to each other by the internal symmetry.

\subsection{Mass}\label{sec:Maj_mass}

After carefully examining the Clifford algebra theory for massless Majorana fermion, we'll step forward to massive theory by extending mass terms. In this section, we'll discuss mass extensions and mass domain wall reductions.

\subsubsection{Mass Extension}

Given a \textit{massless} Majorana fermion theory specified by \(\mathcal{C}\ell (d,0)\), the \textbf{mass extension} concerns the ability to add
mass terms to the theory without enlarging the representation dimension of \(\chi\).
\bea
\mathcal{C}\ell (d,0)\rightarrow \mathcal{C}\ell (d,1)^{(+)}\rightarrow \ldots .\rightarrow \mathcal{C}\ell (d,n)^{(+)}.
\eea

The mass extension for Majorana fermions is demonstrated in Tab.~\ref{tab:Clifford-algebra-extension} as horizontal arrows. If a chiral mass extension is made by promoting the \textbf{Majorana} fermion to a \textbf{Majorana-Weyl} fermion in one of the chiral subalgebras, no further mass can be added for the Majorana-Weyl fermion, so the chiral mass extension is always the \textit{end} of a mass extension
sequence. The mass extension process can also shown by finding explicit mass extensions for \textbf{Majorana fermions}, as demonstrated in Tab.~\ref{tab:explicit-mass-extension}.

\begin{table}[h]
    \centering
    \begin{doublespace}
    \noindent\(\begin{array}{c|ccccccc|cccc}
     \, d & \alpha _1 & \alpha _2 & \alpha _3 & \alpha _4 & \alpha _5 & \alpha _6 & \alpha _7 & \beta _1 & \beta _2 & \beta _3 & \beta _4 \\
    \hline
     0 & \text{} & \text{} & \text{} & \text{} & \text{} & \text{} & \text{} & \text{} & \text{} & \text{} & \text{} \\
     1 & \sigma ^3 & \text{} & \text{} & \text{} & \text{} & \text{} & \text{} & \sigma ^2 & \text{} & \text{} & \text{} \\
     2 & \sigma ^1 & \sigma ^3 & \text{} & \text{} & \text{} & \text{} & \text{} & \sigma ^2 & \text{} & \text{} & \text{} \\
     3 & \sigma ^{10} & \sigma ^{22} & \sigma ^{30} & \text{} & \text{} & \text{} & \text{} & \sigma ^{21} & \sigma ^{23} & \text{} & \text{} \\
     4 & \sigma ^{100} & \sigma ^{212} & \sigma ^{220} & \sigma ^{300} & \text{} & \text{} & \text{} & \sigma ^{211} & \sigma ^{213} & \sigma ^{230}
    & \text{} \\
     5 & \sigma ^{3100} & \sigma ^{3212} & \sigma ^{3220} & \sigma ^{3232} & \sigma ^{3300} & \text{} & \text{} & \sigma ^{1002} & \sigma ^{1021} & \sigma
    ^{1023} & \sigma ^{2000} \\
     6 & \sigma ^{1000} & \sigma ^{3100} & \sigma ^{3212} & \sigma ^{3220} & \sigma ^{3232} & \sigma ^{3300} & \text{} & \sigma ^{2000} & \text{} & \text{}
    & \text{} \\
     7 & \sigma ^{1000} & \sigma ^{2002} & \sigma ^{3100} & \sigma ^{3212} & \sigma ^{3220} & \sigma ^{3232} & \sigma ^{3300} & \text{} & \text{} & \text{}
    & \text{} \\
    \end{array}\)
    \end{doublespace}
    \caption{Explicit representation of the extended mass terms. The explicit mass extensions correspond to the results in Tab.~\ref{tab:Clifford-algebra-extension}.}
    \label{tab:explicit-mass-extension}
\end{table}

\subsubsection{Mass Manifold}

Consider the following mass extension sequence,

\begin{equation}\label{eq:mass-extension}
\mathcal{C}\ell (d,0)\rightarrow \mathcal{C}\ell (d,1)^{(+)}\rightarrow \ldots .\rightarrow \mathcal{C}\ell (d,n)^{(+)}.
\end{equation}

The \textit{number} of mass terms corresponds to the length \(n\) of the sequence \Eq{eq:mass-extension}.

These masses span an \((n-1)\)-dimensional \textbf{mass manifold}, formulated as a Grassmannian manifold:
\bea
M_d=\frac{G(\mathcal{C}\ell (d,0))}{G\left(\mathcal{C}\ell (d,1)^{(+)}\right)}.
\eea
For \textbf{Majorana fermions}, the mass manifolds are listed in Tab.~\ref{tab:mass-manifold}. There are several cases for the manifold $M_d$:

\begin{table}[h]
    \centering
    \begin{doublespace}
    \noindent\(\begin{array}{cc|ccc|cc|c}
     \, d & \, n & \mathcal{C}\ell (d,0) & \mathcal{C}\ell (d,1)^{(+)} & \mathcal{C}\ell (d,n)^+ & G(\mathcal{C}\ell (d,0)) & G\left(\mathcal{C}\ell
    (d,1)^{(+)}\right) & M_d \\
    \hline
     0 & \text{} & \mathbb{R}(1) & \text{} & \text{} & \Z_2 & \text{} & 0 \\
     1 & 1 & \mathbb{R}(1)\oplus \mathbb{R}(1) & \mathbb{R}(2) & \text{} & \mathbb{Z}_2\times \mathbb{Z}_2 & \mathbb{Z}_2 & \mathbb{Z}_2 \\
     2 & 1 & \mathbb{R}(2) & \mathbb{R}(2)^+ & \text{} & \mathbb{Z}_2 & \mathbb{Z}_2 & 1 \\
     3 & 2 & \mathbb{C}(2) & \mathbb{R}(4) & \mathbb{R}(4)^+ & \U(1) & \mathbb{Z}_2 & \frac{\U(1)}{\mathbb{Z}_2}\cong S^1 \\
     4 & 3 & \mathbb{H}(2) & \mathbb{C}(4) & \mathbb{R}(8)^+ & \text{Sp}(1) & \U(1) & \frac{\text{Sp}(1)}{\U(1)}\cong S^2 \\
     5 & 4 & \mathbb{H}(2)\oplus \mathbb{H}(2) & \mathbb{H}(4) & \mathbb{R}(16)^+ & \text{Sp}(1)\times \text{Sp}(1) & \text{Sp}(1) & \frac{\text{Sp}(1)\times
    \text{Sp}(1)}{\text{Sp}(1)}\cong S^3 \\
     6 & 1 & \mathbb{H}(4) & \mathbb{H}(4)^+ & \text{} & \text{Sp}(1) & \text{Sp}(1) & 1 \\
     7 & \text{} & \mathbb{C}(8) & \text{} & \text{} & \U(1) & \text{} & 0 \\
    \end{array}\)
    \end{doublespace}
    \caption{Mass manifold $M_d$ spanned by multiple mass terms. We can at most extend $n$ independent mass terms.}
    \label{tab:mass-manifold}
\end{table}

\(M_d=0\): not admit any mass term.

\(M_d=1\): only one mass \(m\), and \(m=\pm 1\) belongs different topological order.

\(M_d=\mathbb{Z}_2\): only one mass \(m\), and \(m=\pm 1\) are topologically trivial but may belongs to different SPT phases depending on symmetry
assignments.

\(M_d=S^{n-1}\): admits a mass vector \(\textbf{m}\) of \(n\) components, transforming under the internal and CRT symmetries as an \(\rO(n)\) vector.

\subsubsection{Domain Wall Reduction}

\textbf{Domain wall reduction} reduces a massive (bulk) fermion to the mass domain wall as a massless (boundary) fermion in one lower dimension. This
simultaneously removes a momentum term and a mass term, corresponding to
\bea
\mathcal{C}\ell (d,1)\rightarrow \mathcal{C}\ell (d-1,0).
\eea
Note that \(\mathcal{C}\ell (d,1)\cong \mathcal{C}\ell (d-1,0)\otimes _{\mathbb{R}}\mathbb{R}(2)\) holds for all \(d\), the domain wall reduction
is always feasible.

\textbf{Domain wall projection}. Starting with the Hamiltonian

\begin{equation}
H=\frac{1}{2}\int \dd^dx \chi^\mathsf{T}\left(\sum _{i=1}^d\alpha _i\ii\partial _i+\beta  m\left(x_1\right)\right)\chi,
\end{equation}
we introduce a mass domain wall \(m\left(x_1\right)\sim \pm \tanh  x_1\) and project the bulk fermion to the domain wall.

Given \(m\left(x_1\right)\sim \pm \tanh  x_1\sim \pm x_1\), the zero mode equation is given by

\begin{equation}\label{eq:zero-mode1}
\frac{\delta \, H}{\delta \, \chi }=\left(\sum _{i=1}^d\alpha _i\ii\partial _i\pm  \beta  x_1\right)\chi =0.
\end{equation}

Assuming \(\chi\) is uniform in all directions other than \(x_1\), i.e. \(\partial _i\chi =0\) for \(i=2,\ldots\), \Eq{eq:zero-mode1} reduces to

\begin{equation}\label{eq:zero-mode2}
\left(\alpha _1\ii\partial _1\pm  \beta  x_1\right)\chi =0,\Rightarrow \left(\pm \ii\medspace \beta \medspace \alpha _1\partial _1+x_1\right)\chi =0.
\end{equation}

Define \(D_{\pm }=\frac{1}{\sqrt{2}}\left(x_1\pm \ii\medspace \beta \medspace \alpha _1\partial _1\right)\), \Eq{eq:zero-mode2} can be written as

\begin{equation}
D_{\pm }\chi =0.
\end{equation}

\(D_{\pm }\) satisfies the following commutation relation

\bea
\left[D_{\pm },D_{\pm }^{\dagger }\right]&=&\frac{1}{2}\left[x_1\pm \ii\medspace \beta \medspace \alpha \partial _1,x_1\mp \ii\medspace \beta \medspace
\alpha _1\partial _1\right]\nn\\
&=&\frac{\pm \ii\medspace \beta \medspace \alpha _1}{2}\left(\left[\partial _1,x_1\right]-\left[x_1,\partial _1\right]\right)\nn\\
&=&\pm \ii\medspace \beta \medspace \alpha _1.
\eea

The operator \(D_{\pm }\) admits a zero mode if and only if \(\pm \ii \beta  \alpha _1=1\), such that \(D_{\pm }\) behaves as a boson annihilation
operator and \(\chi\) is the vacuum state to be annihilated. So the domain wall projection operator is given by

\begin{equation}
P_{\text{DW}}=\frac{\mathbf{1}\pm \ii \beta  \alpha _1}{2}.
\end{equation}
which always reduces the fermion spinor dimension by \textit{half}:

\begin{equation}
2_{\mathbb{R}}^k\overset{P}{\rightarrow }2_{\mathbb{R}}^{k-1},
\end{equation}
which means that mass domain wall (defect) on each direction always traps half of the fermions. Similarly, a $d$ dimensional monopole defect $m_i\sim x_i\ (i=1,...,d)$ traps $1/2^d$ of the fermions.

The resulting domain wall fermion is then described by

\begin{equation}
H=\frac{1}{2}\int \dd^{d-1}x \chi^\mathsf{T}P_{\text{DW}}\left(\sum _{i=2}^d\alpha _i\ii\partial _i\right)P_{\text{DW}} \chi .
\end{equation}

The domain wall reduction of the real Clifford algebra and domain wall reduction of Majorana/Majorana-Weyl fermions are listed in Tabs.~\ref{tab:domain-wall-real}-\ref{tab:domain-wall-Majorana}. To understand the table, let's focus on a specific path: the domain wall reduction from 3d bulk Majorana fermion $4_{\mathbb{R}}$ to 2d boundary $2_{\mathbb{R}}$ indicates the gapless states on the surface of 3d topological insulators. We can further extend a mass term to the system and reduce it to the 1d boundary $1_{\mathbb{R}}^+$, and the 1d state spontaneously acquires chirality. This 1d state is the famous integer quantum Hall effect.

\begin{table}[htbp]
    \begin{minipage}{.4\columnwidth}
        \centering
        \begin{tikzpicture}
        \matrix[matrix of math nodes,inner sep=1pt,row sep=1em,column sep=1em] (M)
        {
        {\white\backslash}d{\white\backslash} & \mathcal{C}\ell(d,0) & {\white\backslash}\mathcal{C}\ell(d,1){\white\backslash} \\
        0 & \R(1) & \C(1)\\
        1 & \R(1)\oplus\R(1) & \R(2)\\
        2 & \R(2) & \R(2)\oplus\R(2)\\
        3 & \C(2) & \R(4)\\
        4 & \bH(2) & \C(4) \\
        5 & \bH(2)\oplus\bH(2) & \bH(4)\\
        6 & \bH(4) &  \bH(4)\oplus\bH(4)\\
        7 & \C(8) & \bH(8)\\
        {\white\backslash}8{\white\backslash} & \R(16) &\C(16)\\
        \ \\
        }
        ;
        \draw (M-1-1.south west) -- (M-1-3.south east);
        \draw (M-1-1.north east) -- (M-10-1.south east);
        
        \draw[->] (M-3-3.north west) -- (M-2-2.south east);
        \draw[->] (M-4-3.north) -- (M-3-2.south);
        \draw[->] (M-5-3.north west) -- (M-4-2.south east);
        \draw[->] (M-6-3.north west) -- (M-5-2.south east);
        \draw[->] (M-7-3.north west) -- (M-6-2.south east);
        \draw[->] (M-8-3.north) -- (M-7-2.south);
        \draw[->] (M-9-3.north west) -- (M-8-2.south east);
        \draw[->] (M-10-3.north west) -- (M-9-2.south east);
        
        \draw[->,color=green] (M-3-2.east) -- (M-3-3.west);
        \draw[->,color=green] (M-5-2.east) -- (M-5-3.west);
        \draw[->,color=green] (M-6-2.east) -- (M-6-3.west);
        \draw[->,color=green] (M-7-2.east) -- (M-7-3.west);
        
        \draw[->,color=red] (M-4-2.east) -- (M-4-3.west);
        \draw[->,color=red] (M-8-2.east) -- (M-8-3.west);
        \end{tikzpicture}
        \caption{Domain wall reduction of the real Clifford algebra. $\mathbb{R},\mathbb{C},\mathbb{H}$ indicates that the Clifford algebra is real or complex or quaternionic type , the number indicates the dimension of the Clifford algebra. The black arrow means domain wall reduction. The green arrow means regular mass extension. The red arrow means chiral mass extension.}
        \label{tab:domain-wall-real}
   \end{minipage}
    \hfill
    \begin{minipage}{.5\columnwidth}
      \centering
        \begin{tikzpicture}
        \matrix[matrix of math nodes,inner sep=1pt,row sep=1em,column sep=1em,nodes in empty cells] (M)
        {
        {\white\backslash}\,{\white\backslash} &  & & & \\
        {\white\backslash}\,{\white\backslash}& \text{Boundary} & \text{Bulk} & \text{Boundary} & {\white\backslash}\text{Bulk}{\white\backslash} \\
        \,0 & & & 1_{\R} & 1_{\C}\\
        \,1 & 1_{\R}^+ & & 1_{\R}^+\oplus 1_{\R}^- & 2_{\R}\\
        \,2 & & 2_{\R}^+  & 2_{\R} & 2_{\R}^+\oplus 2_{\R}^- \\
        \,3 & & & 2_{\C} & 4_{\R}\\
        \,4 & & & 2_{\bH} & 4_{\C}\\
        \,5 & 2_{\bH}^+ & &2_{\bH}^+\oplus 2_{\bH}^- & 4_{\bH} \\
        \,6 &  & 4_{\bH}^+ & 4_{\bH} & 4_{\bH}^+\oplus 4_{\bH}^- \\
        \,7 & & & 8_{\C} & 8_{\bH}\\
        \,{\white\backslash}8{\white\backslash} & & & 16_{\R}   & 16_{\C}\\
        }
        ;
        \node[fit=(M-1-1)(M-2-1)]{$d$};
        \node[fit=(M-1-2)(M-1-3)]{Majorana-Weyl};
        \node[fit=(M-1-4)(M-1-5)]{Majorana};
        
        \draw (M-2-1.south west) -- (M-2-5.south east);
        \draw (M-1-1.north east) -- (M-11-1.south east);
        
        \draw[->] (M-5-3.north west) -- (M-4-2.south east);
        \draw[->] (M-9-3.north west) -- (M-8-2.south east);
        
        \draw[->] (M-4-5.north west) -- (M-3-4.south east);
        \draw[->] (M-5-5.north) -- (M-4-4.south);
        \draw[->] (M-6-5.north west) -- (M-5-4.south east);
        \draw[->] (M-7-5.north west) -- (M-6-4.south east);
        \draw[->] (M-8-5.north west) -- (M-7-4.south east);
        \draw[->] (M-9-5.north) -- (M-8-4.south);
        \draw[->] (M-10-5.north west) -- (M-9-4.south east);
        \draw[->] (M-11-5.north west) -- (M-10-4.south east);
        
        \draw[->,color=red] (M-5-4.west) -- (M-5-3.east);
        \draw[->,color=red] (M-9-4.west) -- (M-9-3.east);
        
        \draw[->,color=green] (M-4-4.east) -- (M-4-5.west);
        \draw[->,color=green] (M-6-4.east) -- (M-6-5.west);
        \draw[->,color=green] (M-7-4.east) -- (M-7-5.west);
        \draw[->,color=green] (M-8-4.east) -- (M-8-5.west);
        
        \end{tikzpicture}
  \caption{Domain wall reduction of Majorana/Majorana-Weyl fermions. The number indicates the dimension of the representation of the fermion, the lower index indicates that the representation is real or complex or quaternionic, and the upper index indicates the chirality of the fermion. The black arrow means domain wall reduction. The green arrow means regular mass extension. The red arrow means chiral mass extension.}
  \label{tab:domain-wall-Majorana}
\end{minipage}
\end{table}

\subsection{Mass Term and CRT-Internal Symmetry}\label{sec:Maj_mass&sym}

In this section, we'll examine the interplay between the symmetries and the mass terms. We'll focus on the action of symmetries on the mass manifold, and how to obtain the symmetries on the domain wall using the reduction method.

\subsubsection{CRT-Internal Symmetry Acting on Mass Manifold}

Recall that $h$ defined in \Eq{eq:Hamiltonian} is
\bea
h=\sum_{i=1}^d\alpha_i\ii\partial_i+\sum_{i=1}^n\beta_im_i=h_0+m,
\eea
where $h_0$ stands for the massless part and $m$ is the mass matrix. Since we've already proven that the $h_0$ part is invariant under CRT-internal symmetry in Sec.~\ref{sec:Maj_sym}, we'll now focus on the bilinear mass $\frac{1}{2}\int \dd^dx \chi^{\mathsf{T}}m\chi$ and how CRT-internal symmetry acts on the mass manifold.

The bilinear mass term changes under the $\mathcal{R}_i$, $\mathcal{T}$, and internal $U$ symmetries as follows:
\bea
\frac{1}{2}\int \dd^dx \chi^{\mathsf{T}}m\chi&\xrightarrow{\mathcal{R}_i}&
\frac{1}{2}\int \dd^dx(\alpha_iM_{\mathcal{P}}\chi)^{\mathsf{T}}m(\alpha_iM_{\mathcal{P}}\chi)=\frac{1}{2}\int \dd^dx\chi^T(M_{\mathcal{P}}^{\mathsf{T}}\alpha_i\, m\, \alpha_i M_{\mathcal{P}})\chi,\nn\\
\frac{1}{2}\int \dd^dx\chi^{\mathsf{T}}m\chi&\xrightarrow{\mathcal{T}}&
\frac{1}{2}\int \dd^dx(M_{\mathcal{T}}\chi)^{\mathsf{T}}m^*(M_{\mathcal{T}}\chi)=-\frac{1}{2}\int \dd^dx\chi^{\mathsf{T}}(M_{\mathcal{T}}^{\mathsf{T}}\,m\,M_{\mathcal{T}})\chi,\nn\\
\frac{1}{2}\int \dd^dx\chi^{\mathsf{T}}m\chi&\xrightarrow{U}&
\frac{1}{2}\int \dd^dx(M_{U}\chi)^{\mathsf{T}}m (M_{U}\chi)=\frac{1}{2}\int \dd^dx\chi^{\mathsf{T}}(M_{U}^{\mathsf{T}}\,m\,M_{U})\chi.
\eea

If the bilinear mass term is invariant under the $\mathcal{R}_i$, $\mathcal{T}$, and $U$ symmetries, then
\bea
M_{\mathcal{P}}^{\mathsf{T}}\alpha_i\,m\,\alpha_iM_{\mathcal{P}}&=&m,\nn\\
M_{\mathcal{T}}^{\mathsf{T}}\,m\,M_{\mathcal{T}}&=&-m,\nn\\
M_{U}^{\mathsf{T}}\,m\,M_{U}&=&m.
\eea
Therefore, the matrices $M_{\mathcal{P}}$, $M_{\mathcal{T}}$, and $M_{U}$ should satisfy the following relations (for $i=1,\ldots ,d$):
\bea
M_{\mathcal{P}}^{\mathsf{T}}\beta_i M_{\mathcal{P}}&=&-\beta_i,\nn\\
M_{\mathcal{T}}^{\mathsf{T}}\beta_i M_{\mathcal{T}}&=&-\beta_i,\nn\\
M_{U}^{\mathsf{T}}\beta_i M_{U}&=&\beta_i.
\eea
Any violation of these relations is regarded as the corresponding symmetry breaking.

To be more concrete, in $d$=3,4,5$\bmod 8$, we have a nontrivial mass manifold spanned by multiple mass terms. The given CRT and internal symmetry operators can act on the whole manifold:

In $d$=3 case, we have two mass matrices $\sigma^{21}$ and $\sigma^{23}$, they span a general $S^1$ mass manifold with terms characterized by mass angle $\theta$:

\begin{equation}
    m(\theta)=\frac{1}{2}\int \dd^dx\chi^\mathsf{T}(\cos\theta\sigma^{21}+\sin\theta\sigma^{23})\chi.
\end{equation}

The action of parity $\mathcal{P}$, time-reversion $\mathcal{T}$, reflection $\mathcal{R}_i$, continuous internal symmetry $e^{\ii\phi\mathcal{J}/2}$, and its generator $\mathcal{J}$ on the $S^1$ mass manifold is listed in Tab.~\ref{tab:CRT-mass_d=3}. $\mathcal{P}$ acts as a ``reflection" on the manifold about $m_1=0$, $\mathcal{T}$ and $\mathcal{R}_i$ act as ``reflection" on the manifold about $m_2=0$, and $e^{\ii\phi\mathcal{J}/2}$ acts as a ``rotation" of $\phi$ angle on the manifold.

\begin{table}[h]
    \centering
    \begin{doublespace}
    \noindent\(\begin{array}{c|c|c|ccc}
    \text{} & \mathcal{J} & e^{\ii\phi\mathcal{J}/2} & \mathcal{P} & \mathcal{T} & \mathcal{R}_i \\
    \hline
     \theta' & \pi+\theta & \theta+\phi & -\theta & \pi-\theta & \pi-\theta \\
     \hline
     m_1 & \times & \times & \times & \checkmark & \checkmark \\
     m_2 & \times & \times & \checkmark & \times & \times \\
    \end{array}\)
    \end{doublespace}
    \caption{The action of parity $\mathcal{P}$, time-reversion $\mathcal{T}$, reflection $\mathcal{R}_i$, continuous internal symmetry $e^{\ii\phi\mathcal{J}/2}$, and its generator $\mathcal{J}$ on the $S^1$ mass manifold. $\checkmark$ means the mass manifold preserves the symmetry. $\times$ means the mass term breaks the symmetry and mass angle $\theta$ changes.}
    \label{tab:CRT-mass_d=3}
\end{table}

In $d$=4 case, we have three mass matrices $\sigma^{211}$, $\sigma^{213}$ and $\sigma^{230}$, they span a general $S^2$ mass manifold with terms characterized by mass angles $\theta$ and $\varphi$:

\begin{equation}
    m(\theta,\varphi)=\frac{1}{2}\int \dd^dx\chi^\mathsf{T}(\cos\theta\sigma^{211}+\sin\theta\cos\varphi\sigma^{213}+\sin\theta\sin\varphi\sigma^{230})\chi.
\end{equation}

The action of parity $\mathcal{P}$, time-reversion $\mathcal{T}$, reflection $\mathcal{R}_i$, continuous internal symmetry $e^{\ii\phi\mathcal{J}_i/2}$, and its generator $\mathcal{J}_i$ on the $S^2$ mass manifold is listed in Tab.~\ref{tab:CRT-mass_d=4}.

\begin{table}[h]
    \centering
    \begin{doublespace}
    \noindent\(\begin{array}{c|ccc|ccc|ccc}
    \text{} & \mathcal{J}_1 & \mathcal{J}_2 & \mathcal{J}_3 & e^{\ii\phi_1\mathcal{J}_1/2} & e^{\ii\phi_2\mathcal{J}_2/2} & e^{\ii\phi_3\mathcal{J}_3/2} & \mathcal{P} & \mathcal{T} & \mathcal{R}_i \\
    \hline
     \theta' & \pi-\theta & \pi-\theta & \theta & \theta'_{12}(\phi_1) & \theta'_{13}(\phi_2) & \theta'_{23}(\phi_3)=\theta & \theta & \pi-\theta & \pi-\theta \\
     \varphi' & \pi-\varphi & -\varphi & \pi+\varphi & \varphi'_{12}(\phi_1) & \varphi'_{13}(\phi_2) & \varphi'_{23}(\phi_3)=\varphi+\phi_3 & \varphi & \pi+\varphi & \pi+\varphi \\
     \hline
     m_1 & \times & \times & \checkmark & \times & \times & \checkmark & \times & \checkmark & \checkmark \\
     m_2 & \times & \checkmark & \times & \times & \checkmark & \times & \times & \checkmark & \checkmark \\
     m_3 & \checkmark & \times & \times & \checkmark & \times & \times & \checkmark & \times & \times \\
    \end{array}\)
    \end{doublespace}
    \caption{The action of parity $\mathcal{P}$, time-reversion $\mathcal{T}$, reflection $\mathcal{R}_i$, continuous internal symmetry $e^{\ii\phi\mathcal{J}_i/2}$, and its generator $\mathcal{J}_i$ on the $S^2$ mass manifold. $\checkmark$ means the mass manifold preserves the symmetry. $\times$ means the mass term breaks the symmetry and mass angle $\theta$ changes. $\theta'_{ij}(\phi_k)$ means the $\theta'$ angle after a $\phi_k$ rotation in the $m_i-m_j$ plane. Same for $\varphi'_{ij}(\phi_k)$.}
    \label{tab:CRT-mass_d=4}
\end{table}

In $d$=5 case, we have four mass terms $\sigma^{1002}$, $\sigma^{1021}$, $\sigma^{1023}$ and $\sigma^{2000}$, they span a general $S^3$ mass manifold with terms characterized by mass angles $\theta$, $\varphi$ and $\psi$:

\begin{equation}
    m(\theta,\varphi,\psi)=\frac{1}{2}\int \dd^dx\chi^\mathsf{T}(\cos\theta\sigma^{1002}+\sin\theta\cos\varphi\sigma^{1021}+\sin\theta\sin\varphi\cos\psi\sigma^{1023}+\sin\theta\sin\varphi\sin\psi\sigma^{2000})\chi.
\end{equation}

The action of parity $\mathcal{P}$, time-reversion $\mathcal{T}$, reflection $\mathcal{R}_i$, continuous internal symmetry $e^{\ii\phi\mathcal{J}_i/2}$, its generator $\mathcal{J}_i$, and chiral symmetry operator $(-)^\chi$ on the $S^3$ mass manifold is listed in Tab.~\ref{tab:CRT-mass_d=5}.

\begin{table}[h]
    \centering
    \begin{doublespace}
    \noindent\(\begin{array}{c|cccc|ccc|ccc}
    \text{} & (-)^\chi & \mathcal{J}_1 & \mathcal{J}_2 & \mathcal{J}_3 & e^{\ii\phi_1\mathcal{J}_1/2} & e^{\ii\phi_2\mathcal{J}_2/2} & e^{\ii\phi_3\mathcal{J}_3/2} & \mathcal{P} & \mathcal{T} & \mathcal{R}_i \\
    \hline
     \theta' & \pi-\theta & \theta & \pi-\theta & \pi-\theta & \theta'_{23}(\phi_1)=\theta & \theta'_{13}(\phi_2) & \theta'_{12}(\phi_3) & \pi-\theta & \theta & \theta \\
     \varphi' & \pi-\varphi & \pi-\varphi & \varphi & \pi-\varphi & \varphi'_{23}(\phi_1) & \varphi'_{13}(\phi_2) & \varphi'_{12}(\phi_3) & \pi-\varphi & \varphi & \varphi \\
     \psi' & \pi+\psi & \pi-\psi & \pi-\psi & \psi & \psi'_{23}(\phi_1) & \psi'_{13}(\phi_2) & \psi'_{12}(\phi_3)=\psi & \pi-\psi & -\psi & -\psi \\
     \hline
     m_1 & \times & \checkmark & \times & \times & \checkmark & \times & \times & \times & \checkmark & \checkmark \\
     m_2 & \times & \times & \checkmark & \times & \times & \checkmark & \times & \times & \checkmark & \checkmark \\
     m_3 & \times & \times & \times & \checkmark & \times & \times & \checkmark & \times & \checkmark & \checkmark \\
     m_4 & \times & \checkmark & \checkmark & \checkmark & \checkmark & \checkmark & \checkmark & \checkmark & \times & \times \\
    \end{array}\)
    \end{doublespace}
    \caption{The action of parity $\mathcal{P}$, time-reversion $\mathcal{T}$, reflection $\mathcal{R}_i$, continuous internal symmetry $e^{\ii\phi\mathcal{J}_i/2}$, its generator $\mathcal{J}_i$, and chiral symmetry operator $(-)^\chi$ on the $S^3$ mass manifold. $\checkmark$ means the mass manifold preserves the symmetry. $\times$ means the mass term breaks the symmetry and mass angle $\theta$ changes. $\theta'_{ij}(\phi_k)$ means the $\theta'$ angle after a $\phi_k$ rotation in the $m_i-m_j$ plane. Same for $\varphi'_{ij}(\phi_k),\psi'_{ij}(\phi_k)$.}
    \label{tab:CRT-mass_d=5}
\end{table}

\subsubsection{CRT-Internal Symmetry Reduction under Domain Wall}\label{sec:Maj_CPT_dom_wall}

By \textbf{domain wall reduction}, we can reduce a \textit{bulk} Majorana fermion to \textit{boundary} Majorana (or Majorana-Weyl) fermion in a lower dimension. Surprisingly, the CRT-internal symmetry group in different dimensions (see Tabs.~\ref{tab:CRT-internal_Maj}-\ref{tab:CRT-internal_Maj_Weyl}), though exhibit distinct fractionalization properties, is related by domain wall reduction. Once a mass term exist for $d+1$ spacetime dimensional Majorana theory, we can reproduce the CRT-internal symmetry group in $(d-1)+1$ dimension by projecting corresponding symmetry operators to the domain wall.

This method relies on the mass extension and works in spatial dimension $d=1,2,\dots,6\bmod 8$. Once a mass term can be extended, we can randomly add a mass and reduce the fermion to the mass domain wall in either direction (say $m\sim \pm x_d$). Note that a well-defined (i.e., not mixed with broken internal symmetries) reflection $\mathcal{R}_d$ is always preserved under the mass domain wall on the $d$-th direction, since the reflection simultaneously swaps the ground state in the $P_+$ and $P_-$ projection space, and flips the mass profile $m\sim \pm x_d\to\mp x_d$. Under domain wall reduction, $\mathcal{R}_d$ always becomes an internal symmetry on the domain wall. The reduction of CRT and internal symmetries follow the \textit{rules} below:

\begin{itemize}
    \item If the reflection $\mathcal{R}_i$, time-reversion $\mathcal{T}$, or internal symmetry $U$ is \textit{preserved under mass extension}, then these symmetries are directly projected to the $(d-1)$-dimensional CRT-internal symmetry by projection operator $P_{DW}=\dfrac{\mathbf{1}\pm \ii \beta  \alpha _1}{2}$:

    \bea\label{eq:rule1}
        d\text{-dimension} &\xrightarrow{DW}& (d-1)\text{-dimension}\nn\\
        \mathcal{R}_{d}  &\xrightarrow{P_{DW}}& X\nn\\
        \mathcal{R}_i  &\xrightarrow{P_{DW}}&  \mathcal{R}_i\ (\forall i=1,...,d-1)\nn\\
        \mathcal{T}  &\xrightarrow{P_{DW}}&  \mathcal{T}\nn\\
        U  &\xrightarrow{P_{DW}}&  U,
    \eea
    where $X$ is an internal symmetry in the $(d-1)$-dimensional theory.

    \item If the reflection $\mathcal{R}_i$, time-reversion $\mathcal{T}$, or internal symmetry $U$ is \textit{broken under mass extension}, then these symmetries should be combined with the space-orientation-reversing symmetry $\mathcal{R}_d\mathcal{T}$ to obtain a new symmetry on the domain wall~\cite{2312.17126Wan:2023nqe,10.21468/SciPostPhys.8.4.062,1910.14046,Wang:2019obe1910.14664}:

    \bea\label{eq:rule2}
        d\text{-dimension} &\xrightarrow{DW}& (d-1)\text{-dimension}\nn\\
        \mathcal{R}_{d}  &\xrightarrow{P_{DW}}& X\nn\\
        \mathcal{R}_i  &\xrightarrow{\cdot\mathcal{R}_d\mathcal{T}}&  X\mathcal{R}_i\mathcal{T}=\mathcal{R}_i'\mathcal{T}'\ (\forall i=1,...,d-1)\nn\\
        \mathcal{T}  &\xrightarrow{\cdot\mathcal{R}_d\mathcal{T}}&  X\nn\\
        U  &\xrightarrow{\cdot\mathcal{R}_d\mathcal{T}}&  XU\mathcal{T}=\mathcal{T}'.
    \eea
\end{itemize}

To be concrete, the CRT-internal symmetry groups in spacetime $d+1$ and $(d-1)+1$ dimension, and distinct mass terms are listed in Tab.~\ref{tab:sym_reduction&mass}. On each domain wall mass, the explicit result of the domain-wall projection for symmetry operators are listed in Tab.~\ref{tab:sym_operator_reduction}.

    \begin{table}[h]
    \centering
    \begin{doublespace}
    \noindent\(\begin{array}{c|cc|cccccc}
        d         & G_{M,d}     & G_{M/\cblue{MW},d-1}    & m_1   & m_2  & m_3  & m_4                 \\ \hline
        1         & \mathbb{D}_8^{\mathcal{T},\chi}\times\mathbb{Z}_2^{\mathcal{R}_1\mathcal{T}\chi}                    & \mathbb{Z}_2^F\times\mathbb{Z}_2^{\mathcal{T}}                                   & \chi^{\mathsf{T}}\sigma^2\chi &  &  &               \\
        2         & \mathbb{D}_8^{\mathcal{T},\mathcal{R}_1} & \cblue{\mathbb{Z}_2^F\times \mathbb{Z}_2^{\mathcal{R}_1 \mathcal{T}}} & \chi^{\mathsf{T}}\sigma^2\chi   & & &   \\
        3         & \text{Pin}_-^{\mathcal{T}}(2)\times_{\mathbb{Z}_2^F}\mathbb{Z}_4^{\mathcal{JR}_1\mathcal{T}}                    & \mathbb{D}_8^{\mathcal{T},\mathcal{R}_1}                                         & \chi^{\mathsf{T}}\sigma^{21}\chi & \chi^{\mathsf{T}}\sigma^{23}\chi & &     \\
        4         & \text{Spin}(3)\times_{\mathbb{Z}_2^F}\mathbb{D}_8^{\mathcal{J}_1\mathcal{R}_1,\mathcal{J}_1\mathcal{T}}                    & \text{Pin}_-^{\mathcal{T}}(2)\times_{\mathbb{Z}_2^F}\mathbb{Z}_4^{\mathcal{JR}_1\mathcal{T}}      & \chi^{\mathsf{T}}\sigma^{211}\chi & \chi^{\mathsf{T}}\sigma^{213}\chi & \chi^{\mathsf{T}}\sigma^{230}\chi  &    \\
        5         & \text{Spin}(4)\rtimes_{\mathbb{Z}_2^F}\mathbb{Z}_4^{\mathcal{T}}\times\mathbb{Z}_2^{\mathcal{R}_1\mathcal{T}\chi}                    & \text{Spin}(3)\times_{\mathbb{Z}_2^F}\mathbb{D}_8^{\mathcal{J}_1\mathcal{R}_1,\mathcal{J}_1\mathcal{T}}    & \chi^{\mathsf{T}}\sigma^{1002}\chi & \chi^{\mathsf{T}}\sigma^{1021}\chi & \chi^{\mathsf{T}}\sigma^{1023}\chi & \chi^{\mathsf{T}}\sigma^{2000}\chi   \\
        6         & \text{Spin}(3)\times_{\mathbb{Z}_2^F}\mathbb{D}_8^{\mathcal{T},\mathcal{R}_1} & \cblue{\text{Spin}(3)\times \mathbb{Z}_2^{\mathcal{R}_1 \mathcal{T}}}   & \chi^{\mathsf{T}}\sigma^{2000}\chi & & & \\
        7         & \text{Pin}_+^{\mathcal{T}}(2)\times\mathbb{Z}_4^{\mathcal{JR}_1\mathcal{T}}                    &     & & & & \\
        8         & \mathbb{D}_8^{\mathcal{R}_1,\mathcal{T}}                    &   & & & &  \\
    \end{array}\)
    \end{doublespace}
    \caption{The CRT-internal symmetry group for Majorana fermion in $d+1$ spacetime dimension $G_{M,d}$ can be reduced to the CRT-internal symmetry group for Majorana or Majorana-Weyl fermion in $(d-1)+1$ spacetime dimension $G_{M/MW,d-1}$ on the mass domain wall. The domain wall mass term $m_i$ can be chosen in the mass manifold. The CRT-invariant group includes fermion parity $\mathbb{Z}_2^F$, fermion chiral symmetry $\mathbb{Z}_2^\chi$, continuous symmetry generated by Lie algebra ($\mathcal{J}_i$), reflection $\mathbb{Z}_2^{\mathcal{R}_1}$, and time-reversal symmetry $\mathbb{Z}_2^{\mathcal{T}}$.}
    \label{tab:sym_reduction&mass}
    \end{table}

    \begin{table}[h]
    \centering
    \begin{doublespace}
    \noindent\(\begin{array}{c|c|cccccccc}
        d         & m_{DW} & P_{DW}((-)^F)     & P_{DW}((-)^\chi)    & P_{DW}(\mathcal{J}_i)   & P_{DW}(\mathcal{R}_{i<d})  & P_{DW}(\mathcal{R}_d)  & P_{DW}(\mathcal{T})                  \\ \hline
        1    & m_1     & (-)^F    & 0                                   &  \text{}            &      \text{}         &    1                                    &   \mathcal{T}                   \\
        2    & m_1     & (-)^F & \text{} & \text{} & \mathcal{R}_1\mathcal{T}       &  1          & 1        \\
        \multirow{2}{*}{3}   & m_1      & (-)^F   &   &  0 & \mathcal{R}_i  & 1   & \mathcal{T}  \\
              & m_2   & (-)^F   &   &  0 & \mathcal{R}_i  & 1  & \mathcal{T}  \\
        \multirow{3}{*}{4}   & m_1      & (-)^F   &   &  (0,0,\mathcal{J}) & \mathcal{R}_i  & \mathcal{J}   & \mathcal{T}  \\
              & m_2   & (-)^F   &   &  (0,\mathcal{J},0) & \mathcal{R}_i  & \mathcal{J}  & \mathcal{T}  \\
              & m_3   & (-)^F   &   &  (\mathcal{J},0,0) & \mathcal{R}_i  & 1  & \mathcal{T}  \\
        \multirow{4}{*}{5}   & m_1      & (-)^F   & 0  &  (\mathcal{J}_3,\mathcal{J}_2,\mathcal{J}_1) & \mathcal{R}_i  & \mathcal{J}_3   & \mathcal{T}  \\
              & m_2   & (-)^F   &  0 &  (\mathcal{J}_3,\mathcal{J}_2,\mathcal{J}_1) & \mathcal{R}_i  & \mathcal{J}_2  & \mathcal{T}  \\
              & m_3   & (-)^F   &  0 &  (\mathcal{J}_3,\mathcal{J}_1,\mathcal{J}_2) & \mathcal{R}_i  & \mathcal{J}_2  & \mathcal{T}  \\
              & m_4   & (-)^F   & 0  &  (\mathcal{J}_3,\mathcal{J}_2,\mathcal{J}_1) & \mathcal{R}_i  & 1  & \mathcal{T}  \\
        6     & m_1  & (-)^F & & (\mathcal{J}_3,\mathcal{J}_2,\mathcal{J}_1) & \mathcal{R}_i\mathcal{T} & 1 & 1 \\
    \end{array}\)
    \end{doublespace}
    \caption{The projected symmetry operators $P_{DW}(\cdot)$ with the domain wall mass $m_{DW}$ from spatial dimension $d$ to $d-1$. Symmetries include fermion parity $\mathbb{Z}_2^F$, fermion chiral symmetry $\mathbb{Z}_2^\chi$, continuous symmetry generated by Lie algebra ($\mathcal{J}_i$), reflection $\mathbb{Z}_2^{\mathcal{R}_i}$, and time-reversal symmetry $\mathbb{Z}_2^{\mathcal{T}}$.}
    \label{tab:sym_operator_reduction}
    \end{table}

\section{Dirac Fermion}\label{sec:Dirac}

Following the similar process in the discussion of Majorana fermion $\chi$, we'll move on to the complex Dirac fermion $\psi$ case, where other important symmetries, complex conjugation symmetry $\mathbb{Z}_2^{\mathcal{C}}$ and $U(1)$ symmetry, are well-defined and discussed. In Sec.~\ref{sec:Dirac_field_model}, we'll define the Dirac fermion as a complex Grassmannian field, acting as an irreducible representation of complex Clifford algebra, and we'll construct the field theory model for Dirac fermion. In Sec.~\ref{sec:Dirac_sym}, we'll introduce CRT and internal symmetries intrinsic to our minimal Dirac fermion model. The mass terms also form a non-trivial $S^1$ manifold in odd spatial dimensions as we'll give a discussion in Sec.~\ref{sec:Dirac_mass}. The existence of mass terms will break some internal symmetries, and we'll show that the CRT and internal symmetries together are enough to rule out all possible mass bilinear terms in Sec.~\ref{sec:Dirac_mass&sym}. Finally, in Sec.~\ref{sec:Dirac_CPT_dom_wall}, we'll use the domain wall reduction method to give the relation between symmetry groups in different dimensions.

\subsection{Field Theory Models}\label{sec:Dirac_field_model}

By combining real Majorana fermions (i.e. $\psi=(\chi_1+\ii\chi_2)/2$ and $\psi^\dagger=(\chi_1-\ii\chi_2)/2$), we arrive at the field theory model of the complex Dirac fermion $\psi$. To give a general field theory of the complex Dirac fermion in a generalized dimension, we duplicate it to form a spinor $\psi$ with multiple components (or flavors), and the Hamiltonian is the ``square root" of $\sum_i k_i^2+m_i^2$ analogous to the case in 3+1$d$ spacetime.

\subsubsection{Hamiltonian}\label{sec:Dirac_Ham}

A free \textbf{Dirac fermion} theory in \((d+1)\)-dimensional \textit{spacetime} with \(n\)-dimensional \textit{mass manifold} is described by
the field theory Hamiltonian
\begin{equation}
    H=\frac{1}{2}\int \dd^dx \psi^\dagger h\psi,
\end{equation}
where $h$ is the square root of $\sum_i k_i^2+m_i^2$ with $d+n$ anticommuting matrices $\alpha_i$ and $\beta_i$:

\bea\label{eq:Hamiltonian-Dirac}
h=\sum _{i=1}^d\alpha _i\ii\partial _i+\sum _{i=1}^n\beta _i m_i.
\eea

In general, the matrix \(h\) must satisfy

\begin{equation}
h=h^{\dagger},
\end{equation}
for the following reason:

\begin{itemize}
    \item \textit{Hermiticity} of the Hamiltonian: \(H=H^{\dagger }\). Since

\begin{equation}
H^{\dagger }=\frac{1}{2}\int\dd^dx\psi ^{\dagger }h^{\dagger } \psi  ,
\end{equation}

we must have \(h=h^{\dagger }\). 
\end{itemize}

Another significant property for Dirac fermion is the \textit{fermion} statistics: \(\psi _i\psi _j=-\psi _j\psi _i\). This gives a useful relation:

\begin{equation}\label{eq:fermion-statistics}
\frac{1}{2}\int\dd^dx\psi^\dagger h \psi =-\frac{1}{2}\int\dd^dx\psi^\mathsf{T}h^\mathsf{T} \psi^* .
\end{equation}

Now given the specific form of $h$ in \Eq{eq:Hamiltonian-Dirac}, in which momentum operators \(\ii\partial _i\) and real scalar mass terms \(m_i\) satisfy

\begin{equation}\label{eq:momentum-mass}
\ii\partial _i=-\left(\ii\partial _i\right)^\mathsf{T}=-\left(\ii\partial _i\right){}^*=\left(\ii\partial _i\right){}^{\dagger },\quad m_i=m_i^\mathsf{T}=m_i^*=m_i^{\dagger
},
\end{equation}
we must require

\begin{equation}
\alpha _i=\alpha _i^{\dagger },\quad \beta _i=\beta _i^{\dagger }.
\end{equation}
in order for the condition \Eq{eq:momentum-mass} to hold for $h$ in \Eq{eq:Hamiltonian-Dirac}.

In conclusion, the Hamiltonian \(H\) is fully specified by a \textbf{complex Clifford algebra} \(\mathcal{C}\ell (d+n)\), that defines \(\alpha _i\) and \( \beta _i\)
as its \textit{Hermitian generators}, satisfying:

\begin{itemize}
    \item \textit{Anticommutation} relations

\begin{equation}
\left\{\alpha _i,\alpha _j\right\}=2\delta _{i\, j},\quad \left\{ \beta _i, \beta _j\right\}=2\delta _{i\, j},\quad \left\{\alpha _i, \beta _j\right\}=0,
\end{equation}
\end{itemize}

\subsubsection{Complex Clifford Algebra}\label{sec:Dirac_Clif}

After defining the free Dirac fermion field theory specified by complex Clifford algebra $\mathcal{C}\ell(d+n)$, it's conducive for us to further analyze its algebraic structure and give explicit representation.

Complex Clifford algebras have the following recursive relation:

\begin{equation}\label{eq:Bott-complex}
    \mathcal{C}\ell(d+n)\cong \mathcal{C}\ell(d+n-2)\otimes_{\mathbb{C}}\mathbb{C}(2),
\end{equation}
which gives 2-fold Bott periodicity for complex Clifford algebra $\mathcal{C}\ell(d+n)$.

For further analytical derivation, it's intuitive to choose a specific representation for these matrices $\alpha_i$ and $\beta_i$ in the complex Clifford algebra. To be concrete, we may choose the following explicit representations in massless ($n=0$) and massive ($n=1$) cases that we're most interested in~\footnote{Intriguingly, we'll find that the CRT-internal symmetry exhibits 8-fold periodicity, which motivates us to write the explicit representation for $d=0,...,7$.}:

\begin{itemize}
    \item \(\mathcal{C}\ell (d)\) - \textbf{massless} Dirac fermions (boundary)

    The Dirac fermion on the boundary (or domain wall) is described by massless Dirac fermion Hamiltonian, specified by matrices $\alpha_i$ in complex Clifford algebra $\mathcal{C}\ell(d)$. The explicit representation for these matrices is given in Tab.~\ref{tab:Cl(d)}.
    
    \begin{table}[h]
        \centering
        \begin{doublespace}
        \noindent\(\begin{array}{c|ccccccc|ccc}
         d & \alpha _1 & \alpha _2 & \alpha _3 & \alpha _4 & \alpha _5 & \alpha _6 & \alpha _7 & \mathcal{C}\ell(d)  & \psi  & \text{dim}_{\mathbb{C}}\psi \\
        \hline
         0 &   &   &   &   &   &   &   & \mathbb{C}(1)  & 1_{\mathbb{C}} & 1  \\
         1 & \sigma ^3 &   &   &   &   &   &   & \mathbb{C}(1)\oplus \mathbb{C}(1)  & 1_{\mathbb{C}}^+\oplus 1_{\mathbb{C}}^- & 2 \\
         2 & \sigma ^1 & \sigma ^2 &   &   &   &   &   & \mathbb{C}(2)  & 2_{\mathbb{C}} & 2  \\
         3 & \sigma ^{01} & \sigma ^{02} & \sigma^{33}  &   &   &   &   & \mathbb{C}(2)\oplus \mathbb{C}(2)  & 2_{\mathbb{C}}^+\oplus 2_{\mathbb{C}}^- & 4  \\
         4 & \sigma^{01} & \sigma^{02} & \sigma^{13}  & \sigma^{23}  &   &   &   & \mathbb{C}(4)  & 4_{\mathbb{C}} &  4 \\
         5 & \sigma^{001} & \sigma^{002} & \sigma^{013}  & \sigma^{023}  & \sigma^{333}  &   &   & \mathbb{C}(4)\oplus \mathbb{C}(4)  & 4_{\mathbb{C}}^+\oplus 4_{\mathbb{C}}^- &  8 \\
         6 & \sigma^{001} & \sigma^{002} & \sigma^{013}  & \sigma^{023}  & \sigma^{133}& \sigma^{233}  &   & \mathbb{C}(8)  & 8_{\mathbb{C}} &  8 \\
         7 & \sigma^{0001} & \sigma^{0002} & \sigma^{0013}  & \sigma^{0023}  & \sigma^{0133}& \sigma^{0233}  & \sigma^{3333}  & \mathbb{C}(8)\oplus \mathbb{C}(8)  & 8_{\mathbb{C}}^+\oplus 8_{\mathbb{C}}^- & 16  \\
        \end{array}\)
        \end{doublespace}
        \caption{Explicit representation for $\mathcal{C}\ell(d)$ describing massless Dirac fermions on the boundary. dim$_{\mathbb{C}}\,\psi$ is the smallest flavor number of Dirac fermions $\psi_i$ we need to write down a Hamiltonian in Eq.~(\ref{eq:Hamiltonian-Dirac}), which can be calculated through the Clifford algebra structure.}
        \label{tab:Cl(d)}
    \end{table}

    \item \(\mathcal{C}\ell (d+1)\) - \textbf{massive} Dirac fermions (bulk)

    The Dirac fermion in the bulk is described by massive Dirac fermion Hamiltonian, specified by matrices $\alpha_i$ and $\beta_1$ in complex Clifford algebra $\mathcal{C}\ell(d+1)$. The explicit representation for these matrices is given in Tab.~\ref{tab:Cl(d+1)}.

    \begin{table}[h]
        \centering
        \begin{doublespace}
        \noindent\(\begin{array}{c|cccccccc|ccc}
         d & \alpha _1 & \alpha _2 & \alpha _3 & \alpha _4 & \alpha _5 & \alpha _6 & \alpha _7 & \beta_1 & \mathcal{C}\ell(d+1)  & \psi  & \text{dim}_{\mathbb{C}}\psi \\
        \hline
         0 &   &   &   &   &   &   &   &  \sigma ^3 & \mathbb{C}(1)\oplus \mathbb{C}(1)  & 1_{\mathbb{C}}^+\oplus 1_{\mathbb{C}}^- & 2 \\
         1 & \sigma ^1 &   &   &   &   &   &   & \sigma ^2  & \mathbb{C}(2)  & 2_{\mathbb{C}} & 2  \\
         2 & \sigma ^{01} & \sigma ^{02} &   &   &   &   &   & \sigma^{33}  & \mathbb{C}(2)\oplus \mathbb{C}(2)  & 2_{\mathbb{C}}^+\oplus 2_{\mathbb{C}}^- & 4  \\
         3 & \sigma^{01} & \sigma^{02} & \sigma^{13}  &   &   &   &   & \sigma^{23}  & \mathbb{C}(4)  & 4_{\mathbb{C}} &  4 \\
         4 & \sigma^{001} & \sigma^{002} & \sigma^{013}  & \sigma^{023}  &   &   &   & \sigma^{333}  & \mathbb{C}(4)\oplus \mathbb{C}(4)  & 4_{\mathbb{C}}^+\oplus 4_{\mathbb{C}}^- &  8 \\
         5 & \sigma^{001} & \sigma^{002} & \sigma^{013}  & \sigma^{023}  & \sigma^{133}&   &   & \sigma^{233}  & \mathbb{C}(8)  & 8_{\mathbb{C}} &  8 \\
         6 & \sigma^{0001} & \sigma^{0002} & \sigma^{0013}  & \sigma^{0023}  & \sigma^{0133}& \sigma^{0233}  &   & \sigma^{3333}  & \mathbb{C}(8)\oplus \mathbb{C}(8)  & 8_{\mathbb{C}}^+\oplus 8_{\mathbb{C}}^- & 16  \\
         7 & \sigma^{0001} & \sigma^{0002} & \sigma^{0013}  & \sigma^{0023}  & \sigma^{0133}& \sigma^{0233}  & \sigma^{1333}  &  \sigma^{2333} & \mathbb{C}(16)  & 16_{\mathbb{C}} & 16  \\
        \end{array}\)
        \end{doublespace}
        \caption{Explicit representation for $\mathcal{C}\ell(d+1)$ describing massive Dirac fermions in the bulk. dim$_{\mathbb{C}}\,\psi$ is the smallest flavor number of Dirac fermions $\psi_i$ we need to write down a Hamiltonian in Eq.~(\ref{eq:Hamiltonian-Dirac}), which can be calculated through the Clifford algebra structure.}
        \label{tab:Cl(d+1)}
    \end{table}
\end{itemize}

\subsubsection{Dirac Field}\label{sec:Dirac_field}

\(\psi\) is defined as a \textit{complex Grassmannian} (Dirac) field, acting as an \textbf{irreducible representation} of the Clifford algebra \(\mathcal{C}\ell
(d+n)\). Furthermore, we notice that when (and only when) \(d+n=1 \bmod 2\), the Clifford algebra \(\mathcal{C}\ell (d+n)\) \textit{splits}
into two isomorphic Cartan subalgebras, 

\begin{equation}
\mathcal{C}\ell (d+n)=\mathcal{C}\ell (d+n)^+\oplus \mathcal{C}\ell (d+n)^-.
\end{equation}

The subalgebras \(\mathcal{C}\ell (d+n)^{\pm }\) are split by the following projection operator defined by

\begin{equation}
\eta \text{:=}\prod _{i=1}^d\alpha _i\prod _{j=1}^n\beta _j=\pm 1,
\end{equation}
where \(\eta\) is the \textit{pseudo scalar} in \(\mathcal{C}\ell (d+n)\). In this case, the Dirac field \(\psi\) can be further projected into
the representation space of each subalgebra \(\mathcal{C}\ell (d+n)^{\pm }\),
\bea
\psi ^{\pm }\text{:=}\frac{1\pm \eta }{2}\psi ,
\eea
defined as the \textit{Weyl} fermion, which only occurs at \(d+n=1 \bmod 2\).

The component number of the Dirac fermion $\psi$ is calculated by the \textbf{complex representation dimension} \(\dim _{\mathbb{C}}\psi\), chosen to be the (minimal) dimension of a complex vector space in which the representation \(\psi\) can be faithfully embedded. The dimension is counted from the corresponding vector space of $\psi$ analogous to the Clifford algebra structure, following the rules:

\bea
\dim _{\mathbb{C}}2^k{}_{\mathbb{C}}&=&2^k,
\eea
where $n_{\mathbb{C}}$ denotes the complex vector space characterized by $n$ component vector filled with complex numbers.

The 2-fold \textbf{Bott periodicity} of the complex Clifford algebra \(\mathcal{C}\ell (d+n)\) (given in Eq.~(\ref{eq:Bott-complex})) directly implies the 2-fold periodicity in the corresponding vector space:

\begin{equation}
\psi _{\mathcal{C}\ell (d+n+2)}=\psi _{\mathcal{C}\ell (d+n)}\otimes 2_{\mathbb{C}}.
\end{equation}

Due to the Bott periodicity, it will be sufficient to enumerate complex Clifford algebras and their corresponding vector spaces for \(d=0,1\) in the massless Dirac fermion case and Weyl fermion case:

\begin{itemize}
    \item \textbf{Dirac fermion} (massless case, i.e. \(n=0\))

    The Dirac fermion on the boundary (or domain-wall) is represented in a vector space, where different component corresponds to different Dirac flavor (copy). The structure of the vector space corresponds to the Clifford algebra $\mathcal{C}\ell (d)$, and is listed in Tab.~\ref{tab:chi_Cl(d)}.
    
    \item \textbf{Weyl fermion} (massless case, i.e. \(n=0\))

    The Weyl fermion on the boundary (or domain-wall) is also represented in a vector space. The structure of the vector space corresponds to the Clifford subalgebra $\mathcal{C}\ell (d)^+$, and is listed in Tab.~\ref{tab:chi_Cl(d)+}.
    
    \begin{table}[htbp]
        \begin{minipage}{.5\columnwidth}
          \centering
            \begin{equation*}
            \begin{array}{c|ccc}
             d & \mathcal{C}\ell (d) & \psi  & \dim _{\mathbb{C}}\psi  \\
            \hline
             0 & \mathbb{C}(1) & 1_{\mathbb{C}} & 1 \\
             1 & \mathbb{C}(1)\oplus \mathbb{C}(1) & 1_{\mathbb{C}}^+\oplus 1_{\mathbb{C}}^- & 2 \\
            \end{array}
            \end{equation*}
            \caption{Clifford algebra $\mathcal{C}\ell(d)$, vector space of massless Dirac fermion $\psi$ and its complex representation dimension dim$_{\mathbb{C}}\,\psi$.}
            \label{tab:chi_Cl(d)}
        \end{minipage}
        \hfill
        \begin{minipage}{.4\columnwidth}
          \centering
            \begin{equation*}
            \begin{aligned}
            &\begin{array}{c|ccc}
             d & \mathcal{C}\ell (d)^+ & \psi ^+ & \dim _{\mathbb{C}}\psi ^+ \\
            \hline
             1 & \mathbb{C}(1)^+ & 1_{\mathbb{C}}^+ & 1 \\
            \end{array}
            \end{aligned}
            \end{equation*}
            \vspace{6pt}
      \caption{Clifford subalgebra $\mathcal{C}\ell(d)^+$, vector space of massless Weyl fermion $\psi^+$ and its complex representation dimension dim$_{\mathbb{C}}\,\psi^+$.}
      \label{tab:chi_Cl(d)+}
    \end{minipage}
\end{table}

\end{itemize}

\subsection{Symmetries}\label{sec:Dirac_sym}

In the previous sections, we've listed the explicit representation of the massless (boundary) Dirac fermions in Tab.~\ref{tab:Cl(d)}, and we can clearly observe that the $\alpha$ matrices cannot generate the complete $\mathbb{C}(\text{dim}_{\mathbb{C}}\psi_{\mathcal{C}\ell(d)})$ algebra in some dimensions. For example, for $d=1\bmod2$, $\alpha$ matrices span a diagonal matrix space with chiral symmetry operator $(-)^{\chi}=\sigma^3$. These symmetries are called internal symmetries for the Clifford algebra, as we'll define more strictly later. Apart from internal symmetries, we'll also include the most well-known symmetries for physicists: Lorentz symmetry and CRT symmetries.

The Lorentz symmetry of Dirac fermion aligns perfectly with that of Majorana fermion discussed in Sec.~\ref{sec:Lorentz}, so we'll start our discussion from the internal symmetries below.

\subsubsection{Internal Symmetry}

The \textbf{internal symmetry} of the Dirac fermion corresponds to the \textit{invariant group} \(G(\mathcal{C}\ell (d+n))\) of the Clifford algebra
\(\mathcal{C}\ell (d+n)\), defined by the \textit{short exact sequence}:
\bea\label{eq:invariant-group-Dirac}
1\to G(\mathcal{C}\ell (d+n))\to \U\left(\dim _{\mathbb{C}}\psi \right)\to\text{Aut}(\mathcal{C}\ell (d+n))\to1,
\eea
where

\begin{itemize}
    \item \(\U\left(\dim _{\mathbb{C}}\psi \right)\) - the \textbf{maximal} \textbf{unitary group} of \(\psi\) preserving its anticommutation relation \(\{\psi
    ,\psi^\dagger\}=\mathbf{1}\).
    
    \item \(\text{Aut}(\mathcal{C}\ell (d+n))\) - the \textbf{automorphism group} of \(\mathcal{C}\ell (d+n)\). Each automorphism is induced by a \textit{group
    conjugation} for \(g\in \U\left(\dim _{\mathbb{C}}\psi \right)\) and \(h\in \mathcal{C}\ell (d+n)\),

    \begin{equation}
    h\rightarrow g^{-1}h g.
    \end{equation}
\end{itemize}

The short exact sequence \Eq{eq:invariant-group-Dirac} indicates that \(G(\mathcal{C}\ell (d+n))\) is the \textbf{normal subgroup} of \(\U\left(\dim _{\mathbb{C}}\psi \right)\)
that leaves \(\mathcal{C}\ell (d+n)\) invariant.

\begin{itemize}
    \item \textbf{Dirac fermion} (massless case, i.e. \(n=0\))

    The internal symmetry of massless Dirac fermion is listed in Tab.~\ref{tab:internal_Dirac}, which includes vector $\U(1)^F$ and axial $\U(1)^\chi$ symmetries.

    \item \textbf{Weyl fermion} (massless case, i.e. \(n=0\))

    The internal symmetry of massless Weyl fermion is listed in Tab.~\ref{tab:internal_Weyl}, which includes vector $\U(1)^F$ symmetry.
    
    \begin{table}[htbp]
        \begin{minipage}{.47\columnwidth}
          \centering
            \begin{doublespace}
            \noindent\(\begin{array}{c|cc|cc}
             \, d & \mathcal{C}\ell (d) & G(\mathcal{C}\ell (d)) & \U(1)^F & \U(1)^{\chi } \\
            \hline
             0 & \mathbb{C}(1) & \U(1) & \mathcal{Q} & \text{} \\
             1 & \mathbb{C}(1)\oplus \mathbb{C}(1) & \U(1)\times \U(1) & \mathcal{Q}\sigma ^0 & \mathcal{Q}\sigma ^3 \\
            \end{array}\)
            \end{doublespace}
            \caption{The generators of the internal symmetries including vector $\U(1)^F$ and axial $\U(1)^\chi$ for massless Dirac fermion. $\mathcal{Q}$ is the charge operator defined as $\mathcal{Q}\psi=\psi$ and $\mathcal{Q}\psi^*=-\psi^*$.}
            \label{tab:internal_Dirac}
        \end{minipage}
        \hfill
        \begin{minipage}{.47\columnwidth}
          \centering
            \begin{doublespace}
            \noindent\(\begin{array}{c|cc|c}
             \, d & \mathcal{C}\ell (d)^+ & G\left(\mathcal{C}\ell (d)^+\right) & \U(1)^F \\
            \hline
             1 & \mathbb{C}(1)^+ & \U(1) & \mathcal{Q} \\
            \end{array}\)
            \end{doublespace}
      \caption{The generator of the internal vector $\U(1)^F$ symmetry for massless Weyl fermion. $\mathcal{Q}$ is the charge operator defined as $\mathcal{Q}\psi=\psi$ and $\mathcal{Q}\psi^*=-\psi^*$.}
      \label{tab:internal_Weyl}
    \end{minipage}
\end{table}

\end{itemize}

\subsubsection{CRT Symmetry}

For Dirac fermions, the charge conjugation \(\mathcal{C}\), the reflection \(\mathcal{R}_i\), and the time reversal \(\mathcal{T}\)
symmetry can be discussed.

We first define charge conjugation $\mathcal{C}$~\cite{Zirnbauer2004.07107, wigner2012group, Freed1604.06527} which transforms a Dirac fermion $\psi$ to its complex conjugate $\psi^*$. If any operator $U$ acts as $U\psi U^{-1}=M_{U}\psi$, then the action becomes $U\psi^* U^{-1}=M_{U}^*\psi^*$ in the complex conjugate space.~\footnote{See more strict definition as bundle map in Ref.~\cite{2312.17126Wan:2023nqe}.}

\textbf{Charge conjugation} \(\mathcal{C}\): a \textit{unitary} symmetry, acting as 
\bea\label{eq:charge-conjugation-Dirac}
\mathcal{C} \partial _i \mathcal{C}^{-1}&=&\partial _i\ (\text{for } i=1,\ldots ,d),\nn\\
\mathcal{C} \psi  \mathcal{C}^{-1}&=&\mathcal{K}_{\mathcal{C}}M_{\mathcal{C}}\psi.
\eea
$\mathcal{K}_{\mathcal{C}}$ is a complex conjugation operator on Dirac fermion, sending $\psi$ to its complex conjugate $\psi^*$ and vice versa. \(M_{\mathcal{C}}\in \U\left(\dim _{\mathbb{C}}\psi \right)\) must be in the \textit{maximal unitary group} of \(\psi\).

The massless Hamiltonian in \Eq{eq:Hamiltonian-Dirac} transforms under the charge conjugation $\mathcal{C}$ as:

\bea
    \frac{1}{2}\int\dd^dx\psi^{\dagger}h\psi
    &\to&\frac{1}{2}\int\dd^dx(M_{\mathcal{C}}\psi^*)^{\dagger}h (M_{\mathcal{C}}\psi^*)\nn\\
    &=&\frac{1}{2}\int\dd^dx\psi^{\mathsf{T}}(M_{\mathcal{C}}^{\dagger}h M_{\mathcal{C}})\psi^*\nn\\
    &=&-\frac{1}{2}\int\dd^dx\psi^{\dagger}(M_{\mathcal{C}}^{\dagger}h M_{\mathcal{C}})^{\mathsf{T}}\psi,
\eea
where the last step uses the property in \Eq{eq:fermion-statistics}.

To keep the Hamiltonian invariant, $h$ should transform under $M_{\mathcal{P}}$ as

\begin{equation}
    M_{\mathcal{C}}^{\dagger}hM_{\mathcal{C}}=-h^{\mathsf{T}},
\end{equation}
which indicates that $M_{\mathcal{C}}$ should act on the Clifford
algebra as

\begin{equation}\label{eq:charge-conjugation}
M_{\mathcal{C}}^\dagger\alpha _iM_{\mathcal{C}}=\alpha _i^{\mathsf{T}}\ (\text{for } i=1,\ldots ,d).
\end{equation}

To define reflection $\mathcal{R}_i$, it's convenient for us first to define parity $\mathcal{P}$ and then use it to define reflections $\mathcal{R}_i$ in different directions:

\textbf{Parity} \(\mathcal{P}\): a \textit{unitary} symmetry, acting as 
\bea\label{eq:parity-Dirac}
\mathcal{P} \partial _i \mathcal{P}^{-1}&=&-\partial _i\ (\text{for } i=1,\ldots ,d),\nn\\
\mathcal{P} \psi  \mathcal{P}^{-1}&=&M_{\mathcal{P}}\psi .
\eea
\(M_{\mathcal{P}}\in \U\left(\dim _{\mathbb{C}}\psi \right)\) must be in the \textit{maximal unitary group} of \(\psi\).

The massless Hamiltonian in \Eq{eq:Hamiltonian-Dirac} changes under the parity transformation $\mathcal{P}$ as:

\bea
    \frac{1}{2}\int\dd^dx\psi^{\dagger}h\psi
    &\to&-\frac{1}{2}\int\dd^dx(M_{\mathcal{P}}\psi)^{\dagger}h (M_{\mathcal{P}}\psi)\nn\\
    &=&-\frac{1}{2}\int\dd^dx\psi^{\dagger}(M_{\mathcal{P}}^{\dagger}h M_{\mathcal{P}})\psi.
\eea

To keep the Hamiltonian invariant, $h$ should transform under $M_{\mathcal{P}}$ as

\begin{equation}
    M_{\mathcal{P}}^{\dagger}hM_{\mathcal{P}}=-h,
\end{equation}
which indicates that $M_{\mathcal{P}}$ should act on the Clifford
algebra as

\begin{equation}\label{eq:parity1-Dirac}
M_{\mathcal{P}}^\dagger\alpha _iM_{\mathcal{P}}=-\alpha _i\ (\text{for } i=1,\ldots ,d).
\end{equation}

\textbf{Reflection} \(\mathcal{R}_i\): a \textit{unitary} symmetry, acting as 
\bea
\mathcal{R}_i\partial _j\mathcal{R}_i^{-1}&=&
\left\{
\begin{array}{ll}
 -\partial _i & j=i, \\
 \partial _j & j\neq i, \\
\end{array}
 \right.
\nn\\
\mathcal{R}_i\psi  \mathcal{R}_i^{-1}&=&\alpha _iM_{\mathcal{P}}\psi .
\eea
Given \Eq{eq:parity1-Dirac}, one can easily prove that \(\alpha _iM_{\mathcal{P}}\) is also a unitary operator and acts on the Clifford algebra as expected:

\begin{equation}\label{eq:parity2-Dirac}
\left(\alpha _iM_{\mathcal{P}}\right)^\dagger\alpha _j \left(\alpha _iM_{\mathcal{P}}\right)=
\left\{
\begin{array}{ll}
 -\alpha _i & j=i, \\
 \alpha _j & j\ne i. \\
\end{array}
\right.
\end{equation}

Under this construction, we always have \(\forall i:\mathcal{R}_i^2=(-)^F\mathcal{P}^2\), where $F$ denotes the fermion number and is even for boson ($\partial_\mu$).

Similarly, time-reversal $\mathcal{T}$ can be defined as:

\textbf{Time reversal} \(\mathcal{T}\): an \textit{antiunitary} symmetry, acting as
\bea\label{eq:time-reversal-Dirac}
\mathcal{T} \ii \mathcal{T}^{-1}&=&-\ii,\nn\\
\mathcal{T} \psi  \mathcal{T}^{-1}&=&\mathcal{K} M_{\mathcal{T}}\psi .
\eea
$\mathcal{K}$ is the complex conjugation operator. \(M_{\mathcal{T}}\in \U\left(\dim _{\mathbb{C}}\psi \right)\) must be in the \textit{maximal unitary group} of \(\psi\).

The massless Hamiltonian in \Eq{eq:Hamiltonian-Dirac} transforms under the time-reversion $\mathcal{T}$ as:

\bea
    \frac{1}{2}\int\dd^dx\psi^{\dagger}h\psi
    &\to&\frac{1}{2}\int\dd^dx(M_{\mathcal{T}}\psi)^{\dagger}h^* (M_{\mathcal{T}}\psi)\nn\\
    &=&\frac{1}{2}\int\dd^dx\psi^{\dagger}(M_{\mathcal{T}}^{\dagger}h^* M_{\mathcal{T}})\psi.
\eea

To keep the Hamiltonian invariant, $h$ should transform under $M_{\mathcal{T}}$ as

\begin{equation}
    M_{\mathcal{T}}^{\dagger}hM_{\mathcal{T}}=h^*,
\end{equation}
which indicates that $M_{\mathcal{C}}$ should act on the Clifford
algebra as

\begin{equation}\label{eq:time-reversal1-Dirac}
M_{\mathcal{T}}^\dagger\alpha _iM_{\mathcal{T}}=-\alpha _i^*\ (\text{for } i=1,\ldots ,d).
\end{equation}

To find explicit representations for matrices \(M_{\mathcal{C}}\), \(M_{\mathcal{P}}\), and \(M_{\mathcal{T}}\), we notice that the choices are ambiguous up to internal symmetry transformations \(g_{\mathcal{C}},g_{\mathcal{P}},g_{\mathcal{T}}\in
G(\mathcal{C}\ell (d))\),

\begin{equation}
M_{\mathcal{C}}\rightarrow g_{\mathcal{C}}M_{\mathcal{C}},\quad 
M_{\mathcal{P}}\rightarrow g_{\mathcal{P}}M_{\mathcal{P}},\quad M_{\mathcal{T}}\rightarrow g_{\mathcal{T}}M_{\mathcal{T}}.
\end{equation}

To give further constraints on the representation, it's intuitive to assume \textbf{canonical CRT} conditions~\cite{10.21468/SciPostPhys.8.4.062} given by:

\begin{equation}
    (\mathcal{CR}_i\mathcal{T})^2=1,\quad \mathcal{C}(\mathcal{CR}_i\mathcal{T})=(\mathcal{CR}_i\mathcal{T})\mathcal{C},\quad \mathcal{T}(\mathcal{CR}_i\mathcal{T})=(-)^F(\mathcal{CR}_i\mathcal{T})\mathcal{T}\ (\text{for } i=1,\ldots ,d).
\end{equation}

To realize these conditions, one convenient \textit{choice} is

 \begin{equation}
 M_{\mathcal{P}}=M_{\mathcal{C}}M_{\mathcal{T}}.
 \end{equation}

 Under this choice, we always have \(\mathcal{P}^2=(\mathcal{C}\mathcal{T})^2\). In conclusion, for Dirac fermions, we can consistently assume the following:
\bea
&\mathcal{P}^2=(\mathcal{C}\mathcal{T})^2=(-)^F\mathcal{R}_i^2,\quad (\mathcal{C}\mathcal{R}_i \mathcal{T})^2=1,\nn\\
&\mathcal{C}(\mathcal{CR}_i\mathcal{T})=(\mathcal{CR}_i\mathcal{T})\mathcal{C},\quad \mathcal{T}(\mathcal{CR}_i\mathcal{T})=(-)^F(\mathcal{CR}_i\mathcal{T})\mathcal{T}\ (\text{for } i=1,\ldots ,d).
\eea

Intuitively, one may assume that the invariant group is 2-fold periodic parallel to the Bott periodicity of the complex Clifford algebra. Intriguingly the CRT-internal symmetry group actually exhibits 8-fold periodicity as in the Majorana case due to the canonical constraints. The corresponding CRT and internal symmetries for Dirac and Weyl fermions are summarized in Tabs.~\ref{tab:CRT-internal_Dirac} and \ref{tab:CRT-internal_Weyl}, where we've chosen a specific direction for reflection $\mathcal{R}_1$. Other reflections can be generated through rotation in the Lorentz symmetry group, which we've not included in $G_{\text{CRTinternal}}$ for brevity.

\begin{itemize}
    \item \textbf{Dirac fermion} (massless case, i.e. \(n=0\))

    The internal symmetries $\U(1)^F$ and $\U(1)^\chi$, along with CRT symmetries generate the invariant group $G_{\text{CRTinternal}}$ for Dirac fermions, which is independent of explicit representation basis. [See Tab.~\ref{tab:CRT-internal_Dirac}] The invariant group $G_{\text{CRTinternal}}$ is complicated, and the detailed presentation is collected in App.~\ref{app:presentation}.

    \begin{table}[h]
        \centering
        \begin{doublespace}
        \resizebox{\textwidth}{!}{
        \noindent\(\begin{array}{c|ccc|ccccccc}
         \, d & \mathcal{C}\ell (d) & G_{\text{CRT}} & G_{\text{CRT}\text{internal}} & \U(1)^F & \U(1)^{\chi } & \mathbb{Z}_2^{\mathcal{C}} & \mathbb{Z}_2^{\mathcal{P}} &  \mathbb{Z}_2^{\mathcal{R}_1} 
        & \mathbb{Z}_2^{\mathcal{T}} & \mathbb{Z}_2^{\mathcal{C}\mathcal{R}_1\mathcal{T}}  \\
        \hline
         0 & \mathbb{C}(1) &\mathbb{Z}_2^\mathcal{C}\times\mathbb{Z}_2^\mathcal{T}  &\U(1)^F\rtimes G_{\text{CRT}}  & \mathcal{Q} & \text{} & \mathcal{K}_{\mathcal{C}} &  \text{}  &  \text{}  & \mathcal{K}\text{} & \mathcal{K}_{\mathcal{C}}\mathcal{K} \\
         1 & \mathbb{C}(1)\oplus \mathbb{C}(1) & \mathbb{D}_8^{\mathcal{CT,C}}\times\mathbb{Z}_2^{\mathcal{R}_1\mathcal{T}} & (\U(1)^F\times \U(1)^\chi)\rtimes_{\mathbb{Z}_2^F} G_{\text{CRT}} & \mathcal{Q}\sigma ^0 & \mathcal{Q}\sigma ^3
         & \mathcal{K}_{\mathcal{C}}\sigma^3  & \ii\sigma^2 & \sigma^1 & \mathcal{K} \sigma ^1 & \mathcal{K}_{\mathcal{C}}\mathcal{K}\sigma^3 \\
         2 & \mathbb{C}(2) & \mathbb{D}_8^{\mathcal{T,R}_1}\times\mathbb{Z}_2^{\mathcal{C}} & \U(1)^F\rtimes_{\mathbb{Z}_2^F} G_{\text{CRT}} & \mathcal{Q}\sigma ^0 & \text{} & \mathcal{K}_{\mathcal{C}}\sigma^1 &     \ii\sigma^3 &     \sigma^2 & \mathcal{K} \sigma ^2 & \mathcal{K}_{\mathcal{C}}\mathcal{K}\sigma^1\\
         3 & \mathbb{C}(2)\oplus \mathbb{C}(2) & \mathbb{D}_8^{\mathcal{T,R}_1}\times\mathbb{Z}_2^{\mathcal{C}} & (\U(1)^F\times \U(1)^\chi)\rtimes_{\mathbb{Z}_2^F} G_{\text{CRT}} & \mathcal{Q}\sigma ^{00} & \mathcal{Q}\sigma^{30} & \mathcal{K}_{\mathcal{C}}\sigma^{11} &    \ii\sigma^{13} &    \sigma^{12} & \mathcal{K} \sigma ^{02} & \mathcal{K}_{\mathcal{C}}\mathcal{K}\sigma^{01}\\
         4 & \mathbb{C}(4) & \mathbb{D}_8^{\mathcal{T},\mathcal{CR}_1}\times_{\mathbb{Z}_2^F}\mathbb{Z}_4^{\mathcal{C}F} & \U(1)^F\rtimes_{\mathbb{Z}_2^F} G_{\text{CRT}} & \mathcal{Q}\sigma ^{00} &  & \mathcal{K}_{\mathcal{C}}\sigma^{21} &    \sigma^{33} &    \ii\sigma^{32} & \mathcal{K} \sigma ^{12} & \mathcal{K}_{\mathcal{C}}\mathcal{K}\sigma^{01} \\
         5 & \mathbb{C}(4)\oplus \mathbb{C}(4) & \mathbb{D}_8^{\mathcal{R}_1,\mathcal{CR}_1\mathcal{T}}\times_{\mathbb{Z}_2^F}\mathbb{Z}_4^{\mathcal{C}F} & (\U(1)^F\times \U(1)^\chi)\rtimes_{\mathbb{Z}_2^F} G_{\text{CRT}} & \mathcal{Q}\sigma ^{000} &  \mathcal{Q}\sigma^{300} & \mathcal{K}_{\mathcal{C}}\sigma^{021} &   \sigma ^{133}  &   \ii\sigma ^{132}  & \mathcal{K} \sigma ^{112} & \mathcal{K}_{\mathcal{C}}\mathcal{K}\sigma ^{001} \\
           6 & \mathbb{C}(8) & \mathbb{D}_8^{\mathcal{CR}_1,\mathcal{T}}\times_{\mathbb{Z}_2^F}\mathbb{Z}_4^{\mathcal{C}F} & \U(1)^F\rtimes_{\mathbb{Z}_2^F} G_{\text{CRT}} & \mathcal{Q}\sigma ^{000} &   & \mathcal{K}_{\mathcal{C}}\sigma^{121} &   \ii\sigma ^{333}  &   \sigma ^{332}  & \mathcal{K} \sigma ^{212} & \mathcal{K}_{\mathcal{C}}\mathcal{K}\sigma ^{001} \\
           7 & \mathbb{C}(8)\oplus \mathbb{C}(8) & \mathbb{D}_8^{\mathcal{CR}_1,\mathcal{T}}\times_{\mathbb{Z}_2^F}\mathbb{Z}_4^{\mathcal{C}F} & (\U(1)^F\times \U(1)^\chi)\rtimes_{\mathbb{Z}_2^F} G_{\text{CRT}} & \mathcal{Q}\sigma ^{0000} &  \mathcal{Q}\sigma^{3000} & \mathcal{K}_{\mathcal{C}}\sigma^{1121} &   \ii\sigma ^{1333}  &   \sigma ^{1332}  & \mathcal{K} \sigma ^{0212} & \mathcal{K}_{\mathcal{C}}\mathcal{K}\sigma ^{0001} \\
            8 & \mathbb{C}(16) & \mathbb{D}_8^{\mathcal{CR}_1,\mathcal{CT}}\times\mathbb{Z}_2^\mathcal{C} & \U(1)^F\rtimes_{\mathbb{Z}_2^F} G_{\text{CRT}} & \mathcal{Q}\sigma ^{0000} &   & \mathcal{K}_{\mathcal{C}}\sigma^{2121} &   \sigma ^{3333}  &   \ii\sigma ^{3332}  & \mathcal{K} \sigma ^{1212} & \mathcal{K}_{\mathcal{C}}\mathcal{K}\sigma ^{0001} \\
        \end{array}\)}
        \end{doublespace}
        \caption{The invariant group of massless Dirac fermion in different dimensions, including vector $\U(1)^F$ symmetry, axial $\U(1)^\chi$ symmetry, charge conjugation $\mathbb{Z}_2^{\mathcal{C}}$, reflection $\mathbb{Z}_2^{\mathcal{R}_1}$, and time-reversal symmetry $\mathbb{Z}_2^{\mathcal{T}}$.}
        \label{tab:CRT-internal_Dirac}
    \end{table}

    \item \textbf{Weyl fermion} (massless case, i.e. \(n=0\))

    The internal symmetry $\U(1)^F$, along with CRT symmetries generates the invariant group $G_{\text{CRTinternal}}$ for Weyl fermions, which is independent of explicit representation basis. [See Tab.~\ref{tab:CRT-internal_Weyl}] The invariant group $G_{\text{CRTinternal}}$ is complicated, and the detailed presentation is demonstrated in App.~\ref{app:presentation_W}.

    \begin{table}[h]
        \centering
        \begin{doublespace}
        \noindent\(\begin{array}{c|ccc|cccc}
         \, d & \mathcal{C}\ell (d)^+ & G_{\text{CRT}} & G_{\text{CRT}\text{internal}} & \U(1)^F &  \mathbb{Z}_2^{\mathcal{C}} & \mathbb{Z}_2^{\mathcal{T}} & \mathbb{Z}_2^{\mathcal{CR}_1\mathcal{T}}
        \\
        \hline
         1 & \mathbb{C}(1)^+ & \mathbb{Z}_2^{\mathcal{C}}\times\mathbb{Z}_2^{\mathcal{CR}_1\mathcal{T}}  & \U(1)^F\rtimes G_{\text{CRT}}  & \mathcal{Q} & \mathcal{K}_{\mathcal{C}} & & \mathcal{K}_{\mathcal{C}}\mathcal{K} \\
        3 & \mathbb{C}(2)^+ & \mathbb{Z}_4^{\mathcal{T}}\times\mathbb{Z}_2^{\mathcal{CR}_1\mathcal{T}} & \U(1)^F\rtimes_{\mathbb{Z}_2^F} G_{\text{CRT}}  & \mathcal{Q}\sigma^0 & & \mathcal{K}\sigma^2 & \mathcal{K}_{\mathcal{C}}\mathcal{K}\sigma^1 \\
         5 & \mathbb{C}(4)^+ & \mathbb{D}_8^{\mathcal{C},\mathcal{CR}_1\mathcal{T}} & \U(1)^F\rtimes_{\mathbb{Z}_2^F} G_{\text{CRT}}  & \mathcal{Q}\sigma^{00} & \mathcal{K}_{\mathcal{C}}\sigma^{21} & & \mathcal{K}_{\mathcal{C}}\mathcal{K}\sigma^{01} \\
        7 & \mathbb{C}(8)^+ & \mathbb{D}_8^{\mathcal{CR}_1,\mathcal{T}} & \U(1)^F\rtimes_{\mathbb{Z}_2^F} G_{\text{CRT}}  & \mathcal{Q}\sigma^{000} & & \mathcal{K}\sigma^{212} & \mathcal{K}_{\mathcal{C}}\mathcal{K}\sigma^{001} \\
        \end{array}\)
        \end{doublespace}
        \caption{The invariant group of massless Weyl fermion in different dimensions, including vector $\U(1)^F$ symmetry, charge conjugation $\mathbb{Z}_2^{\mathcal{C}}$, time-reversal symmetry $\mathbb{Z}_2^{\mathcal{T}}$, and combined $\mathbb{Z}_2^{\mathcal{CR}_1\mathcal{T}}$ symmetry.}
        \label{tab:CRT-internal_Weyl}
    \end{table}

\end{itemize}

Under the canonical CRT condition, the choice of $M_{\mathcal{C}}$, $M_{\mathcal{P}}$, $M_{\mathcal{T}}$ is still ambiguous up to internal symmetry transformations. It's proved in Ref.~\cite{2312.17126Wan:2023nqe} that $\mathcal{P}^2$ can be either $1$ or $(-)^F$ in each dimension, and $\mathcal{C}^2$, $\mathcal{T}^2$ is fixed by canonical conditions.

\subsection{Mass}\label{sec:Dirac_mass}

After carefully examining the Clifford algebra theory for massless Dirac fermion, we'll step forward to massive theory by extending mass terms. In this section, we'll discuss mass extensions and mass domain wall reductions.

\subsubsection{Mass Extension and Mass Manifold}

Given a \textit{massless} Dirac fermion theory specified by \(\mathcal{C}\ell (d)\), the \textbf{mass extension} concerns the ability to add
mass terms to the theory without enlarging the representation dimension of \(\psi\).
\bea
\mathcal{C}\ell (d)\rightarrow \mathcal{C}\ell (d+1)^{(+)}\rightarrow \ldots .\rightarrow \mathcal{C}\ell (d+n)^{(+)}.
\eea

The mass extension for Dirac fermion is demonstrated in Tab.~\ref{tab:mass-manifold_Dirac}(a). There are two possible extensions:

\begin{itemize}
    \item \textbf{Regular mass extension}: if

    \begin{equation}
    \dim _{\mathbb{C}}\psi_{\mathcal{C}\ell (d+n)}=\dim _{\mathbb{C}}\psi_{\mathcal{C}\ell (d+n+1)},
    \end{equation}
    a mass term can be added directly.

    \item \textbf{Chiral mass extension}: if \(\mathcal{C}\ell (d+n+1)\cong \mathcal{C}\ell (d+n+1)^+\oplus \mathcal{C}\ell (d+n+1)^-\) splits and

    \begin{equation}
    \dim _{\mathbb{C}}\psi_{\mathcal{C}\ell (d+n)}=\dim _{\mathbb{C}}\psi^{\pm }_{\mathcal{C}\ell (d+n+1)^{\pm }}=\frac{1}{2}\dim _{\mathbb{C}}\psi_{\mathcal{C}\ell
    (d+n+1)}.
    \end{equation}
    a mass term can be added by promoting the \textbf{Dirac} fermion to a \textbf{Weyl} fermion in one of the chiral subalgebras (say \(\mathcal{C}\ell
    (d+n+1)^+\)). No further mass can be added for a Weyl fermion, so the chiral mass extension is always the \textit{end} of a mass extension
    sequence.

\end{itemize}

The mass extension process can also shown by finding explicit mass extensions for \textbf{Dirac fermions}, as demonstrated in Tab.~\ref{tab:mass-manifold_Dirac}. These masses span an $(n-1)-$dimensional \textbf{mass manifold}, formulated as a Grassmannian manifold:

\begin{equation}
    M_d=\frac{G(\mathcal{C}\ell(d))}{G(\mathcal{C}\ell(d+1)^{(+)})}.
\end{equation}

\begin{table}[htbp]
    \begin{minipage}{.44\columnwidth}
      \centering
        \begin{tikzpicture}
        \matrix[matrix of math nodes,inner sep=1pt,row sep=1em,column sep=1em] (M)
        {
        {\white\backslash}d{\white\backslash} & \mathcal{C}\ell(d) & \mathcal{C}\ell(d+1) & {\white\backslash}\mathcal{C}\ell(d+2){\white\backslash}  \\
        0 & \C(1) & \C(1)\oplus\C(1) & \C(2) \\
        {\white\backslash}1{\white\backslash} & \C(1)\oplus\C(1) & \C(2) & \C(2)\oplus\C(2) \\
        }
        ;
        
        \draw (M-1-1.south west) -- (M-1-4.south east);
        \draw (M-1-1.north east) -- (M-3-1.south east);
        
        \draw[->,color=green] (M-3-2.east) -- (M-3-3.west);
        \draw[->,color=red] (M-3-3.east) -- (M-3-4.west);

        \draw[->,color=red] (M-2-2.east) -- (M-2-3.west);
        
        \end{tikzpicture}
        \caption{Mass extension. The green arrow means regular mass extension. The red arrow means chiral mass extension.}
        \label{fig:mass-extension-Dirac}
    \end{minipage}
    \hfill
    \begin{minipage}{.55\columnwidth}
      \centering
        \begin{doublespace}
        \noindent\(\begin{array}{c|c|cc|cc|c}
         \, d & \alpha _1  & \beta _1 & \beta _2 & G(\mathcal{C}\ell(d)) & G(\mathcal{C}\ell(d+1)^{(+)} & M_d  \\
        \hline
         0 & \text{}  & \sigma^3 & \text{} & \U(1) & \U(1) & 1 \\
         1 & \sigma ^3  & \sigma ^1 & \sigma^2 & \U(1)\times\U(1) & \U(1) & \frac{\U(1)\times\U(1)}{\U(1)}\cong S^1 \\
        \end{array}\)
        \end{doublespace}
  \caption{Explicit mass extension and the mass manifold $M_d$.}
  \label{tab:mass-manifold_Dirac}
\end{minipage}
\end{table}

For \textbf{Dirac fermions}, there are two cases for the manifold $M_d$:

$M_d=1:$ only one mass $m$, and $m=\pm 1$ belongs to different topological order.

$M_d=S^1$: admits a mass vector $\bm$ of 2 components, transforming under the internal and CRT symmetries as an $\rO(2)$ vector.

\subsubsection{Domain Wall Reduction}

\textbf{Domain wall reduction} reduces a massive (bulk) fermion to the mass domain wall as a massless (boundary) fermion in one lower dimension. This
simultaneously removes a momentum term and a mass term, corresponding to
\bea
\mathcal{C}\ell (d+1)\rightarrow \mathcal{C}\ell (d-1).
\eea
Note that \(\mathcal{C}\ell (d+1)\cong \mathcal{C}\ell (d-1)\otimes _{\mathbb{C}}\mathbb{C}(2)\) holds for all \(d\), the domain wall reduction
is always feasible.

Similar to Majorana fermion, the $\textbf{domain wall projection}$ of Dirac fermion is also given by the projection operator

\begin{equation}
P_{\text{DW}}=\frac{\mathbf{1}\pm \ii \beta  \alpha _1}{2}.
\end{equation}
which always reduces the fermion spinor dimension by $\textit{half}$:

\begin{equation}
    2^k_{\mathbb{C}}\xrightarrow{P}2^{k-1}_{\mathbb{C}}.
\end{equation}

The domain wall reduction of the complex Clifford algebra and domain wall reduction of Dirac/Weyl fermions are listed in Tabs.~\ref{fig:domain-wall-complex}-\ref{fig:domain-wall-Dirac}.

\begin{table}[htbp]
    \begin{minipage}{.4\columnwidth}
      \centering
        \begin{tikzpicture}
        \matrix[matrix of math nodes,inner sep=1pt,row sep=1em,column sep=1em] (M)
        {
        {\white\backslash}d{\white\backslash} & \mathcal{C}\ell(d) & {\white\backslash}\mathcal{C}\ell(d+1){\white\backslash} \\
        0 & \C(1) & \C(1)\oplus\C(1)\\
        1 & \C(1)\oplus\C(1) & \C(2)\\
        {\white\backslash}2{\white\backslash} & \C(2) & \C(2)\oplus\C(2)\\
        }
        ;
        
        \draw (M-1-1.south west) -- (M-1-3.south east);
        \draw (M-1-1.north east) -- (M-4-1.south east);
        
        \draw[->] (M-3-3.north west) -- (M-2-2.south east);
        \draw[->] (M-4-3.north) -- (M-3-2.south);
        
        \draw[->,color=green] (M-3-2.east) -- (M-3-3.west);
        
        \draw[->,color=red] (M-4-2.east) -- (M-4-3.west);
        \draw[->,color=red] (M-2-2.east) -- (M-2-3.west);
        \end{tikzpicture}
        \vspace{10pt}
        \caption{Domain wall reduction of the complex Clifford algebra. The black arrow means domain wall reduction. The green arrow means regular mass extension. The red arrow means projective mass extension.}
        \label{fig:domain-wall-complex}
    \end{minipage}
    \hfill
    \begin{minipage}{.55\columnwidth}
      \centering
        \begin{tikzpicture}
        \matrix[matrix of math nodes,inner sep=1pt,row sep=1em,column sep=1em,nodes in empty cells] (M)
        {
        {\white\backslash}\ \,{\white\backslash} &  & & & \\
        {\white\backslash}\ \,{\white\backslash}& \text{Boundary} & \text{Bulk} & \text{Boundary} & {\white\backslash}\text{Bulk}{\white\backslash} \\
        0 & & 1_{\C}^+ & 1_{\C} & 1_{\C}^+\oplus 1_{\C}^-\\
        1 & 1_{\C}^+ & & 1_{\C}^+\oplus 1_{\C}^- & 2_{\C}\\
        {\white\backslash}2{\white\backslash} & & 2_{\C}^+  & 2_{\C} & 2_{\C}^+\oplus 2_{\C}^- \\
        }
        ;
        \node[fit=(M-1-1)(M-2-1)]{$d$};
        \node[fit=(M-1-2)(M-1-3)]{Weyl};
        \node[fit=(M-1-4)(M-1-5)]{Dirac};

        \draw (M-2-1.south west) -- (M-2-5.south east);
        \draw (M-1-1.north east) -- (M-5-1.south east);
        
        \draw[->] (M-5-3.north west) -- (M-4-2.south east);
        
        \draw[->] (M-4-5.north west) -- (M-3-4.south east);
        \draw[->] (M-5-5.north) -- (M-4-4.south);
        
        \draw[->,color=red] (M-3-4.west) -- (M-3-3.east);
        \draw[->,color=red] (M-5-4.west) -- (M-5-3.east);
        
        \draw[->,color=green] (M-4-4.east) -- (M-4-5.west);
        \end{tikzpicture}
  \caption{Domain wall reduction of Dirac/Weyl fermions. The number indicates the dimension of the representation of the fermion, the lower index indicates that the representation is complex, and the upper index indicates the chirality of the fermion.}
  \label{fig:domain-wall-Dirac}
\end{minipage}
\end{table}

\subsection{Mass Term and CRT-Internal Symmetry}\label{sec:Dirac_mass&sym}

In this section, we'll examine the interplay between the symmetries and the mass terms. We'll focus on the action of symmetries on the manifold, and how to obtain the symmetries on the domain wall using the reduction method.

\subsubsection{CRT-Internal Symmetry Acting on Mass Manifold}

Recall that $h$ defined in \Eq{eq:Hamiltonian-Dirac} is
\bea
h=\sum_{i=1}^d\alpha_i\ii\partial_i+\sum_{i=1}^n\beta_im_i=h_0+m.
\eea
where $h_0$ stands for the massless part and $m$ is the mass matrix. Since we've already proven that the $h_0$ part is invariant under CRT-internal symmetry in Sec.~\ref{sec:Dirac_sym}, we'll now focus on the bilinear mass $\frac{1}{2}\int\dd^dx \psi^\dagger m\psi$ and how CRT-internal symmetry acts on the mass manifold.

The bilinear mass term changes under the $\mathcal{C}$, $\mathcal{R}_i$, $\mathcal{T}$, and internal $U$ symmetries as follows:
\bea
\frac{1}{2}\int\dd^dx\psi^{\dagger}m\psi&\xrightarrow{\mathcal{C}}&\frac{1}{2}\int\dd^dx(M_{\mathcal{C}}\psi^*)^{\dagger}m(M_{\mathcal{C}}\psi^*)=\frac{1}{2}\int\dd^dx\psi^{\mathsf{T}}(M_{\mathcal{C}}^{\dagger}mM_{\mathcal{C}})\psi^*\nn\\
&=&-\frac{1}{2}\int\dd^dx\psi^{\dagger}(M_{\mathcal{C}}^{\dagger}mM_{\mathcal{C}})^{\mathsf{T}}\psi,\nn\\
\frac{1}{2}\int\dd^dx\psi^{\dagger}m\psi&\xrightarrow{\mathcal{R}_i}&
\frac{1}{2}\int\dd^dx(\alpha_iM_{\mathcal{P}}\psi)^{\dagger}m(\alpha_iM_{\mathcal{P}}\psi)=\frac{1}{2}\int\dd^dx\psi^{\dagger}(M_{\mathcal{P}}^\dagger\alpha_i m\alpha_i M_{\mathcal{P}})\psi,\nn\\
\frac{1}{2}\int\dd^dx\psi^{\dagger}m\psi&\xrightarrow{\mathcal{T}}&
\frac{1}{2}\int\dd^dx(M_{\mathcal{T}}\psi)^{\dagger}m^*(M_{\mathcal{T}}\psi)=\frac{1}{2}\int\dd^dx\psi^{\dagger}(M_{\mathcal{T}}^\dagger m^*M_{\mathcal{T}})\psi,\nn\\
\frac{1}{2}\int\dd^dx\psi^{\dagger}m\psi&\xrightarrow{U}&
\frac{1}{2}\int\dd^dx(M_{U}\psi)^{\dagger}m(M_{U}\psi)=\frac{1}{2}\int\dd^dx\psi^{\dagger}(M_{U}^\dagger mM_{U})\psi.
\eea

If the bilinear mass term is invariant under the $\mathcal{C}$, $\mathcal{R}_i$, $\mathcal{T}$, and internal $U$ symmetries, then
\bea
M_{\mathcal{C}}^{\dagger}mM_{\mathcal{C}}&=&-m^{\mathsf{T}},\nn\\
M_{\mathcal{P}}^{\dagger}\alpha_im\alpha_iM_{\mathcal{P}}&=&m,\nn\\
M_{\mathcal{T}}^{\dagger}mM_{\mathcal{T}}&=&m^*,\nn\\
M_{U}^{\dagger}mM_{U}&=&m.
\eea
Therefore, the matrices $M_{\mathcal{C}}$, $M_{\mathcal{P}}$, $M_{\mathcal{T}}$, and $U$ should satisfy the following relations (for $i=1,\ldots ,d$):
\bea
M_{\mathcal{C}}^{\dagger}\beta_iM_{\mathcal{C}}&=&-\beta_i^*,\nn\\
M_{\mathcal{P}}^{\dagger}\beta_iM_{\mathcal{P}}&=&-\beta_i,\nn\\
M_{\mathcal{T}}^{\dagger}\beta_iM_{\mathcal{T}}&=&\beta_i^*,\nn\\
M_{U}^{\dagger}\beta_iM_{U}&=&\beta_i.
\eea
Any violation of these relations is regarded as the corresponding symmetry breaking.

To be more specific, in $d$=odd, we have a nontrivial $S^1$ mass manifold spanned by two mass terms. The given CRT and internal symmetry operators can act on the whole manifold:

In $d$=1 case, we have two mass matrices $\sigma^{1}$ and $\sigma^{2}$, they span a general $S^1$ mass manifold with terms characterized by mass angle $\theta$:

\begin{equation}
    m(\theta)=\frac{1}{2}\int\dd^dx\psi^\dagger(\cos\theta\sigma^{1}+\sin\theta\sigma^{2})\psi.
\end{equation}

The action of $\mathcal{P},\mathcal{T},\mathcal{R},(-)^\chi,\mathcal{U}^F(\phi_F),\mathcal{U}^\chi(\phi_\chi)$ on the $S^1$ mass manifold is listed in Tab.~\ref{tab:CRT-mass_d=1_Dirac}. $\mathcal{U}^F(\phi_F)$ and $\mathcal{T}$ act trivially on the manifold, $\mathcal{P}$ acts as a ``reflection" on the manifold about $m_1=0$, $\mathcal{C}$ and $\mathcal{R}$ act as ``reflection" on the manifold about $m_2=0$, and $\mathcal{U}^\chi(\phi_\chi)$ acts as a ``rotation" of $\phi_\chi$ angle on the manifold.

\begin{table}[h]
    \centering
    \begin{doublespace}
    \noindent\(\begin{array}{c|c|cc|cccc}
    \text{} & (-)^\chi & \mathcal{U}^F(\phi_F) & \mathcal{U}^\chi(\phi_\chi) & \mathcal{C} & \mathcal{P} & \mathcal{T} & \mathcal{R} \\
    \hline
     \theta' & \pi+\theta & \theta & \theta+\phi_\chi & -\theta & \pi-\theta & \theta & -\theta \\
     \hline
     m_1 & \times & \checkmark & \times & \checkmark & \times & \checkmark & \checkmark \\
     m_2 & \times & \checkmark & \times & \times & \checkmark & \checkmark & \times \\
    \end{array}\)
    \end{doublespace}
    \caption{The action of $\mathcal{P},\mathcal{T},\mathcal{R},(-)^\chi,\mathcal{U}^F(\phi_F),\mathcal{U}^\chi(\phi_\chi)$ on the $S^1$ mass manifold. $\checkmark$ means the mass manifold preserves the symmetry. $\times$ means the mass term breaks the symmetry and mass angle $\theta$ changes.}
    \label{tab:CRT-mass_d=1_Dirac}
\end{table}

In $d$=3 case, we have two mass matrices $\sigma^{13}$ and $\sigma^{23}$, they span a general $S^1$ mass manifold with terms characterized by mass angle $\theta$:

\begin{equation}
    m(\theta)=\frac{1}{2}\int\dd^dx\psi^\dagger(\cos\theta\sigma^{13}+\sin\theta\sigma^{23})\psi.
\end{equation}

The action of $\mathcal{P},\mathcal{T},\mathcal{R}_i,(-)^\chi,\mathcal{U}^F(\phi_F),\mathcal{U}^\chi(\phi_\chi)$ on the $S^1$ mass manifold is listed in Tab.~\ref{tab:CRT-mass_d=3_Dirac}. $\mathcal{U}^F(\phi_F)$ and $\mathcal{C}$ act trivially on the manifold, $\mathcal{T}$ and $\mathcal{R}_i$ act as ``reflection" on the manifold about $m_1=0$, $\mathcal{P}$ acts as a ``reflection" on the manifold about $m_2=0$, and $\mathcal{U}^\chi(\phi_\chi)$ acts as a ``rotation" of $\phi_\chi$ angle on the manifold.

\begin{table}[h]
    \centering
    \begin{doublespace}
    \noindent\(\begin{array}{c|c|cc|cccc}
    \text{} & (-)^\chi & \mathcal{U}^F(\phi_F) & \mathcal{U}^\chi(\phi_\chi) & \mathcal{C} & \mathcal{P} & \mathcal{T} & \mathcal{R}_i \\
    \hline
     \theta' & \pi+\theta & \theta & \theta+\phi_\chi & \theta & -\theta & \pi-\theta & \pi-\theta \\
     \hline
     m_1 & \times & \checkmark & \times & \checkmark & \checkmark & \times & \times \\
     m_2 & \times & \checkmark & \times & \checkmark & \times & \checkmark & \checkmark \\
    \end{array}\)
    \end{doublespace}
    \caption{The action of $\mathcal{P},\mathcal{T},\mathcal{R}_i,(-)^\chi,\mathcal{U}^F(\phi_F),\mathcal{U}^\chi(\phi_\chi)$ on the $S^1$ mass manifold. $\checkmark$ means the mass manifold preserves the symmetry. $\times$ means the mass term breaks the symmetry and mass angle $\theta$ changes.}
    \label{tab:CRT-mass_d=3_Dirac}
\end{table}

In $d$=5 case, we have two mass matrices $\sigma^{133}$ and $\sigma^{233}$, they span a general $S^1$ mass manifold with terms characterized by mass angle $\theta$:

\begin{equation}
    m(\theta)=\frac{1}{2}\int\dd^dx\psi^\dagger(\cos\theta\sigma^{133}+\sin\theta\sigma^{233})\psi.
\end{equation}

The action of $\mathcal{P},\mathcal{T},\mathcal{R}_i,(-)^\chi,\mathcal{U}^F(\phi_F),\mathcal{U}^\chi(\phi_\chi)$ on the $S^1$ mass manifold is listed in Tab.~\ref{tab:CRT-mass_d=5_Dirac}. $\mathcal{U}^F(\phi_F)$ and $\mathcal{T}$ act trivially on the manifold, $\mathcal{C}$ and $\mathcal{R}_i$ act as ``reflection" on the manifold about $m_1=0$, $\mathcal{P}$ acts as a ``reflection" on the manifold about $m_2=0$, and $\mathcal{U}^\chi(\phi_\chi)$ acts as a ``rotation" of $\phi_\chi$ angle on the manifold.

\begin{table}[h]
    \centering
    \begin{doublespace}
    \noindent\(\begin{array}{c|c|cc|cccc}
    \text{} & (-)^\chi & \mathcal{U}^F(\phi_F) & \mathcal{U}^\chi(\phi_\chi) & \mathcal{C} & \mathcal{P} & \mathcal{T} & \mathcal{R}_i \\
    \hline
     \theta' & \pi+\theta & \theta & \theta+\phi_\chi & \pi-\theta & -\theta & \theta & \pi-\theta \\
     \hline
     m_1 & \times & \checkmark & \times & \times & \checkmark & \checkmark & \times \\
     m_2 & \times & \checkmark & \times & \checkmark & \times & \checkmark & \checkmark \\
    \end{array}\)
    \end{doublespace}
    \caption{The action of $\mathcal{P},\mathcal{T},\mathcal{R}_i,(-)^\chi,\mathcal{U}^F(\phi_F),\mathcal{U}^\chi(\phi_\chi)$ on the $S^1$ mass manifold. $\checkmark$ means the mass manifold preserves the symmetry. $\times$ means the mass term breaks the symmetry and mass angle $\theta$ changes.}
    \label{tab:CRT-mass_d=5_Dirac}
\end{table}

In $d$=7 case, we have two mass matrices $\sigma^{1333}$ and $\sigma^{2333}$, they span a general $S^1$ mass manifold with terms characterized by mass angle $\theta$:

\begin{equation}
    m(\theta)=\frac{1}{2}\int\dd^dx\psi^\dagger(\cos\theta\sigma^{1333}+\sin\theta\sigma^{2333})\psi.
\end{equation}

The action of $\mathcal{P},\mathcal{T},\mathcal{R}_i,(-)^\chi,\mathcal{U}^F(\phi_F),\mathcal{U}^\chi(\phi_\chi)$ on the $S^1$ mass manifold is listed in Tab.~\ref{tab:CRT-mass_d=7_Dirac}. $\mathcal{U}^F(\phi_F)$ and $\mathcal{C}$ act trivially on the manifold, $\mathcal{T}$ and $\mathcal{R}_i$ act as ``reflection" on the manifold about $m_1=0$, $\mathcal{P}$ acts as a ``reflection" on the manifold about $m_2=0$, and $\mathcal{U}^\chi(\phi_\chi)$ acts as a ``rotation" of $\phi_\chi$ angle on the manifold.

\begin{table}[h]
    \centering
    \begin{doublespace}
    \noindent\(\begin{array}{c|c|cc|cccc}
    \text{} & (-)^\chi & \mathcal{U}^F(\phi_F) & \mathcal{U}^\chi(\phi_\chi) & \mathcal{C} & \mathcal{P} & \mathcal{T} & \mathcal{R}_i \\
    \hline
     \theta' & \pi+\theta & \theta & \theta+\phi_\chi & \theta & -\theta & \pi-\theta & \pi-\theta \\
     \hline
     m_1 & \times & \checkmark & \times & \checkmark & \checkmark & \times & \times \\
     m_2 & \times & \checkmark & \times & \checkmark & \times & \checkmark & \checkmark \\
    \end{array}\)
    \end{doublespace}
    \caption{The action of $\mathcal{P},\mathcal{T},\mathcal{R}_i,(-)^\chi,\mathcal{U}^F(\phi_F),\mathcal{U}^\chi(\phi_\chi)$ on the $S^1$ mass manifold. $\checkmark$ means the mass manifold preserves the symmetry. $\times$ means the mass term breaks the symmetry and mass angle $\theta$ changes.}
    \label{tab:CRT-mass_d=7_Dirac}
\end{table}

\subsubsection{CRT-Internal Symmetry Reduction under Domain Wall}\label{sec:Dirac_CPT_dom_wall}

By \textbf{domain wall reduction}, we can reduce a \textit{bulk} Dirac fermion to \textit{boundary} Dirac (or Weyl) fermion in a lower dimension. Surprisingly, the CRT-internal symmetry group in different dimensions (see Tabs.~\ref{tab:CRT-internal_Dirac}-\ref{tab:CRT-internal_Weyl}), though exhibit distinct fractionalization properties, is related by domain wall reduction. We can always reproduce the CRT-internal symmetry group in $(d-1)+1$ dimension by projecting corresponding symmetry operators to the mass domain wall.

We can randomly add a mass and reduce the fermion to the mass domain wall in either direction (say $m\sim \pm x_d$). Note that a well-defined (i.e., not modified by broken internal symmetries) reflection $\mathcal{R}_d$ is always preserved under the mass domain wall on the $d$-th direction, since the reflection simultaneously swaps the ground state in the $P_+$ and $P_-$ projection space, and flips the mass profile $m\sim \pm x_d\to\mp x_d$. Under domain wall reduction, $\mathcal{R}_d$ always becomes an internal symmetry on the domain wall. The reduction of CRT and internal symmetries follow the \textit{rules} below:

\begin{itemize}
    \item If the charge conjugation $\mathcal{C}$, reflection $\mathcal{R}_i$, time-reversion $\mathcal{T}$, or internal symmetry $U$ is \textit{preserved under mass extension}, then these symmetries are directly projected to the $(d-1)$-dimensional CRT-internal symmetry by projection operator $P_{DW}=\dfrac{\mathbf{1}\pm \ii \beta  \alpha _1}{2}$:

    \bea\label{eq:rule1_Dirac}
        d\text{-dimension} &\xrightarrow{DW}& (d-1)\text{-dimension}\nn\\
        \mathcal{C}  &\xrightarrow{P_{DW}}& \mathcal{C}\nn\\
        \mathcal{R}_{d}  &\xrightarrow{P_{DW}}& X\nn\\
        \mathcal{R}_i  &\xrightarrow{P_{DW}}&  \mathcal{R}_i\ (\forall i=1,...,d-1)\nn\\
        \mathcal{T}  &\xrightarrow{P_{DW}}&  \mathcal{T}\nn\\
        U  &\xrightarrow{P_{DW}}&  U,
    \eea
    where $X$ is an internal symmetry in the $(d-1)$-dimensional theory.

    \item If the charge conjugation $\mathcal{C}$, reflection $\mathcal{R}_i$, time-reversion $\mathcal{T}$, or internal symmetry $U$ is \textit{broken under mass extension}, then these symmetries should be combined with the space-orientation-reversing symmetry $\mathcal{CR}_d\mathcal{T}$ to obtain a new symmetry on the domain wall~\cite{2312.17126Wan:2023nqe,10.21468/SciPostPhys.8.4.062,1910.14046,Wang:2019obe1910.14664}:

    \bea\label{eq:rule2_Dirac}
        d\text{-dimension} &\xrightarrow{DW}& (d-1)\text{-dimension}\nn\\
        \mathcal{C}  &\xrightarrow{\cdot\mathcal{CR}_d\mathcal{T}}&  X\mathcal{T}=\mathcal{T}'\nn\\
        \mathcal{R}_{d}  &\xrightarrow{P_{DW}}& X\nn\\
        \mathcal{R}_i  &\xrightarrow{\cdot\mathcal{CR}_d\mathcal{T}}&  X\mathcal{CR}_i\mathcal{T}=\mathcal{C'R'}_i\mathcal{T}'\ (\forall i=1,...,d-1)\nn\\
        \mathcal{T}  &\xrightarrow{\cdot\mathcal{CR}_d\mathcal{T}}&  X\mathcal{C}=\mathcal{C}'\nn\\
        U  &\xrightarrow{\cdot\mathcal{CR}_d\mathcal{T}}&  XU\mathcal{CT}=\mathcal{C}'\mathcal{T}'.
    \eea
\end{itemize}

To be specific, the CRT-internal symmetry groups in spacetime $d+1$ and $(d-1)+1$ dimension, and distinct mass terms are listed in Tab.~\ref{tab:sym_reduction&mass_D}. On each domain wall mass, the explicit result of the domain-wall projection for symmetry operators are listed in Tab.~\ref{tab:sym_operator_reduction_D}.

    \begin{table}[h]
    \centering
    \begin{doublespace}
    \noindent\(\begin{array}{c|cccc|cccc}
        d         & G^{\text{CRT}}_{D,d}     &  G_{D,d}   & G^{\text{CRT}}_{D/\cblue{W},d-1}   & G_{D/\cblue{W},d-1}  & m_1  & m_2                 \\ \hline
        1         & \mathbb{D}_8^{\mathcal{CT,C}}\times\mathbb{Z}_2^{\mathcal{R}_1\mathcal{T}} & (\U(1)^F\times \U(1)^\chi)\rtimes_{\mathbb{Z}_2^F} G^{\text{CRT}}_{D,d}    &\mathbb{Z}_2^\mathcal{C}\times\mathbb{Z}_2^\mathcal{T}  &\U(1)^F\rtimes G^{\text{CRT}}_{D,d-1}  & \psi^\dagger\sigma^1\psi  &   \psi^\dagger\sigma^2\psi            \\
        2         & \mathbb{D}_8^{\mathcal{T,R}_1}\times\mathbb{Z}_2^{\mathcal{C}} & \U(1)^F\rtimes_{\mathbb{Z}_2^F} G^{\text{CRT}}_{D,d}    & \cblue{\mathbb{Z}_2^{\mathcal{C}}\times\mathbb{Z}_2^{\mathcal{CR}_1\mathcal{T}}}  & \cblue{\U(1)^F\rtimes G^{\text{CRT}}_{W,d-1}}  & \psi^\dagger\sigma^3\psi  &             \\
        3         & \mathbb{D}_8^{\mathcal{T,R}_1}\times\mathbb{Z}_2^{\mathcal{C}} & (\U(1)^F\times \U(1)^\chi)\rtimes_{\mathbb{Z}_2^F} G^{\text{CRT}}_{D,d}    & \mathbb{D}_8^{\mathcal{T,R}_1}\times\mathbb{Z}_2^{\mathcal{C}} & \U(1)^F\rtimes_{\mathbb{Z}_2^F} G^{\text{CRT}}_{D,d-1}  & \psi^\dagger\sigma^{13}\psi  &  \psi^\dagger\sigma^{23}\psi           \\
        4         & \mathbb{D}_8^{\mathcal{T},\mathcal{CR}_1}\times_{\mathbb{Z}_2^F}\mathbb{Z}_4^{\mathcal{C}F} & \U(1)^F\rtimes_{\mathbb{Z}_2^F} G^{\text{CRT}}_{D,d}    & \cblue{\mathbb{Z}_4^{\mathcal{T}}\times\mathbb{Z}_2^{\mathcal{CR}_1\mathcal{T}}}  & \cblue{\U(1)^F\rtimes_{\mathbb{Z}_2^F} G^{\text{CRT}}_{W,d-1}}  & \psi^\dagger\sigma^{33}\psi  &             \\
        5         & \mathbb{D}_8^{\mathcal{R}_1,\mathcal{CR}_1\mathcal{T}}\times_{\mathbb{Z}_2^F}\mathbb{Z}_4^{\mathcal{C}F} & (\U(1)^F\times \U(1)^\chi)\rtimes_{\mathbb{Z}_2^F} G^{\text{CRT}}_{D,d}    & \mathbb{D}_8^{\mathcal{T},\mathcal{CR}_1}\times_{\mathbb{Z}_2^F}\mathbb{Z}_4^{\mathcal{C}F} & \U(1)^F\rtimes_{\mathbb{Z}_2^F} G^{\text{CRT}}_{D,d-1}  & \psi^\dagger\sigma^{133}\psi  &   \psi^\dagger\sigma^{233}\psi            \\
        6         & \mathbb{D}_8^{\mathcal{CR}_1,\mathcal{T}}\times_{\mathbb{Z}_2^F}\mathbb{Z}_4^{\mathcal{C}F} & \U(1)^F\rtimes_{\mathbb{Z}_2^F} G^{\text{CRT}}_{D,d}    & \cblue{\mathbb{D}_8^{\mathcal{C},\mathcal{CR}_1\mathcal{T}}}  & \cblue{\U(1)^F\rtimes_{\mathbb{Z}_2^F} G^{\text{CRT}}_{W,d-1}}  & \psi^\dagger\sigma^{333}\psi  &             \\
        7         & \mathbb{D}_8^{\mathcal{CR}_1,\mathcal{T}}\times_{\mathbb{Z}_2^F}\mathbb{Z}_4^{\mathcal{C}F} & (\U(1)^F\times \U(1)^\chi)\rtimes_{\mathbb{Z}_2^F} G^{\text{CRT}}_{D,d}    & \mathbb{D}_8^{\mathcal{CR}_1,\mathcal{T}}\times_{\mathbb{Z}_2^F}\mathbb{Z}_4^{\mathcal{C}F} & \U(1)^F\rtimes_{\mathbb{Z}_2^F} G^{\text{CRT}}_{D,d-1}  & \psi^\dagger\sigma^{1333}\psi  &  \psi^\dagger\sigma^{2333}\psi           \\
        8         & \mathbb{D}_8^{\mathcal{CR}_1,\mathcal{CT}}\times\mathbb{Z}_2^\mathcal{C} & \U(1)^F\rtimes_{\mathbb{Z}_2^F} G^{\text{CRT}}_{D,d}    & \cblue{\mathbb{D}_8^{\mathcal{CR}_1,\mathcal{T}}}  & \cblue{\U(1)^F\rtimes_{\mathbb{Z}_2^F} G^{\text{CRT}}_{W,d-1}}  & \psi^\dagger\sigma^{3333}\psi  &             \\
    \end{array}\)
    \end{doublespace}
    \caption{The CRT-internal symmetry group for Dirac fermion in $d+1$ spacetime dimension $G_{D,d}$ can be reduced to the CRT-internal symmetry group for Dirac or Weyl fermion in $(d-1)+1$ spacetime dimension $G_{D/W,d-1}$ on the mass domain wall. The domain wall mass term $m_i$ can be chosen in the mass manifold. The CRT-invariant group includes vector $\U(1)^F$ symmetry, axial $\U(1)^\chi$ symmetry, charge conjugation $\mathbb{Z}_2^{\mathcal{C}}$, reflection $\mathbb{Z}_2^{\mathcal{R}_1}$, and time-reversal symmetry $\mathbb{Z}_2^{\mathcal{T}}$.}
    \label{tab:sym_reduction&mass_D}
    \end{table}

    \begin{table}[h]
    \centering
    \begin{doublespace}
    \noindent\(\begin{array}{c|c|cccccccc}
        d         & m_{DW} & P_{DW}((-)^F)     & P_{DW}((-)^\chi)    & P_{DW}(\mathcal{C})   & P_{DW}(\mathcal{R}_{i<d})  & P_{DW}(\mathcal{R}_d)  & P_{DW}(\mathcal{T})                  \\ \hline
        \multirow{2}{*}{1}    & m_1     & (-)^F    & 0                                   &              &   \mathcal{C}            &    (-)^F                                    &   \mathcal{T}                   \\
            & m_2     & (-)^F    & 0                                   &  \text{}            &      \mathcal{C}         &    1                                    &   \mathcal{T}                   \\
        2    & m_1     & (-)^F & \text{} & \mathcal{C} & \mathcal{CR}_1\mathcal{T}       &  1          & \mathcal{C}        \\
        \multirow{2}{*}{3}   & m_1      & (-)^F   & 0  &  \mathcal{C} & \mathcal{R}_i  & 1   & \mathcal{T}  \\
              & m_2   & (-)^F   & 0  &  \mathcal{C} & \mathcal{R}_i  & (-)^F  & \mathcal{T}  \\
        4   & m_1      & (-)^F   &   &  \mathcal{T} & \mathcal{CR}_i\mathcal{T}  & 1   & \mathcal{T}  \\
        \multirow{2}{*}{5}   & m_1      & (-)^F   & 0  &  \mathcal{C} & \mathcal{R}_i  & 1   & \mathcal{T}  \\
              & m_2   & (-)^F   &  0 &  \mathcal{C} & \mathcal{R}_i  & (-)^F  & \mathcal{T}  \\
        6     & m_1  & (-)^F & & \mathcal{C} & \mathcal{CR}_i\mathcal{T} & 1 & \mathcal{C} \\
        \multirow{2}{*}{7}   & m_1      & (-)^F   & 0  &  \mathcal{C} & \mathcal{R}_i  & 1   & \mathcal{T}  \\
              & m_2   & (-)^F   &  0 &  \mathcal{C} & \mathcal{R}_i  & (-)^F  & \mathcal{T}  \\
        8     & m_1  & (-)^F & & \mathcal{T} & \mathcal{CR}_i\mathcal{T} & 1 & \mathcal{T} \\
    \end{array}\)
    \end{doublespace}
    \caption{The projected symmetry operators $P_{DW}(\cdot)$ with the domain wall mass $m_{DW}$ from spatial dimension $d$ to $d-1$. Symmetries include vector $\U(1)^F$ symmetry generated by $(-)^F$, axial $\U(1)^\chi$ symmetry generated by $(-)^\chi$, charge conjugation $\mathbb{Z}_2^{\mathcal{C}}$, reflection $\mathbb{Z}_2^{\mathcal{R}_i}$, and time-reversal symmetry $\mathbb{Z}_2^{\mathcal{T}}$.}
    \label{tab:sym_operator_reduction_D}
    \end{table}

\section{Conclusion}

In this work, we've systematically analyzed the CRT fractionalization in a single-particle Hamiltonian theory.

In Sec.~\ref{sec:Maj_field_model}, we've defined the Majorana field as a real Grassmannian field acting as an irreducible representation of the Clifford algebra $\mathcal{C}\ell(d,n)$. This definition is compatible with the conventional definition~\cite{2312.17126Wan:2023nqe} as a single Dirac fermion with trivial charge conjugation and can extend to $d+1=5,6,7\bmod 8$ where symplectic Majorana fermion emerges. We've also reviewed the Clifford algebra and its 8-fold periodicity. 

In Sec.~\ref{sec:Maj_sym}, we've defined Lorentz symmetries, internal symmetries and RT symmetries for the massless Majorana (and Majorana-Weyl) fermion, and specified the invariant group with 8-fold periodicity. The results are listed in Tabs.~\ref{tab:CRT-internal_Maj} and \ref{tab:CRT-internal_Maj_Weyl}. To further analyze the ambiguity of the explicit representation basis of the CRT operators, we introduce Clifford algebra extension in Tab.~\ref{tab:Clifford-algebra-extension}, the explicit choices are listed in Tab.~\ref{tab:explicit-mass-extension}.

In Sec.~\ref{sec:Maj_mass}, we've studied the massive Majorana theory by extending mass terms. In different dimensions, we can add different numbers of mass terms, and they reflect distinct topological properties by expanding mass manifolds. The resulting mass manifold in different dimensions is demonstrated in Tab.~\ref{tab:mass-manifold}. By further introducing domain wall reduction, we can connect Majorana (and Majorana-Weyl) fermions in different dimensions by mass extension and domain wall reduction, as listed in Tabs.~\ref{tab:domain-wall-real}-\ref{tab:domain-wall-Majorana}.

In Sec.~\ref{sec:Maj_mass&sym}, we've studied the interplay between mass terms and CRT-internal symmetries. Intriguingly, when a mass manifold is formed by multiple mass terms, CRT-internal symmetries can act on the manifold as either flipping or reflection, as demonstrated in Tabs.~\ref{tab:CRT-mass_d=3}-\ref{tab:CRT-mass_d=5}. The CRT-internal symmetries together suffices to rule out all possible mass terms. Furthermore, we've studied the symmetry reduction under domain wall by extending a mass and then apply domain wall reduction. The reduction follows the rules in Eqs.~(\ref{eq:rule1},\ref{eq:rule2}) and bridges the CRT-internal symmetries in different dimensions (see Tabs.~\ref{tab:sym_reduction&mass}-\ref{tab:sym_operator_reduction}).

\begin{table}[h]
    {\centering
    \begin{tikzpicture}
    \matrix[matrix of math nodes,inner sep=1pt,row sep=1em,column sep=1em] (M)
    {
    {\white\backslash}\ \,\,{\white\backslash} &\text{Majorana-Weyl (MW)} &\text{Majorana (M)} &\qquad\qquad\quad &\qquad\qquad\quad \\
    {\white\backslash}\,{\white\backslash} &  
    \text{\cblue{\& Symplectic MW}}
    & 
    \text{\cblue{\& Symplectic M}}
    & \, & {\white\backslash}\,{\white\backslash}\\
    0 & & 1_{\R} & & 1_{\C}\\
    1 & 1_{\R}^+ & 1_{\R}^+\oplus 1_{\R}^- & 1_{\C}^+ & 1_{\C}^+\oplus 1_{\C}^-\\
    2 & & 2_{\R} &  & 2_{\C}\\
    3 & & 2_{\C} & 2_{\C}^+ &  2_{\C}^+\oplus 2_{\C}^-\\
    4 & &\cblue{2_{\bH}} & & 4_{\C} \\
    5 &\cblue{2_{\bH}^+} & \cblue{2_{\bH}^+\oplus 2_{\bH}^-} & 4_{\C}^+ & 4_{\C}^+\oplus 4_{\C}^-\\
    6 & & \cblue{4_{\bH}} & & 8_{\C} \\
    7 & &  8_{\C} & 8_{\C}^+ & 8_{\C}^+\oplus 8_{\C}^-\\
    8 & & 16_{\R} &  & 16_{\C}\\
    9 &  16_{\R}^+ & 16_{\R}^+\oplus 16_{\R}^- & 16_{\C}^+ & 16_{\C}^+\oplus 16_{\C}^-\\
    10 & & 32_{\R} &  & 32_{\C}\\
    {\white\backslash}11{\white\backslash} & & 32_{\C} & 32_{\C}^+ & 32_{\C}^+\oplus 32_{\C}^-\\
    }
    ;
    \node[fit=(M-1-1)(M-2-1)]{$d$};
    \node[fit=(M-1-4)(M-2-4)]{Weyl (W)};
    \node[fit=(M-1-5)(M-2-5)]{Dirac (D)};

    \draw (M-2-1.south west) -- (M-2-5.south east);
    \draw (M-1-1.north east) -- (M-14-1.south east);
    
    \draw[dashed] (M-4-2.east) -- (M-4-3.west);
    \draw[dashed] (M-4-4.east) -- (M-4-5.west);
    
    \draw (M-6-3.east) -- (M-6-4.west);
    \draw[dashed] (M-6-4.east) -- (M-6-5.west);
    
    \draw[dashed] (M-8-4.east) -- (M-8-5.west);
    
    \draw[dashed] (M-10-4.east) -- (M-10-5.west);
    
    \draw[dashed] (M-12-2.east) -- (M-12-3.west);
    \draw[dashed] (M-12-4.east) -- (M-12-5.west);
    
    \draw (M-14-3.east) -- (M-14-4.west);
    \draw[dashed] (M-14-4.east) -- (M-14-5.west);
    
    \draw[->] (M-4-3.north) -- (M-3-3.south);
    \draw[->] (M-4-5.north) -- (M-3-5.south);
    \draw[->] (M-5-3.north west) -- (M-4-2.south east);
    \draw[->] (M-5-5.north west) -- (M-4-4.south east);
    \draw[->] (M-6-3.north) -- (M-5-3.south);
    \draw[->] (M-6-5.north) -- (M-5-5.south);
    \draw[->] (M-7-5.north west) -- (M-6-4.south east);
    \draw[->] (M-8-5.north) -- (M-7-5.south);
    \draw[->] (M-9-5.north west) -- (M-8-4.south east);
    \draw[->] (M-10-5.north) -- (M-9-5.south);
    \draw[->] (M-11-5.north west) -- (M-10-4.south east);
    \draw[->] (M-12-3.north) -- (M-11-3.south);
    \draw[->] (M-12-5.north) -- (M-11-5.south);
    \draw[->] (M-13-3.north west) -- (M-12-2.south east);
    \draw[->] (M-13-5.north west) -- (M-12-4.south east);
    \draw[->] (M-14-3.north) -- (M-13-3.south);
    \draw[->] (M-14-5.north) -- (M-13-5.south);
    
    \draw[->] (M-7-3.north) -- (M-6-3.south);
    \draw[->][blue] (M-8-3.north) -- (M-7-3.south);
    \draw[->][blue] (M-9-3.north west) -- (M-8-2.south east);
    \draw[dashed][blue] (M-8-2.east) -- (M-8-3.west);
    \end{tikzpicture}}
    \caption{Summary of domain wall reduction of fermions. The target of the arrow is the domain wall of the source of the arrow. The number indicates the dimension of the representation of the fermion, the lower index indicates that the representation is real or complex or quaternionic, and the upper index indicates the chirality of the fermion.
    The dashed lines (- - -) connecting Majorana-Weyl and Majorana fermions or 
    connecting Weyl and Dirac fermions mean that the left and right Majorana-Weyl or Weyl fermions can be combined into a Majorana or Dirac fermion. The solid lines (---) connecting Majorana and Weyl fermions in spacetime dimensions $d+1=4\bmod8$ mean that Majorana and Weyl fermions in spacetime dimensions $d+1=4\bmod8$ can be identified~\cite{2312.17126Wan:2023nqe}. For instance, in $3+1d$, we can write Majorana fermion in Weyl basis as shown in App.~\ref{app:Maj_1}. However, for $d+1=0\bmod 8$, though Majorana fermion and Weyl fermion share the Clifford algebra $\mathbb{C}(2^{\frac{d-1}{2}})$, the Majorana fermion cannot be written in Weyl fermion~\cite{2312.17126Wan:2023nqe}.}
    \label{tab:domain-wall-fermion-summary}
\end{table}

In Sec.~\ref{sec:Dirac_field_model}, we've similarly defined the Dirac field as a complex Grassmannian field acting as an irreducible representation of the Clifford algebra $\mathcal{C}\ell(d+n)$. By examining the complex Clifford algebra, we've found that the Dirac field exhibits 2-fold periodicity.

In Sec.~\ref{sec:Dirac_sym}, we've defined the Lorentz symmetries, internal symmetries and CRT symmetries for the massless Dirac (and Weyl) fermion. Surprisingly, by assuming the canonical CRT conditions, the invariant group is 8-fold periodic rather than 2-fold (as the periodicity of complex Clifford algebra). The results are listed in Tabs.~\ref{tab:CRT-internal_Dirac} and \ref{tab:CRT-internal_Weyl}.

In Sec.~\ref{sec:Dirac_mass}, we've studied the massive Dirac theory by extending mass terms. In even spatial dimensions, we can only add one mass term, while in odd cases, two distinct mass terms generate a $S^1$ mass manifold. The resulting mass extension and mass manifold are listed in Tabs.~\ref{fig:mass-extension-Dirac}-\ref{tab:mass-manifold_Dirac}. By further introducing domain wall reduction, we can connect Dirac (and Weyl) fermions in different dimensions by mass extension and domain wall reduction, as listed in Tabs.~\ref{fig:domain-wall-complex}-\ref{fig:domain-wall-Dirac}.

In Sec.~\ref{sec:Dirac_mass&sym}, we've studied the interplay between mass terms and CRT-internal symmetries. We've found that the axial $\U(1)$ symmetry can rotate the mass manifold and other symmetries can flip the manifold, as demonstrated in Tabs.~\ref{tab:CRT-mass_d=1_Dirac}-\ref{tab:CRT-mass_d=7_Dirac}. The CRT-internal symmetries together suffices to rule out all possible mass terms. Furthermore, we've studied the symmetry reduction under domain wall by extending a mass and then apply domain wall reduction. The reduction follows the rules in Eqs.~(\ref{eq:rule1_Dirac},\ref{eq:rule2_Dirac}) and bridges the CRT-internal symmetries in different dimensions (see Tabs.~\ref{tab:sym_reduction&mass_D}-\ref{tab:sym_operator_reduction_D}).

The domain wall reduction of Majorana/Majorana-Weyl and Dirac/Weyl fermions is summarized in Tab.~\ref{tab:domain-wall-fermion-summary}. This map is conducive to the derivation of relation between $N_f = 3$ families of 16 Weyl fermions of
the Standard Model in 4d to the 48 Majorana-Weyl fermions in 2d~\cite{Wang:2023tbjFamily2312.14928}. Moreover, the domain wall reduction method has a profound impact on the classification of interacting fermions~\cite{QueirozKhalafStern1601.01596}. Another intriguing phenomenon is that Majorana fermion and Weyl fermion can be identified only when $d+1=4\bmod 8$. For instance, in $3+1d$, we can write Majorana fermion in Weyl basis as shown in App.~\ref{app:Maj_1}. However, for $d+1=0\bmod 8$, though Majorana fermion and Weyl fermion share the Clifford algebra $\mathbb{C}(2^{\frac{d-1}{2}})$, the Majorana fermion cannot be written in Weyl fermion~\cite{2312.17126Wan:2023nqe}.

\begin{acknowledgments}
ZW is supported by the NSFC Grant No. 12405001. JW is supported by Harvard University CMSA and 
LIMS fellow fund. YZY is supported by the National Science Foundation Grant No. DMR-2238360.
\end{acknowledgments}

\appendix

\section{Brief Review of Majorana Fermion in $3+1d$ Case}\label{app:Maj_1}

In this appendix, we briefly review the process of deriving conventional Majorana fermion from Dirac fermion.

\subsection{Basis Independent Discussion}

In $3+1d$ spacetime, the Dirac fermion is characterized by the following Lagrangian:

\begin{equation}
    L=\frac{1}{2}\int \dd^3 x \bar{\psi}(\ii\gamma^\mu\partial_\mu-m)\psi,
\end{equation}
and corresponding Dirac equation:

\begin{equation}
    (\ii\gamma^\mu\partial_\mu-m)\psi=0,
\end{equation}
where $\gamma^\mu$ is chosen in $\mathcal{C}\ell_{1,3}(\mathbb{C})$, i.e. $\gamma^\mu$ are complex matrices satisfying $(\gamma^0)^2=1$ and $(\gamma^i)^2=-1$ for $i=1,2,3$.

The charge conjugation for the Dirac spinor is defined by

\begin{equation}
    \psi^c\overset{def}{=\!=}\mathcal{C}\psi\mathcal{C}^{-1}=M_{\mathcal{C}}\psi^*.
\end{equation}

Define another matrix $C$ by

\begin{equation}
    M_{\mathcal{C}}=C^{-1}(\gamma^0)^{\mathsf{T}}.
\end{equation}

Therefore, by choosing $C$ symmetric or antisymmetric ($C^{\mathsf{T}}=\pm C$), we obtain

\begin{equation}
    \bar{\psi}=(\psi^c)^{\mathsf{T}}C^{\mathsf{T}}=\pm(\psi^c)^{\mathsf{T}}C.
\end{equation}

$M_{\mathcal{C}}$ is defined to satisfy:

\begin{equation}
    M_{\mathcal{C}}^\dagger\gamma^0M_{\mathcal{C}}=-(\gamma^0)^{\mathsf{T}},\ M_{\mathcal{C}}^\dagger\gamma^iM_{\mathcal{C}}=(\gamma^i)^{\mathsf{T}}\ (\text{for }i=1,2,3).
\end{equation}

Thus, $C$ needs to satisfy:

\begin{equation}
    C\gamma^\mu C^{-1}=-(\gamma^\mu)^{\mathsf{T}}.
\end{equation}

In order to find a Majorana solution, we assume $\psi^c=\psi$:

\begin{equation}
    \psi^c=C^{-1}(\gamma^0)^{\mathsf{T}}\psi^*=\psi,
\end{equation}
or written in the form

\begin{equation}
    \bar{\psi}=\pm\psi^{\mathsf{T}}C.
\end{equation}

Then the Lagrangian becomes ($\pm$ is neglected since it'll not affect the equation of motion)

\begin{equation}
    L=\frac{1}{2}\int \dd^3 x\psi^{\mathsf{T}}C(\ii\gamma^\mu\partial_\mu-m)\psi.
\end{equation}

The equation of motion (EOM) is again the Dirac equation

\begin{equation}
    (\ii\gamma^\mu\partial_\mu-m)\psi=0,
\end{equation}
with additional constraint

\begin{equation}
    \psi=C^{-1}(\gamma^0)^{\mathsf{T}}\psi^*.
\end{equation}

To relax the constraint, we need to introduce an explicit basis and minimize the degree of freedom.

\subsection{Weyl Basis}

To faithfully write the Majorana equation in the representation of real Clifford algebra $\mathcal{C}\ell(3,0)\cong \mathbb{C}(2)$, it's intuitive to relate the Majorana fermion to one of the Weyl fermions. The Weyl basis is given by:

\begin{equation}
\begin{aligned}
    \gamma^0&=&\sigma^{10}&=&\begin{pmatrix}
        0&\sigma^0\\\sigma^0&0
    \end{pmatrix},\ \gamma^1&=&\ii\sigma^{21}&=&\begin{pmatrix}
        0&\sigma^1\\-\sigma^1&0
    \end{pmatrix},\\
    \gamma^2&=&\ii\sigma^{22}&=&\begin{pmatrix}
        0&\sigma^2\\-\sigma^2&0
    \end{pmatrix},\ \gamma^3&=&\ii\sigma^{23}&=&\begin{pmatrix}
        0&\sigma^3\\-\sigma^3&0
    \end{pmatrix}.
\end{aligned}
\end{equation}

To satisfy

\bea
    M_{\mathcal{C}}^\dagger\gamma^0M_{\mathcal{C}}&=&-(\gamma^0)^{\mathsf{T}}=-\gamma^0,\nn\\
    M_{\mathcal{C}}^\dagger\gamma^1M_{\mathcal{C}}&=&(\gamma^1)^{\mathsf{T}}=-\gamma^1,\nn\\
    M_{\mathcal{C}}^\dagger\gamma^2M_{\mathcal{C}}&=&(\gamma^2)^{\mathsf{T}}=\gamma^2,\nn\\
    M_{\mathcal{C}}^\dagger\gamma^3M_{\mathcal{C}}&=&(\gamma^3)^{\mathsf{T}}=-\gamma^3,
\eea
we can choose $M_{\mathcal{C}}=-\ii\gamma^2=\sigma^{22}$ and $C=(\gamma^0)^TM_{\mathcal{C}}^\dagger=\ii\sigma^{32}$.

The Majorana constraint becomes

\begin{equation}
    \psi=M_{\mathcal{C}}\psi^*=\sigma^{22}\psi^*,
\end{equation}
which gives

\begin{equation}
    \psi_L=-\ii\sigma^2\psi_R^*,\ \psi_R=\ii\sigma^2\psi_L^*.
\end{equation}

This means that we can construct the Majorana fermion through only one Weyl fermion, say $\psi_L$:

\begin{equation}
    \psi=\begin{pmatrix}
        \psi_L\\\ii\sigma^2\psi_L^*.
    \end{pmatrix}
\end{equation}

The Lagrangian

\begin{equation}
    L=\frac{1}{2}\int \dd^3 x\psi^{\mathsf{T}}C(\ii\gamma^\mu\partial_\mu-m)\psi
\end{equation}
can be reduced to the Lagrangian for $\psi_L$ and $\psi_L^*$:

\begin{equation}
\begin{aligned}
    L=&\frac{1}{2}\int \dd^3 x\psi^{\mathsf{T}}C(\ii\gamma^\mu\partial_\mu-m)\psi\\
    =&\frac{1}{2}\int \dd^3 x(\psi_L^{\mathsf{T}},-\ii\sigma^2\psi_L^\dagger)\ii\sigma^{32}(\ii\sigma^{10}\partial_0-\sum_i\sigma^{2i}\partial_i-m)\begin{pmatrix}
        \psi_L\\\ii\sigma^2\psi_L^*
    \end{pmatrix}\\
    =&\frac{1}{2}\int \dd^3 x(\psi_L^{\mathsf{T}},-\ii\sigma^2\psi_L^\dagger)(-\ii\sigma^{22}\partial_0+\ii\sigma^{13}\partial_1-\sigma^{10}\partial_2-\ii\sigma^{11}\partial_3-\ii m\sigma^{32})\begin{pmatrix}
    \psi_L\\\ii\sigma^2\psi_L^*
    \end{pmatrix}\\
    =&\frac{1}{2}\int \dd^3 x\psi_L^{\mathsf{T}}(-\ii\partial_0+\ii\partial_1\sigma^1-\ii\partial_2\sigma^2+\ii\partial_3\sigma^3)\psi_L^*+\psi_L^\dagger(-\ii\partial_0+\ii\partial_1\sigma^1+\ii\partial_2\sigma^2+\ii\partial_3\sigma^3)\psi_L\\
    +&\psi_L^{\mathsf{T}}(-\ii m\sigma^2)\psi_L+\psi_L^\dagger(\ii m\sigma^2)\psi_L^*\\
    =&\frac{1}{2}\int \dd^3 x[-\ii\partial_0(\psi_L^{\mathsf{T}}\psi_L^*)+\ii\partial_1(\psi_L^{\mathsf{T}}\sigma^1\psi_L^*)-\ii\partial_2(\psi_L^{\mathsf{T}}\sigma^2\psi_L^*)+\ii\partial_3(\psi_L^{\mathsf{T}}\sigma^3\psi_L^*)]\\
    -&\int \dd^3 x\psi_L^\dagger(\ii\bar{\sigma}^\mu\partial_\mu)\psi_L+\frac{\ii m}{2}(\psi_L^{\mathsf{T}}\sigma^2\psi_L-\psi_L^\dagger\sigma^2\psi_L^*)\\
    \sim&\int \dd^3 x\psi_L^\dagger(\ii\bar{\sigma}^\mu\partial_\mu)\psi_L+\frac{\ii m}{2}(\psi_L^{\mathsf{T}}\sigma^2\psi_L-\psi_L^\dagger\sigma^2\psi_L^*).
\end{aligned}
\end{equation}

Therefore, the EOM becomes

\begin{equation}
    \ii\bar{\sigma}^\mu\partial_\mu\psi_L-\ii m\sigma^2\psi_L^*=0.
\end{equation}

\subsection{Embedded in $\mathbb{R}(4)$: Majorana Basis}

Instead of throwing away half of the component, we can also relax the constraint by finding a basis where all $i\gamma^\mu$ matrices are real so that the Dirac equation (and Lagrangian) is real and spontaneously allows a real solution. The Majorana basis is given by:

\begin{equation}
\begin{aligned}
    \gamma^0&=&\sigma^{12}&=&\begin{pmatrix}
        0&\sigma^2\\\sigma^2&0
    \end{pmatrix},\ \gamma^1&=&\ii\sigma^{03}&=&\begin{pmatrix}
        \ii\sigma^3&0\\0&\ii\sigma^3
    \end{pmatrix},\\
    \gamma^2&=&-\ii\sigma^{22}&=&\begin{pmatrix}
        0&-\sigma^2\\\sigma^2&0
    \end{pmatrix},\ \gamma^3&=&-\ii\sigma^{01}&=&\begin{pmatrix}
        -\ii\sigma^1&0\\0&-\ii\sigma^1
    \end{pmatrix}.
\end{aligned}
\end{equation}

In this case, the charge conjugation matrix $M_{\mathcal{C}}$ needs to satisfy:

\bea
    M_{\mathcal{C}}^\dagger\gamma^0M_{\mathcal{C}}&=&-(\gamma^0)^{\mathsf{T}}=\gamma^0,\nn \\
    M_{\mathcal{C}}^\dagger\gamma^iM_{\mathcal{C}}&=&(\gamma^i)^{\mathsf{T}}=\gamma^i,\ (\text{for }i=1,2,3),
\eea
which means we can simply set $M_{\mathcal{C}}=I_{4\times 4}$ and $\mathcal{C}=(\gamma^0)^{\mathsf{T}}=-\sigma^{12}$.

Now the Majorana constraint becomes:

\begin{equation}
    \psi=M_{\mathcal{C}}\psi^*=\psi^*,
\end{equation}
which means $\psi$ is real and we can throw the imaginary part of $\psi$ away. Majorana fermion is just a real fermion satisfying Dirac equation on the Majorana basis. This embedding process can be done in all dimensions, and by embedding to $\mathbb{R}(n)$, we can always find a real solution of $\psi$, the trivial charge conjugation becomes trivial complex conjugation in this sense.

\section{Brief Review of Symplectic Majorana Fermion in $4+1d$ Case}\label{app:Maj_2}

In this appendix, we review the explicit process of deriving symplectic Majorana fermion from a \emph{pair} of two Dirac fermions. Also, we show how the $\mathbb{H}(2)$ theory can be embedded in $\mathbb{R}(8)$.

\subsection{Basis Independent Discussion}

For spacetime dimension $d+1=0,1,2,3,4\bmod 8$, the representation algebra for Majorana fermion $\mathcal{C}\ell(d,0)$ and Dirac fermion $\mathcal{C}\ell(d)$ satisfy the relation:

\begin{equation}
    \text{dim}_{\mathbb{R}}\psi_{\mathcal{C}\ell(d)}=2\,\text{dim}_{\mathbb{R}}\chi_{\mathcal{C}\ell(d,0)},
\end{equation}
therefore we can always set a ``trivial charge conjugation" constraint to get a Majorana fermion from the corresponding Dirac fermion similar to the discussion above for $3+1d$.

However, for spacetime dimension $d+1=5,6,7\bmod 8$, the representation algebra for Majorana fermion $\mathcal{C}\ell(d,0)$ and Dirac fermion $\mathcal{C}\ell(d)$ satisfy the relation:

\begin{equation}
    \text{dim}_{\mathbb{R}}\psi_{\mathcal{C}\ell(d)}=\text{dim}_{\mathbb{R}}\chi_{\mathcal{C}\ell(d,0)},
\end{equation}
which motivates us to use a \emph{pair} of two Dirac fermions with trivial charge conjugation to construct the symplectic Majorana fermion.

\subsubsection{Complexification of Quaternion}

For $d+1=5\bmod 8$

\begin{equation}
    \mathcal{C}\ell(d,0)=\mathbb{H}(2^{\frac{d-2}{2}}),
\end{equation}
for $d+1=6\bmod 8$

\begin{equation}
    \mathcal{C}\ell(d,0)=\mathbb{H}(2^{\frac{d-3}{2}})\oplus\mathbb{H}(2^{\frac{d-3}{2}}),
\end{equation}
for $d+1=7\bmod 8$

\begin{equation}
    \mathcal{C}\ell(d,0)=\mathbb{H}(2^{\frac{d-2}{2}}),
\end{equation}
they're all quaternion type. In order to get a quaternion type spinor, we need to do ``trivial complex conjugation" from a biquaternion (complexified quaternion) type spinor. Complexification means that a complex number substitutes the coefficients for the corresponding algebra. It can be denoted as tensor product $\otimes_{\mathbb{R}}\mathbb{C}$. For $\mathbb{R},\mathbb{C},\mathbb{H}$, the result of complexification is:

\begin{equation}
    \mathbb{R}\otimes_{\mathbb{R}}\mathbb{C}\cong\mathbb{C},\quad \mathbb{C}\otimes_{\mathbb{R}}\mathbb{C}\cong\mathbb{C}\oplus\mathbb{C},\quad \mathbb{H}\otimes_{\mathbb{R}}\mathbb{C}\cong\mathbb{C}(2),
\end{equation}
and the last result is what we need. We can give a specific map from $\mathbb{H}\otimes_{\mathbb{R}}\mathbb{C}$ to $\mathbb{C}(2)$. Note that

\begin{equation}
    \begin{pmatrix}
        \ii&0\\0&-\ii
    \end{pmatrix}\begin{pmatrix}
        0&1\\-1&0
    \end{pmatrix}=\begin{pmatrix}
        0&\ii\\\ii&0
    \end{pmatrix},
\end{equation}
and these 3 matrices have squares equal to $-1$ while anticommuting with each other. We can map them to the units of quaternion:

\begin{equation}
    \ii'\sim \begin{pmatrix}
        \ii&0\\0&-\ii
    \end{pmatrix},\ \jj'\sim \begin{pmatrix}
        0&1\\-1&0
    \end{pmatrix},\ \kk'\sim \begin{pmatrix}
        0&\ii\\\ii&0
    \end{pmatrix},
\end{equation}
therefore a complexified quaternion $q=u+v\ii'+w\jj'+x\kk'$ can be mapped to

\begin{equation}
    q\sim\begin{pmatrix}
        u+\ii v&w+\ii x\\-w+\ii x&u-\ii v
    \end{pmatrix},
\end{equation}
where $u,v,w,x$ are complex numbers. From a general $2\times 2$ complex matrix

\begin{equation}
    M=\begin{pmatrix}
        c_{11}&c_{12}\\c_{21}&c_{22}
    \end{pmatrix},
\end{equation}
we can similarly map back to a complexified quaternion:

\begin{equation}
    M\sim \frac{c_{11}+c_{22}}{2}+\frac{c_{11}-c_{22}}{2\ii}\ii'+\frac{c_{12}-c_{21}}{2}\jj'+\frac{c_{12}+c_{21}}{2\ii}\kk'.
\end{equation}

Equipped with this mapping, we can return to the analysis of the symplectic Majorana fermion.

\subsubsection{Mapping from 2 Dirac Fermions to 1 Symplectic Majorana Fermion}

In order to go through the same ``trivial charge conjugation" process, we need a pair of Dirac fermions now.

Let's first examine the $5+1d$ case, where Dirac fermion is the representation of $\mathbb{C}(4)$ and symplectic Majorana fermion is the representation of $\mathbb{H}(2)$.

The process goes as follows:

\begin{itemize}
    \item Introduce 2 Dirac fermions $\psi_1$ and $\psi_2$, each $\psi_a$ has 4 complex components.

    \item Rearrange their components to form a $4\times 2$ matrix.

    \item Map each 4 components to a compexificated quaternion $\psi'^{\mathbb{C}}_i$. The $4\times 2$ matrix becomes a 2-component spinor with compexified quaternion value for each component.

    \item Introduce trivial charge conjugation constraint.
\end{itemize}

To be specific, the mapping is given by:

\begin{equation}
    \begin{pmatrix}
    (\psi_1)_1\\
    (\psi_1)_2\\
    (\psi_1)_3\\
    (\psi_1)_4
  \end{pmatrix}\text{ and }\begin{pmatrix}
    (\psi_2)_1\\
    (\psi_2)_2\\
    (\psi_2)_3\\
    (\psi_2)_4
  \end{pmatrix}\to
    \begin{pmatrix}
    (\psi_1)_1 &\vline\  (\psi_2)_1 \\
    (\psi_1)_2 &\vline\  (\psi_2)_2 \\
    (\psi_1)_3 &\vline\  (\psi_2)_3 \\
    (\psi_1)_4 &\vline\  (\psi_2)_4 
  \end{pmatrix}\to \left(\begin{array}{cc}
    (\psi_1)_1 & (\psi_2)_2 \\
    (\psi_1)_2 & (\psi_2)_1 \\
    \hline
    (\psi_1)_3 & (\psi_2)_4 \\
    (\psi_1)_4 & (\psi_2)_3 
  \end{array}\right)\to \left(\begin{array}{c}
    \psi'^{\mathbb{C}}_1 \\
    \hline
    \psi'^{\mathbb{C}}_2 
  \end{array}\right)\to \begin{pmatrix}
      \psi'_1\\\psi'_2
  \end{pmatrix}
\end{equation}
where

\bea\label{eq:map}
    \psi'^{\mathbb{C}}_1&=&\frac{(\psi_1)_1+(\psi_2)_1}{2}+\frac{(\psi_1)_1-(\psi_2)_1}{2\ii}\ii'+\frac{(\psi_2)_2-(\psi_1)_2}{2}\jj'+\frac{(\psi_2)_2+(\psi_1)_2}{2\ii}\kk',\\
    \psi'^{\mathbb{C}}_2&=&\frac{(\psi_1)_3+(\psi_2)_3}{2}+\frac{(\psi_1)_3-(\psi_2)_3}{2\ii}\ii'+\frac{(\psi_2)_4-(\psi_1)_4}{2}\jj'+\frac{(\psi_2)_4+(\psi_1)_4}{2\ii}\kk'.
\eea

The trivial charge conjugation constraint can be set analogous to the $3+1d$ case as:

\begin{equation}
    \psi'=M'_{\mathcal{C}}\psi'^*,
\end{equation}
or

\begin{equation}
    \bar{\psi}'=\pm \psi'^{\mathsf{T}} C',
\end{equation}
where

\begin{equation}
    M'_{\mathcal{C}}=C'^{-1}(\gamma^0)^{\mathsf{T}},
\end{equation}
and

\begin{equation}
    \bar{\psi}'\overset{def}{=\!=}\psi'^\dagger\gamma^0,
\end{equation}
where the complex conjugation act only on $i$ but not on $\ii',\jj',\kk'$.

Now the quaternion spinor $\psi'$ has the following Lagrangian

\begin{equation}
    L=\frac{1}{2}\int \dd^4 x \bar{\psi}'(\ii\gamma^\mu\partial_\mu-m)\psi'=\frac{1}{2}\int \dd^4 x \psi'^{\mathsf{T}}C'(\ii\gamma^\mu\partial_\mu-m)\psi',
\end{equation}
and the EOM is given by

\begin{equation}
    (\ii\gamma^\mu\partial_\mu-m)\psi'=0
\end{equation}
with trivial charge conjugation constraint

\begin{equation}
    \bar{\psi}'=\pm \psi'^{\mathsf{T}} C'.
\end{equation}

\subsubsection{Summary}

In representation space, the combination process is given by

\begin{equation}
    4_{\mathbb{C}}\text{ and }4_{\mathbb{C}}\to M_{4\times 2}(\mathbb{C})\to M_{2\times 2}(\mathbb{C})\oplus M_{2\times 2}(\mathbb{C})\to 2_{\mathbb{H}\otimes_{\mathbb{R}}\mathbb{C}}\to 2_{\mathbb{H}}.
\end{equation}

In the Clifford algebra sense, the combination process is given by

\begin{equation}
    \mathbb{C}(4)\text{ and }\mathbb{C}(4)\to\mathbb{C}(4)\to\mathbb{C}(4)\to\mathbb{H}(2)\otimes_{\mathbb{R}}\mathbb{C}\to\mathbb{H}(2).
\end{equation}

\subsection{Symplectic Majorana Basis}

Similar to the Majorana basis, we want to find a symplectic Majorana basis in $4+1d$ so that the matrices $\ii\gamma^\mu$ do not contain the imaginary unit $\ii$, and are quaternion-valued (rather than biquaternion). Once the symplectic Majorana basis is found, the matrix $M'_{\mathcal{C}}$ can be set to identity, and the trivial charge conjugation constraint becomes $\psi'^*=\psi'$. Under this basis, $\psi'$ can be quaternion-valued instead of biquaternion-valued.

Unfortunately, for $d+1=5\bmod 8$, this symplectic Majorana basis does not exist. However, if $m=0$, we can still find symplectic Majorana basis such that the matrices $\gamma^\mu$ do not contain the imaginary unit $\ii$, and the EOM for massless symplectic Majorana fermion is imaginary.

To simplify the notation, we define:

\begin{equation}
    \sigma^0=\begin{pmatrix}
        1&0\\0&1
    \end{pmatrix},\ \sigma^1=\begin{pmatrix}
        0&1\\1&0
    \end{pmatrix},\ \tilde{\sigma}^2=-\ii\sigma^2=\begin{pmatrix}
        0&-1\\1&0
    \end{pmatrix},\ \sigma^3=\begin{pmatrix}
        1&0\\0&-1
    \end{pmatrix},
\end{equation}
and $\sigma^0,\sigma^1,\tilde{\sigma}^2,\sigma^3$ do not explicitly contain imaginary unit $i$. 

The symplectic Majorana basis is given by:

\begin{equation}
\begin{aligned}
    \gamma^0&=&\sigma^{1}&=&\begin{pmatrix}
        0&1\\
        1&0
    \end{pmatrix},\ \gamma^1&=&\ii'\sigma^{3}&=&\begin{pmatrix}
        \ii'&0\\
        0&-\ii'
    \end{pmatrix}&,\ \gamma^2&=&\jj'\sigma^{3}&=&\begin{pmatrix}
        \jj'&0\\
        0&-\jj'
    \end{pmatrix},\\
    \gamma^3&=&\kk'\sigma^{3}&=&\begin{pmatrix}
        \kk'&0\\
        0&-\kk'
    \end{pmatrix},\ \gamma^4&=&\tilde{\sigma}^{2}&=&\begin{pmatrix}
        0&-1\\
        1&0
    \end{pmatrix}&.&&
\end{aligned}
\end{equation}

When $d+1=6,7$, we can find gamma matrices satisfying $\ii\gamma^\mu$ is quaternion-valued. We give the symplectic Majorana basis for $d+1=6$ and $d+1=7$.

For $d+1=6$, symplectic Majorana basis is given by:

\begin{equation}
\begin{aligned}
    \ii\gamma^0&=&\sigma^{0}\otimes\tilde{\sigma}^2&=&\begin{pmatrix}
        0&-1&0&0\\
        1&0&0&0\\
        0&0&0&-1\\
        0&0&1&0
    \end{pmatrix},\ \ii\gamma^1&=&\ii'\tilde{\sigma}^{2}\otimes\sigma^1&=&\begin{pmatrix}
        0&0&0&-\ii'\\
        0&0&-\ii'&0\\
        0&\ii'&0&0\\
        \ii'&0&0&0
    \end{pmatrix},\\
    \ii\gamma^2&=&\jj'\tilde{\sigma}^{2}\otimes\sigma^1&=&\begin{pmatrix}
        0&0&0&-\jj'\\
        0&0&-\jj'&0\\
        0&\jj'&0&0\\
        \jj'&0&0&0
    \end{pmatrix},\ \ii\gamma^3&=&\kk'\tilde{\sigma}^{2}\otimes\sigma^1&=&\begin{pmatrix}
        0&0&0&-\kk'\\
        0&0&-\kk'&0\\
        0&\kk'&0&0\\
        \kk'&0&0&0
    \end{pmatrix},\\
    \ii\gamma^4&=&\sigma^{1}\otimes\sigma^1&=&\begin{pmatrix}
        0&0&0&1\\
        0&0&1&0\\
        0&1&0&0\\
        1&0&0&0
    \end{pmatrix},\ \ii\gamma^5&=&\sigma^{3}\otimes\sigma^1&=&\begin{pmatrix}
        0&1&0&0\\
        1&0&0&0\\
        0&0&0&-1\\
        0&0&-1&0
    \end{pmatrix}.
\end{aligned}
\end{equation}

For $d+1=7$, symplectic Majorana basis is given by:

\begin{equation}
\begin{aligned}
    \ii\gamma^0&=&\sigma^{0}\otimes\tilde{\sigma}^2&=&\begin{pmatrix}
        0&-1&0&0\\
        1&0&0&0\\
        0&0&0&-1\\
        0&0&1&0
    \end{pmatrix},\ \ii\gamma^1&=&\ii'\tilde{\sigma}^{2}\otimes\sigma^1&=&\begin{pmatrix}
        0&0&0&-\ii'\\
        0&0&-\ii'&0\\
        0&\ii'&0&0\\
        \ii'&0&0&0
    \end{pmatrix},\\
    \ii\gamma^2&=&\jj'\tilde{\sigma}^{2}\otimes\sigma^1&=&\begin{pmatrix}
        0&0&0&-\jj'\\
        0&0&-\jj'&0\\
        0&\jj'&0&0\\
        \jj'&0&0&0
    \end{pmatrix},\ \ii\gamma^3&=&\kk'\tilde{\sigma}^{2}\otimes\sigma^1&=&\begin{pmatrix}
        0&0&0&-\kk'\\
        0&0&-\kk'&0\\
        0&\kk'&0&0\\
        \kk'&0&0&0
    \end{pmatrix},\\
    \ii\gamma^4&=&\sigma^{1}\otimes\sigma^1&=&\begin{pmatrix}
        0&0&0&1\\
        0&0&1&0\\
        0&1&0&0\\
        1&0&0&0
    \end{pmatrix},\ \ii\gamma^5&=&\sigma^{3}\otimes\sigma^1&=&\begin{pmatrix}
        0&1&0&0\\
        1&0&0&0\\
        0&0&0&-1\\
        0&0&-1&0
    \end{pmatrix}&,\\
    \ii\gamma^6&=&\sigma^{0}\otimes\sigma^3&=&\begin{pmatrix}
        1&0&0&0\\
        0&-1&0&0\\
        0&0&1&0\\
        0&0&0&-1
    \end{pmatrix}.&\ &&&
\end{aligned}
\end{equation}

They satisfy

\begin{equation}
    (\gamma^0)^2=1,\ (\gamma^i)^2=-1,\ (\text{for }i=1,2,3,...),
\end{equation}
and they anticommute with each other.

\subsection{Embedded in $\mathbb{C}(4)$}

A more convenient choice is to embed the $\mathbb{H}(2)$ theory into $\mathbb{C}(4)$, which is the original Dirac theory. The constraint will rule out half of the DOF. Luckily, we can use only $\psi_1$ to describe the theory.

We can write the theory in the original two Dirac fermions $\psi_1$ and $\psi_2$, each with 4 complex components. Now the Lagrangian is given by

\begin{equation}
    L=\frac{1}{2}\int \dd^4 x \sum_{a=1}^2 \bar{\psi}_a(\ii\gamma^\mu\partial_\mu-m)\psi_a
\end{equation}

The trivial charge conjugation condition is

\begin{equation}
    \bar{\psi}'=\pm \psi'^{\mathsf{T}}C'
\end{equation}
written in $\psi_1$ and $\psi_2$, we need to use the map in Eq.~(\ref{eq:map}). To further simplify the relation, we assume $C'=\ii'C''$ and $C''$ does not contain quaternion unit $\ii',\jj',\kk'$:

\begin{equation}
    (\bar{\psi}_1)_i=\pm ((\psi_1)^{\mathsf{T}}C)_i,\quad (\bar{\psi}_2)_i=\mp ((\psi_2)^{\mathsf{T}}C)_i
\end{equation}
which gives

\begin{equation}
    \bar{\psi}_a=\pm\sum_{b}\varepsilon_{ab}\psi_b^{\mathsf{T}} C
\end{equation}

Therefore, the Lagrangian can be reduced to

\begin{equation}\label{eq:Lagrangian}
    L=\frac{1}{2}\int \dd^4 x\sum_{a,b}\psi_b^{\mathsf{T}}C\varepsilon_{ab}(\ii\gamma^\mu\partial_\mu-m)\psi_a
\end{equation}

The EOM is still Dirac equation

\begin{equation}
    (\ii\gamma^\mu\partial_\mu-m)\psi_1=(\ii\gamma^\mu\partial_\mu-m)\psi_2=0
\end{equation}
with the constraint

\begin{equation}
    \bar{\psi}_a=\pm\sum_{b}\varepsilon_{ab}\psi_b^{\mathsf{T}} C
\end{equation}
or

\begin{equation}\label{eq:ab}
    \psi_a=\sum_{b}\varepsilon_{ab}C^{-1}(\gamma^0)^{\mathsf{T}}\psi_b^*
\end{equation}

With some explicit basis, we can substitute Eq.~(\ref{eq:ab}) into Eq.~(\ref{eq:Lagrangian}) so that we can describe the theory with only $\psi_1$.

\subsection{Embedded in $\mathbb{R}(8)$}

A convenient choice is to embed the symplectic Majorana theory into $\mathbb{R}(8)$. Now we can write $\ii\gamma^\mu$ as real matrices in $\mathbb{R}(8)$ space, and it suffices to set $M_{\mathcal{C}}$ to identity. The Majorana fermion is embedded in $8_\mathbb{R}$, characterized by real numbers.

Now we give one explicit basis to show that this embedding is possible:

For $d+1=5$:

\begin{equation}
    \ii\gamma^0=i\sigma^{211},\ \ii\gamma^1=\sigma^{100},\ \ii\gamma^2=\sigma^{212},\ \ii\gamma^3=\sigma^{220},\ \ii\gamma^4=\sigma^{300}.
\end{equation}

For $d+1=6$:

\begin{equation}
    \ii\gamma^0=i\sigma^{1002},\ \ii\gamma^1=\sigma^{3100},\ \ii\gamma^2=\sigma^{3212},\ \ii\gamma^3=\sigma^{3220},\ \ii\gamma^4=\sigma^{3232},\ \ii\gamma^5=\sigma^{3300}.
\end{equation}

For $d+1=7$:

\begin{equation}
    \ii\gamma^0=i\sigma^{2000},\ \ii\gamma^1=\sigma^{1000},\ \ii\gamma^2=\sigma^{3100},\ \ii\gamma^3=\sigma^{3212},\ \ii\gamma^4=\sigma^{3220},\ \ii\gamma^5=\sigma^{3232},\ \ii\gamma^6=\sigma^{3300}.
\end{equation}

The Lagrangian is given by

\begin{equation}
    L=\frac{1}{2}\int \dd^d x \bar{\psi}(\ii\gamma^\mu\partial_\mu-m)\psi.
\end{equation}

The EOM is given by

\begin{equation}
    (\ii\gamma^\mu\partial_\mu-m)\psi=0,
\end{equation}
without any constraint.

\section{Symplectic Majorana Fermion in $5+1d$ and $6+1d$ Cases}\label{app:Maj_3}

The process of combining 2 Dirac fermions is similar. For $d+1=6$:

In representation space, the combination process is

\begin{equation}
\begin{aligned}
    4_{\mathbb{C}}\oplus4_{\mathbb{C}}\text{ and }4_{\mathbb{C}}\oplus4_{\mathbb{C}}&\to M_{4\times 2}(\mathbb{C})\oplus M_{4\times 2}(\mathbb{C})\to (M_{2\times 2}(\mathbb{C})\oplus M_{2\times 2}(\mathbb{C}))\oplus (M_{2\times 2}(\mathbb{C})\oplus M_{2\times 2}(\mathbb{C}))\\
    &\to 2_{\mathbb{H}\otimes_{\mathbb{R}}\mathbb{C}}\oplus 2_{\mathbb{H}\otimes_{\mathbb{R}}\mathbb{C}}\to 2_{\mathbb{H}}\oplus 2_{\mathbb{H}}.\\
\end{aligned}
\end{equation}

In the Clifford algebra sense, the combination process is

\begin{equation}
    \mathbb{C}(4)\oplus\mathbb{C}(4)\text{ and }\mathbb{C}(4)\oplus\mathbb{C}(4)\to\mathbb{C}(4)\oplus\mathbb{C}(4)\to\mathbb{C}(4)\oplus\mathbb{C}(4)\to(\mathbb{H}(2)\otimes_{\mathbb{R}}\mathbb{C})\oplus (\mathbb{H}(2)\otimes_{\mathbb{R}}\mathbb{C})\to\mathbb{H}(2)\oplus \mathbb{H}(2).
\end{equation}

For $d+1=7$:

In representation space, the combination process is

\begin{equation}
    8_{\mathbb{C}}\text{ and }8_{\mathbb{C}}\to M_{8\times 2}(\mathbb{C})\to M_{2\times 2}(\mathbb{C})\oplus M_{2\times 2}(\mathbb{C})\oplus M_{2\times 2}(\mathbb{C})\oplus M_{2\times 2}(\mathbb{C})\to 4_{\mathbb{H}\otimes_{\mathbb{R}}\mathbb{C}}\to 4_{\mathbb{H}}.
\end{equation}

In the Clifford algebra sense, the combination process is

\begin{equation}
    \mathbb{C}(8)\text{ and }\mathbb{C}(8)\to\mathbb{C}(8)\to\mathbb{C}(8)\to\mathbb{H}(4)\otimes_{\mathbb{R}}\mathbb{C}\to\mathbb{H}(4).
\end{equation}

\section{Lorentz Symmetry in Minkowski Spacetime}\label{app:verif_Minkowski}

In this section, we'll check that the Lorentz boost and rotation in \Eq{eq:Lorentz-Minkowski} indeed keep the Lagrangian in \Eq{eq:Minkowski} (and the corresponding action $S=\int \dd x^0L$) invariant.

We first check Lorentz boost: take the \(\zeta _1\) boost for example, the differential operators transform as

\begin{equation}
\left(
\begin{array}{c}
 \partial _0 \\
 \partial _1 \\
\end{array}
\right)\rightarrow \exp \left(
\begin{array}{cc}
 0 & \zeta _1 \\
 \zeta _1 & 0 \\
\end{array}
\right)\left(
\begin{array}{c}
 \partial _0 \\
 \partial _1 \\
\end{array}
\right)=\left(
\begin{array}{cc}
 \cosh  \zeta _1 & \sinh  \zeta _1 \\
 \sinh  \zeta _1 & \cosh  \zeta _1 \\
\end{array}
\right)\left(
\begin{array}{c}
 \partial _0 \\
 \partial _1 \\
\end{array}
\right),
\end{equation}
and the Clifford algebra transforms as

\bea
\left(
\begin{array}{c}
 \mathbf{1} \\
 \alpha _1 \\
\end{array}
\right)&\to&\left(\e^{\frac{1}{2}\zeta _1\alpha _1}\right)^\mathsf{T}\left(
\begin{array}{c}
 \mathbf{1} \\
 \alpha _1 \\
\end{array}
\right)\e^{\frac{1}{2}\zeta _1\alpha _1}
=\e^{\zeta _1\alpha _1}\left(
\begin{array}{c}
 \mathbf{1} \\
 \alpha _1 \\
\end{array}
\right)\nn\\
&=&\left(\cosh  \zeta _1+\sinh  \zeta _1\alpha _1\right)\left(
\begin{array}{c}
 \mathbf{1} \\
 \alpha _1 \\
\end{array}
\right)\nn\\
&=&\left(
\begin{array}{cc}
 \cosh  \zeta _1 & \sinh  \zeta _1 \\
 \sinh  \zeta _1 & \cosh  \zeta _1 \\
\end{array}
\right)\left(
\begin{array}{c}
 \mathbf{1} \\
 \alpha _1 \\
\end{array}
\right),
\eea
therefore,

\bea
\ii\partial _0-\ii\partial _1\alpha _1&=&\ii \left(
\begin{array}{cc}
 \partial _0 & \partial _1 \\
\end{array}
\right)\left(
\begin{array}{cc}
 1 & 0 \\
 0 & -1 \\
\end{array}
\right)\left(
\begin{array}{c}
 \mathbf{1} \\
 \alpha _1 \\
\end{array}
\right)\nn\\
&\to& \ii \left(
\begin{array}{cc}
 \partial _0 & \partial _1 \\
\end{array}
\right)\left(
\begin{array}{cc}
 \cosh  \zeta _1 & \sinh  \zeta _1 \\
 \sinh  \zeta _1 & \cosh  \zeta _1 \\
\end{array}
\right)\left(
\begin{array}{cc}
 1 & 0 \\
 0 & -1 \\
\end{array}
\right)\left(
\begin{array}{cc}
 \cosh  \zeta _1 & \sinh  \zeta _1 \\
 \sinh  \zeta _1 & \cosh  \zeta _1 \\
\end{array}
\right)\left(
\begin{array}{c}
 \mathbf{1} \\
 \alpha _1 \\
\end{array}
\right)\nn\\
&=&\ii \left(
\begin{array}{cc}
 \partial _0 & \partial _1 \\
\end{array}
\right)\left(
\begin{array}{cc}
 \cosh  \zeta _1 & \sinh  \zeta _1 \\
 \sinh  \zeta _1 & \cosh  \zeta _1 \\
\end{array}
\right)\left(
\begin{array}{cc}
 \cosh  \zeta _1 & -\sinh  \zeta _1 \\
 -\sinh  \zeta _1 & \cosh  \zeta _1 \\
\end{array}
\right)\left(
\begin{array}{cc}
 1 & 0 \\
 0 & -1 \\
\end{array}
\right)\left(
\begin{array}{c}
 \mathbf{1} \\
 \alpha _1 \\
\end{array}
\right)\nn\\
&=&\ii \left(
\begin{array}{cc}
 \partial _0 & \partial _1 \\
\end{array}
\right)\left(
\begin{array}{cc}
 1 & 0 \\
 0 & -1 \\
\end{array}
\right)\left(
\begin{array}{c}
 \mathbf{1} \\
 \alpha _1 \\
\end{array}
\right)\nn\\
&=&\ii\partial _0-\ii\partial _1\alpha _1,
\eea
which is indeed invariant.

Then we check spatial rotation: take the \(\theta _{1\, 2}\) rotation for example, the differential operators transform as

\begin{equation}
\left(
\begin{array}{c}
 \partial _1 \\
 \partial _2 \\
\end{array}
\right)\rightarrow \exp \left(
\begin{array}{cc}
 0 & -\theta _{1\, 2} \\
 \theta _{1\, 2} & 0 \\
\end{array}
\right)\left(
\begin{array}{c}
 \partial _1 \\
 \partial _2 \\
\end{array}
\right)=\left(
\begin{array}{cc}
 \cos  \theta _{1\, 2} & -\sin  \theta _{1\, 2} \\
 \sin  \theta _{1\, 2} & \cos  \theta _{1\, 2} \\
\end{array}
\right)\left(
\begin{array}{c}
 \partial _1 \\
 \partial _2 \\
\end{array}
\right),
\end{equation}
and the Clifford algebra transforms as

\bea
\left(
\begin{array}{c}
 \alpha _1 \\
 \alpha _2 \\
\end{array}
\right)&\to&\left(\e^{\frac{\ii}{2}\theta _{1\, 2}\Sigma _{1\, 2}}\right)^\mathsf{T}\left(
\begin{array}{c}
 \alpha _1 \\
 \alpha _2 \\
\end{array}
\right)\e^{\frac{\ii}{2}\theta _{1\, 2}\Sigma _{1\, 2}}\nn\\
&=&\e^{-\ii \theta _{1\, 2}\Sigma _{1\, 2}}\left(
\begin{array}{c}
 \alpha _1 \\
 \alpha _2 \\
\end{array}
\right)\nn\\
&=&\left(\cos  \theta _{1\, 2}-\ii \sin  \theta _{1\, 2} \Sigma _{1\, 2}\right)\left(
\begin{array}{c}
 \alpha _1 \\
 \alpha _2 \\
\end{array}
\right)\nn\\
&=&\left(\cos  \theta _{1\, 2}+ \sin  \theta _{1\, 2} \alpha _1\alpha _2\right)\left(
\begin{array}{c}
 \alpha _1 \\
 \alpha _2 \\
\end{array}
\right)\nn\\
&=&\left(
\begin{array}{cc}
 \cos  \theta _{1\, 2} & -\sin  \theta _{1\, 2} \\
 \sin  \theta _{1\, 2} & \cos  \theta _{1\, 2} \\
\end{array}
\right)\left(
\begin{array}{c}
 \alpha _1 \\
 \alpha _2 \\
\end{array}
\right),
\eea
therefore,

\bea
-\ii\partial _1\alpha _1-\ii\partial _2\alpha _2&=&-\ii \left(
\begin{array}{cc}
 \partial _1 & \partial _2 \\
\end{array}
\right)\left(
\begin{array}{c}
 \alpha _1 \\
 \alpha _2 \\
\end{array}
\right)\nn\\
&\rightarrow& -\ii \left(
\begin{array}{cc}
 \partial _1 & \partial _2 \\
\end{array}
\right)\left(
\begin{array}{cc}
 \cos  \theta _{1\, 2} & \sin  \theta _{1\, 2} \\
 -\sin  \theta _{1\, 2} & \cos  \theta _{1\, 2} \\
\end{array}
\right)\left(
\begin{array}{cc}
 \cos  \theta _{1\, 2} & -\sin  \theta _{1\, 2} \\
 \sin  \theta _{1\, 2} & \cos  \theta _{1\, 2} \\
\end{array}
\right)\left(
\begin{array}{c}
 \alpha _1 \\
 \alpha _2 \\
\end{array}
\right)\nn\\
&=&-\ii \left(
\begin{array}{cc}
 \partial _1 & \partial _2 \\
\end{array}
\right)\left(
\begin{array}{c}
 \alpha _1 \\
 \alpha _2 \\
\end{array}
\right)\nn\\
&=&-\ii\partial _1\alpha _1-\ii\partial _2\alpha _2,
\eea
which is indeed invariant.

\section{Lorentz Symmetry in Euclidean Spacetime}\label{app:verif_Euclidean}

In this section, we'll check that the Lorentz boost and rotation in \Eq{eq:Lorentz-Euclidean} indeed keep the Lagrangian in \Eq{eq:Euclidean} (and the corresponding action $S=\int \dd x^0L$) invariant.

We first check Lorentz boost: take the \(\zeta _1\) boost for example, the differential operators transform as

Check Lorentz boost: take the \(\zeta _1\) boost for example, the differential operators transform as

\begin{equation}
\left(
\begin{array}{c}
 \partial _0 \\
 \partial _1 \\
\end{array}
\right)\rightarrow \exp \left(
\begin{array}{cc}
 0 & -\zeta _1 \\
 \zeta _1 & 0 \\
\end{array}
\right)\left(
\begin{array}{c}
 \partial _0 \\
 \partial _1 \\
\end{array}
\right)=\left(
\begin{array}{cc}
 \cos  \zeta _1 & -\sin  \zeta _1 \\
 \sin  \zeta _1 & \cos  \zeta _1 \\
\end{array}
\right)\left(
\begin{array}{c}
 \partial _0 \\
 \partial _1 \\
\end{array}
\right),
\end{equation}
and the Clifford algebra transforms as

\bea
\left(
\begin{array}{c}
 \mathbf{1} \\
 \alpha _1 \\
\end{array}
\right)&\to&\left(\e^{-\frac{\ii}{2}\zeta _1\alpha _1}\right)^\mathsf{T}\left(
\begin{array}{c}
 \mathbf{1} \\
 \alpha _1 \\
\end{array}
\right)\e^{-\frac{\ii}{2}\zeta _1\alpha _1}\nn\\
&=&\e^{-\ii \zeta _1\alpha _1}\left(
\begin{array}{c}
 \mathbf{1} \\
 \alpha _1 \\
\end{array}
\right)\nn\\
&=&\left(\cos  \zeta _1-\ii \sin  \zeta _1\alpha _1\right)\left(
\begin{array}{c}
 \mathbf{1} \\
 \alpha _1 \\
\end{array}
\right)\nn\\
&=&\left(
\begin{array}{cc}
 \cos  \zeta _1 & -\ii \sin  \zeta _1 \\
 -\ii \sin  \zeta _1 & \cos  \zeta _1 \\
\end{array}
\right)\left(
\begin{array}{c}
 \mathbf{1} \\
 \alpha _1 \\
\end{array}
\right),
\eea
therefore,

\bea
\partial _0+\ii\partial _1\alpha _1&=&\left(
\begin{array}{cc}
 \partial _0 & \partial _1 \\
\end{array}
\right)\left(
\begin{array}{cc}
 1 & 0 \\
 0 & \ii \\
\end{array}
\right)\left(
\begin{array}{c}
 \mathbf{1} \\
 \alpha _1 \\
\end{array}
\right)\nn\\
&\to&\left(
\begin{array}{cc}
 \partial _0 & \partial _1 \\
\end{array}
\right)\left(
\begin{array}{cc}
 \cos  \zeta _1 & \sin  \zeta _1 \\
 -\sin  \zeta _1 & \cos  \zeta _1 \\
\end{array}
\right)\left(
\begin{array}{cc}
 1 & 0 \\
 0 & \ii \\
\end{array}
\right)\left(
\begin{array}{cc}
 \cos  \zeta _1 & -\ii \sin  \zeta _1 \\
 -\ii \sin  \zeta _1 & \cos  \zeta _1 \\
\end{array}
\right)\left(
\begin{array}{c}
 \mathbf{1} \\
 \alpha _1 \\
\end{array}
\right)\nn\\
&=&\left(
\begin{array}{cc}
 \partial _0 & \partial _1 \\
\end{array}
\right)\left(
\begin{array}{cc}
 \cos  \zeta _1 & \sin  \zeta _1 \\
 -\sin  \zeta _1 & \cos  \zeta _1 \\
\end{array}
\right)\left(
\begin{array}{cc}
 \cos  \zeta _1 & -\sin  \zeta _1 \\
 \sin  \zeta _1 & \cos  \zeta _1 \\
\end{array}
\right)\left(
\begin{array}{cc}
 1 & 0 \\
 0 & \ii \\
\end{array}
\right)\left(
\begin{array}{c}
 \mathbf{1} \\
 \alpha _1 \\
\end{array}
\right)\nn\\
&=&\left(
\begin{array}{cc}
 \partial _0 & \partial _1 \\
\end{array}
\right)\left(
\begin{array}{cc}
 1 & 0 \\
 0 & \ii \\
\end{array}
\right)\left(
\begin{array}{c}
 \mathbf{1} \\
 \alpha _1 \\
\end{array}
\right)\nn\\
&=&\partial _0+\ii\partial _1\alpha _1.
\eea
which is indeed invariant.

The rotation in Euclidean spacetime is the same as in Minkowski spacetime, therefore the verification of the invariance of the Lagrangian in \Eq{eq:Euclidean} rotation is parallel to that in \App{app:verif_Minkowski}.

\section{Presentation of the Invariant Group for Dirac Fermion}\label{app:presentation}

For d=0, the invariant group is given by the presentation:

\begin{equation}
\begin{aligned}
    &\mathcal{C}^2=\mathcal{T}^2=(\mathcal{CT})^2=1,\quad \mathcal{C}\mathcal{U}^F(\theta)=\mathcal{U}^F(-\theta)\mathcal{C},\quad \mathcal{T}\mathcal{U}^F(\theta)=\mathcal{U}^F(-\theta)\mathcal{T}.
\end{aligned}
\end{equation}

For d=1, the invariant group is given by the presentation:

\begin{equation}
\begin{aligned}
    &\mathcal{C}^2=\mathcal{R}_1^2=\mathcal{T}^2=(\mathcal{R}_1\mathcal{T})^2=1,\quad (\mathcal{CR}_1)^2=(\mathcal{CT})^2=(-)^F,\quad \mathcal{C}\mathcal{U}^F(\theta)=\mathcal{U}^F(-\theta)\mathcal{C},\quad \mathcal{R}_1\mathcal{U}^F(\theta)=\mathcal{U}^F(\theta)\mathcal{R}_1,\\
    &\mathcal{T}\mathcal{U}^F(\theta)=\mathcal{U}^F(-\theta)\mathcal{T},\quad \mathcal{C}\mathcal{U}^\chi(\theta)=\mathcal{U}^\chi(-\theta)\mathcal{C},\quad \mathcal{R}_1\mathcal{U}^\chi(\theta)=\mathcal{U}^\chi(-\theta)\mathcal{R}_1,\quad \mathcal{T}\mathcal{U}^\chi(\theta)=\mathcal{U}^\chi(\theta)\mathcal{T} .
\end{aligned}
\end{equation}

For d=2, the invariant group is given by the presentation:

\begin{equation}
\begin{aligned}
    &\mathcal{C}^2=\mathcal{R}_1^2=(\mathcal{CR}_1)^2=(\mathcal{R}_1\mathcal{T})^2=1,\quad \mathcal{T}^2=(\mathcal{CT})^2=(-)^F,\\
    &\mathcal{C}\mathcal{U}^F(\theta)=\mathcal{U}^F(-\theta)\mathcal{C},\quad \mathcal{R}_1\mathcal{U}^F(\theta)=\mathcal{U}^F(\theta)\mathcal{R}_1,\quad \mathcal{T}\mathcal{U}^F(\theta)=\mathcal{U}^F(-\theta)\mathcal{T}.
\end{aligned}
\end{equation}

For d=3, the invariant group is given by the presentation:

\begin{equation}
\begin{aligned}
    &\mathcal{C}^2=\mathcal{R}_1^2=(\mathcal{CR}_1)^2=(\mathcal{R}_1\mathcal{T})^2=1,\quad \mathcal{T}^2=(\mathcal{CT})^2=(-)^F,\quad \mathcal{C}\mathcal{U}^F(\theta)=\mathcal{U}^F(-\theta)\mathcal{C},\quad \mathcal{R}_1\mathcal{U}^F(\theta)=\mathcal{U}^F(\theta)\mathcal{R}_1,\\
    &\mathcal{T}\mathcal{U}^F(\theta)=\mathcal{U}^F(-\theta)\mathcal{T},\quad \mathcal{C}\mathcal{U}^\chi(\theta)=\mathcal{U}^\chi(\theta)\mathcal{C},\quad \mathcal{R}_1\mathcal{U}^\chi(\theta)=\mathcal{U}^\chi(-\theta)\mathcal{R}_1,\quad \mathcal{T}\mathcal{U}^\chi(\theta)=\mathcal{U}^\chi(-\theta)\mathcal{T} .
\end{aligned}
\end{equation}

For d=4, the invariant group is given by the presentation:

\begin{equation}
\begin{aligned}
    &(\mathcal{CR}_1)^2=(\mathcal{CT})^2=1,\quad \mathcal{C}^2=\mathcal{R}_1^2=\mathcal{T}^2=(\mathcal{R}_1\mathcal{T})^2=(-)^F,\\
    &\mathcal{C}\mathcal{U}^F(\theta)=\mathcal{U}^F(-\theta)\mathcal{C},\quad \mathcal{R}_1\mathcal{U}^F(\theta)=\mathcal{U}^F(\theta)\mathcal{R}_1,\quad \mathcal{T}\mathcal{U}^F(\theta)=\mathcal{U}^F(-\theta)\mathcal{T}.
\end{aligned}
\end{equation}

For d=5, the invariant group is given by the presentation:

\begin{equation}
\begin{aligned}
    &\mathcal{C}^2=\mathcal{R}_1^2=\mathcal{T}^2=(\mathcal{R}_1\mathcal{T})^2=1,\quad (\mathcal{CR}_1)^2=(\mathcal{CT})^2=(-)^F,\quad \mathcal{C}\mathcal{U}^F(\theta)=\mathcal{U}^F(-\theta)\mathcal{C},\quad \mathcal{R}_1\mathcal{U}^F(\theta)=\mathcal{U}^F(\theta)\mathcal{R}_1,\\
    &\mathcal{T}\mathcal{U}^F(\theta)=\mathcal{U}^F(-\theta)\mathcal{T},\quad \mathcal{C}\mathcal{U}^\chi(\theta)=\mathcal{U}^\chi(-\theta)\mathcal{C},\quad \mathcal{R}_1\mathcal{U}^\chi(\theta)=\mathcal{U}^\chi(-\theta)\mathcal{R}_1,\quad \mathcal{T}\mathcal{U}^\chi(\theta)=\mathcal{U}^\chi(\theta)\mathcal{T} .
\end{aligned}
\end{equation}

For d=6, the invariant group is given by the presentation:

\begin{equation}
\begin{aligned}
    &\mathcal{R}_1^2=\mathcal{T}^2=1,\quad \mathcal{C}^2=(\mathcal{CR}_1)^2=(\mathcal{R}_1\mathcal{T})^2=(\mathcal{CT})^2=(-)^F,\\
    &\mathcal{C}\mathcal{U}^F(\theta)=\mathcal{U}^F(-\theta)\mathcal{C},\quad \mathcal{R}_1\mathcal{U}^F(\theta)=\mathcal{U}^F(\theta)\mathcal{R}_1,\quad \mathcal{T}\mathcal{U}^F(\theta)=\mathcal{U}^F(-\theta)\mathcal{T}.
\end{aligned}
\end{equation}

For d=7, the invariant group is given by the presentation:

\begin{equation}
\begin{aligned}
    &\mathcal{R}_1^2=\mathcal{T}^2=1,\quad \mathcal{C}^2=(\mathcal{CR}_1)^2=(\mathcal{R}_1\mathcal{T})^2=(\mathcal{CT})^2=(-)^F,\quad \mathcal{C}\mathcal{U}^F(\theta)=\mathcal{U}^F(-\theta)\mathcal{C},\quad \mathcal{R}_1\mathcal{U}^F(\theta)=\mathcal{U}^F(\theta)\mathcal{R}_1,\\
    &\mathcal{T}\mathcal{U}^F(\theta)=\mathcal{U}^F(-\theta)\mathcal{T},\quad \mathcal{C}\mathcal{U}^\chi(\theta)=\mathcal{U}^\chi(\theta)\mathcal{C},\quad \mathcal{R}_1\mathcal{U}^\chi(\theta)=\mathcal{U}^\chi(-\theta)\mathcal{R}_1,\quad \mathcal{T}\mathcal{U}^\chi(\theta)=\mathcal{U}^\chi(-\theta)\mathcal{T} .
\end{aligned}
\end{equation}

For d=8, the invariant group is given by the presentation:

\begin{equation}
\begin{aligned}
    &\mathcal{C}^2=\mathcal{T}^2=(\mathcal{R}_1\mathcal{T})^2=(\mathcal{CT})^2=1,\quad \mathcal{R}_1^2=(\mathcal{CR}_1)^2=(-)^F,\\
    &\mathcal{C}\mathcal{U}^F(\theta)=\mathcal{U}^F(-\theta)\mathcal{C},\quad \mathcal{R}_1\mathcal{U}^F(\theta)=\mathcal{U}^F(\theta)\mathcal{R}_1,\quad \mathcal{T}\mathcal{U}^F(\theta)=\mathcal{U}^F(-\theta)\mathcal{T}.
\end{aligned}
\end{equation}

\section{Presentation of the Invariant Group for Weyl Fermion}\label{app:presentation_W}

For d=1, the invariant group is given by the presentation:

\begin{equation}
\begin{aligned}
    &\mathcal{C}^2=(\mathcal{R}_1\mathcal{T})^2=(\mathcal{CR}_1\mathcal{T})^2=1,\quad \mathcal{C}\mathcal{U}^F(\theta)=\mathcal{U}^F(-\theta)\mathcal{C},\quad (\mathcal{CR}_1\mathcal{T})\mathcal{U}^F(\theta)=\mathcal{U}^F(\theta)(\mathcal{CR}_1\mathcal{T}).
\end{aligned}
\end{equation}

For d=3, the invariant group is given by the presentation:

\begin{equation}
\begin{aligned}
    &(\mathcal{CR}_1\mathcal{T})^2=1,\quad \mathcal{T}^2=(\mathcal{CR}_1)^2=(-)^F,\quad \mathcal{T}\mathcal{U}^F(\theta)=\mathcal{U}^F(-\theta)\mathcal{T},\quad (\mathcal{CR}_1\mathcal{T})\mathcal{U}^F(\theta)=\mathcal{U}^F(\theta)(\mathcal{CR}_1\mathcal{T}).
\end{aligned}
\end{equation}

For d=5, the invariant group is given by the presentation:

\begin{equation}
\begin{aligned}
    &(\mathcal{R}_1\mathcal{T})^2=(\mathcal{CR}_1\mathcal{T})^2=1,\quad \mathcal{C}^2=(-)^F,\quad \mathcal{C}\mathcal{U}^F(\theta)=\mathcal{U}^F(-\theta)\mathcal{C},\quad (\mathcal{CR}_1\mathcal{T})\mathcal{U}^F(\theta)=\mathcal{U}^F(\theta)(\mathcal{CR}_1\mathcal{T}).
\end{aligned}
\end{equation}

For d=7, the invariant group is given by the presentation:

\begin{equation}
\begin{aligned}
    &\mathcal{T}^2=(\mathcal{CR}_1\mathcal{T})^2=1,\quad (\mathcal{CR}_1)^2=(-)^F,\quad \mathcal{T}\mathcal{U}^F(\theta)=\mathcal{U}^F(-\theta)\mathcal{T},\quad (\mathcal{CR}_1\mathcal{T})\mathcal{U}^F(\theta)=\mathcal{U}^F(\theta)(\mathcal{CR}_1\mathcal{T}).
\end{aligned}
\end{equation}

\section{Symmetry Reduction for Majorana Fermion}

To prove that the symmetry of spatial dimension $d$ can be reduced from the symmetry of spatial dimension $d+1$ once a mass term can be found, we give the explicit symmetry reduction in $d=1,2,3,4,5,6$ cases ($d=0,7\bmod 8$ has no mass terms) in the form:

\begin{equation}
    G_{M,d}\sim G'_{M,d}\xrightarrow{m_i}G'_{M/MW,d-1}\sim G_{M/MW,d-1},
\end{equation}
indicating the domain wall reduction of symmetry group $G$ from a $d$ dimensional Majorana fermion to a $d-1$ dimensional Majorana/Majorana-Weyl fermion on the $m_i$ mass domain wall. To do this, we first modify these symmetries by internal symmetries to maximally preserve the group $G'$ under mass domain wall $m_i$. Following the rules in Eqs.~(\ref{eq:rule1},\ref{eq:rule2}), we obtain the reduced symmetries $G'$, and modify the symmetry operators to conventional $G$. In the following content, we denote the generators of $\mathbb{Z}_2^F,\mathbb{Z}_2^\chi$,Lie algebra,$\mathbb{Z}_2^{\mathcal{R}_i},$ and $\mathbb{Z}_2^{\mathcal{T}}$ as $(-)^F,(-)^\chi,\mathcal{J}_i,\mathcal{R}_i,$ and $\mathcal{T}$, respectively.

\bea\label{dom-reduc-1}
    &&G_{M,d=1}((-)^F,(-)^\chi,\mathcal{R},\mathcal{T})\sim G_{M,d=1}((-)^F,(-)^\chi,\mathcal{R},\mathcal{T}'=(-)^\chi\mathcal{T})\nn\\
    &\xrightarrow{m_1}&G_{M,d=0}((-)^F,0,1,\mathcal{T})\sim G_{M,d=0}((-)^F,\mathcal{T}),
\eea
\bea
    &&G_{M,d=2}((-)^F,\mathcal{R}_1,\mathcal{R}_2,\mathcal{T})\nn\\
    &\xrightarrow{m_1}&G_{MW,d=1}((-)^F,\mathcal{R}\mathcal{T},1,1)\sim G_{MW,d=1}((-)^F,\mathcal{R}\mathcal{T}),
\eea
\bea
    &&G_{M,d=3}((-)^F,\mathcal{J},\mathcal{R}_1,\mathcal{R}_2,\mathcal{R}_3,\mathcal{T})\sim G_{M,d=3}((-)^F,\mathcal{J},\mathcal{R}_1,\mathcal{R}_2,\mathcal{R}'_3=\mathcal{JR}_3,\mathcal{T})\nn\\
    &\xrightarrow{m_1}&G_{M,d=2}((-)^F,0,\mathcal{R}_1,\mathcal{R}_2,1,\mathcal{T})\sim G_{M,d=2}((-)^F,\mathcal{R}_1,\mathcal{R}_2,\mathcal{T}),\nn\\
    &&G_{M,d=3}((-)^F,\mathcal{J},\mathcal{R}_1,\mathcal{R}_2,\mathcal{R}_3,\mathcal{T})\sim G_{M,d=3}((-)^F,\mathcal{J},\mathcal{R}_1'=\mathcal{JR}_1,\mathcal{R}_2'=\mathcal{JR}_2,\mathcal{R}_3,\mathcal{T}'=\mathcal{JT})\nn\\
    &\xrightarrow{m_2}&G_{M,d=2}((-)^F,0,\mathcal{R}_1,\mathcal{R}_2,1,\mathcal{T})\sim G_{M,d=2}((-)^F,\mathcal{R}_1,\mathcal{R}_2,\mathcal{T}),
\eea
\bea
    &&G_{M,d=4}((-)^F,\mathcal{J}_1,\mathcal{J}_2,\mathcal{J}_3,\mathcal{R}_1,\mathcal{R}_2,\mathcal{R}_3,\mathcal{R}_4,\mathcal{T})\sim G_{M,d=4}((-)^F,\mathcal{J}_1,\mathcal{J}_2,\mathcal{J}_3,\mathcal{R}_1,\mathcal{R}_2,\mathcal{R}_3,\mathcal{R}_4'=\mathcal{J}_1\mathcal{R}_4,\mathcal{T})\nn\\
    &\xrightarrow{m_1}&G_{M,d=3}((-)^F,0,0,\mathcal{J},\mathcal{R}_1,\mathcal{R}_2,\mathcal{R}_3,\mathcal{J},\mathcal{T})\sim G_{M,d=3}((-)^F,\mathcal{J},\mathcal{R}_1,\mathcal{R}_2,\mathcal{R}_3,\mathcal{T}),\nn\\
    &&G_{M,d=4}((-)^F,\mathcal{J}_1,\mathcal{J}_2,\mathcal{J}_3,\mathcal{R}_1,\mathcal{R}_2,\mathcal{R}_3,\mathcal{R}_4,\mathcal{T})\sim G_{M,d=4}((-)^F,\mathcal{J}_1,\mathcal{J}_2,\mathcal{J}_3,\mathcal{R}_1,\mathcal{R}_2,\mathcal{R}_3,\mathcal{R}_4'=\mathcal{J}_1\mathcal{R}_4,\mathcal{T})\nn\\
    &\xrightarrow{m_2}&G_{M,d=3}((-)^F,0,\mathcal{J},0,\mathcal{R}_1,\mathcal{R}_2,\mathcal{R}_3,\mathcal{J},\mathcal{T})\sim G_{M,d=3}((-)^F,\mathcal{J},\mathcal{R}_1,\mathcal{R}_2,\mathcal{R}_3,\mathcal{T}),\nn\\
    &&G_{M,d=4}((-)^F,\mathcal{J}_1,\mathcal{J}_2,\mathcal{J}_3,\mathcal{R}_1,\mathcal{R}_2,\mathcal{R}_3,\mathcal{R}_4,\mathcal{T})\nn\\
    &\sim &G_{M,d=4}((-)^F,\mathcal{J}_1,\mathcal{J}_2,\mathcal{J}_3,\mathcal{R}_1'=\mathcal{J}_2\mathcal{R}_1,\mathcal{R}_2'=\mathcal{J}_2\mathcal{R}_2,\mathcal{R}_3'=\mathcal{J}_2\mathcal{R}_3,\mathcal{R}_4,\mathcal{T}'=\mathcal{J}_2\mathcal{T})\nn\\
    &\xrightarrow{m_3}&G_{M,d=3}((-)^F,\mathcal{J},0,0,\mathcal{R}_1,\mathcal{R}_2,\mathcal{R}_3,1,\mathcal{T})\sim G_{M,d=3}((-)^F,\mathcal{J},\mathcal{R}_1,\mathcal{R}_2,\mathcal{R}_3,\mathcal{T}),
\eea
\bea
    &&G_{M,d=5}((-)^F,(-)^\chi,\mathcal{J}_1,\mathcal{J}_2,\mathcal{J}_3,\mathcal{R}_1,...,\mathcal{R}_4,\mathcal{R}_5,\mathcal{T})\nn\\
    &\sim &G_{M,d=5}((-)^F,(-)^\chi,\mathcal{J}_1,\mathcal{J}_2'=(-)^\chi\mathcal{J}_2,\mathcal{J}_3'=(-)^\chi\mathcal{J}_3,\mathcal{R}_1,...,\mathcal{R}_4,\mathcal{R}_5'=(-)^\chi\mathcal{R}_5,\mathcal{T})\nn\\
    &\xrightarrow{m_1}&G_{M,d=4}((-)^F,0,\mathcal{J}_3,\mathcal{J}_2,\mathcal{J}_1,\mathcal{J}_2\mathcal{R}_1,...,\mathcal{J}_2\mathcal{R}_4,\mathcal{J}_3,\mathcal{J}_2\mathcal{T})\sim G_{M,d=4}((-)^F,\mathcal{J}_1,\mathcal{J}_2,\mathcal{J}_3,\mathcal{R}_1,...,\mathcal{R}_4,\mathcal{T}),\nn\\
    &&G_{M,d=5}((-)^F,(-)^\chi,\mathcal{J}_1,\mathcal{J}_2,\mathcal{J}_3,\mathcal{R}_1,...,\mathcal{R}_4,\mathcal{R}_5,\mathcal{T})\nn\\
    &\sim &G_{M,d=5}((-)^F,(-)^\chi,\mathcal{J}_1'=(-)^\chi\mathcal{J}_1,\mathcal{J}_2,\mathcal{J}_3'=(-)^\chi\mathcal{J}_3,\mathcal{R}_1,...,\mathcal{R}_4,\mathcal{R}_5'=(-)^\chi\mathcal{R}_5,\mathcal{T})\nn\\
    &\xrightarrow{m_2}&G_{M,d=4}((-)^F,0,\mathcal{J}_3,\mathcal{J}_2,\mathcal{J}_1,\mathcal{J}_3\mathcal{R}_1,...,\mathcal{J}_3\mathcal{R}_4,\mathcal{J}_2,\mathcal{J}_3\mathcal{T})\sim G_{M,d=4}((-)^F,\mathcal{J}_1,\mathcal{J}_2,\mathcal{J}_3,\mathcal{R}_1,...,\mathcal{R}_4,\mathcal{T}),\nn\\
    &&G_{M,d=5}((-)^F,(-)^\chi,\mathcal{J}_1,\mathcal{J}_2,\mathcal{J}_3,\mathcal{R}_1,...,\mathcal{R}_4,\mathcal{R}_5,\mathcal{T})\nn\\
    &\sim &G_{M,d=5}((-)^F,(-)^\chi,\mathcal{J}_1'=(-)^\chi\mathcal{J}_1,\mathcal{J}_2'=(-)^\chi\mathcal{J}_2,\mathcal{J}_3,\mathcal{R}_1,...,\mathcal{R}_4,\mathcal{R}_5'=(-)^\chi\mathcal{R}_5,\mathcal{T})\nn\\
    &\xrightarrow{m_3}&G_{M,d=4}((-)^F,0,\mathcal{J}_3,\mathcal{J}_1,\mathcal{J}_2,\mathcal{J}_3\mathcal{R}_1,...,\mathcal{J}_3\mathcal{R}_4,\mathcal{J}_2,\mathcal{J}_3\mathcal{T})\sim G_{M,d=4}((-)^F,\mathcal{J}_1,\mathcal{J}_2,\mathcal{J}_3,\mathcal{R}_1,...,\mathcal{R}_4,\mathcal{T}),\nn\\
    &&G_{M,d=5}((-)^F,(-)^\chi,\mathcal{J}_1,\mathcal{J}_2,\mathcal{J}_3,\mathcal{R}_1,...,\mathcal{R}_4,\mathcal{R}_5,\mathcal{T})\nn\\
    &\sim &G_{M,d=5}((-)^F,(-)^\chi,\mathcal{J}_1,\mathcal{J}_2,\mathcal{J}_3,\mathcal{R}_1'=(-)^\chi\mathcal{R}_1,...,\mathcal{R}_4'=(-)^\chi\mathcal{R}_4,\mathcal{R}_5,\mathcal{T}'=(-)^\chi\mathcal{T})\nn\\
    &\xrightarrow{m_4}&G_{M,d=4}((-)^F,0,\mathcal{J}_3,\mathcal{J}_2,\mathcal{J}_1,\mathcal{J}_1\mathcal{R}_1,...,\mathcal{J}_1\mathcal{R}_4,1,\mathcal{J}_1\mathcal{T})\sim G_{M,d=4}((-)^F,\mathcal{J}_1,\mathcal{J}_2,\mathcal{J}_3,\mathcal{R}_1,...,\mathcal{R}_4,\mathcal{T}),
\eea
\bea\label{dom-reduc-6}
    &&G_{M,d=6}((-)^F,\mathcal{J}_1,\mathcal{J}_2,\mathcal{J}_3,\mathcal{R}_1,...,\mathcal{R}_5,\mathcal{R}_6,\mathcal{T})\nn\\
    &\xrightarrow{m_1}&G_{MW,d=5}((-)^F,\mathcal{J}_3,\mathcal{J}_2,\mathcal{J}_1,\mathcal{R}_1\mathcal{T},...,\mathcal{R}_5\mathcal{T},1,1)\sim G_{MW,d=5}((-)^F,\mathcal{J}_1,\mathcal{J}_2,\mathcal{J}_3,\mathcal{R}_1\mathcal{T},...,\mathcal{R}_5\mathcal{T}).
\eea

For higher spatial dimensions, the same procedure is applied to prove the symmetry reduction.

\section{Symmetry Reduction for Dirac Fermion}

To prove that the symmetry of spatial dimension $d$ can be reduced from the symmetry of spatial dimension $d+1$, we give the explicit symmetry reduction in $d=1,\dots,8$ cases in the form:

\begin{equation}
    G_{D,d}\sim G'_{D,d}\xrightarrow{m_i}G'_{D/W,d-1}\sim G_{D/W,d-1},
\end{equation}
indicating the domain wall reduction of symmetry group $G$ from a $d$ dimensional Dirac fermion to a $d-1$ dimensional Dirac/Weyl fermion on the $m_i$ mass domain wall. To do this, we first modify these symmetries by internal symmetries to maximally preserve the group $G'$ under mass domain wall $m_i$. Following the rules in Eqs.~(\ref{eq:rule1_Dirac},\ref{eq:rule2_Dirac}), we obtain the reduced symmetries $G'$, and modify the symmetry operators to conventional $G$. In the following content, we denote the generators of $\mathbb{Z}_2^F,\mathbb{Z}_2^\chi$,$\mathbb{Z}_2^{\mathcal{C}}$,$\mathbb{Z}_2^{\mathcal{R}_i},$ and $\mathbb{Z}_2^{\mathcal{T}}$ as $(-)^F,(-)^\chi,\mathcal{C},\mathcal{R}_i,$ and $\mathcal{T}$, respectively.

\bea\label{dom-reduc-Dirac-1}
    &&G_{D,d=1}((-)^F,(-)^\chi,\mathcal{C},\mathcal{R},\mathcal{T})\sim G_{D,d=1}((-)^F,(-)^\chi,\mathcal{C},\mathcal{R}'=(-)^\chi\mathcal{R},\mathcal{T})\nn\\
    &\xrightarrow{m_1}&G_{D,d=0}((-)^F,0,\mathcal{C},(-)^F,\mathcal{T})\sim G_{D,d=0}((-)^F,\mathcal{C},\mathcal{T}),\nn\\
    &&G_{D,d=1}((-)^F,(-)^\chi,\mathcal{C},\mathcal{R},\mathcal{T})\sim G_{D,d=1}((-)^F,(-)^\chi,\mathcal{C}'=(-)^\chi\mathcal{C},\mathcal{R},\mathcal{T})\nn\\
    &\xrightarrow{m_2}&G_{D,d=0}((-)^F,0,(-)^F\mathcal{C},1,\mathcal{T})\sim G_{D,d=0}((-)^F,\mathcal{C},\mathcal{T}),
\eea

\bea
    &&G_{D,d=2}((-)^F,\mathcal{C},\mathcal{R}_1,\mathcal{R}_2,\mathcal{T})\nn\\
    &\xrightarrow{m_1}&G_{DW,d=1}((-)^F,\mathcal{C},\mathcal{CRT},1,\mathcal{C})\sim G_{DW,d=1}((-)^F,\mathcal{C},\mathcal{CRT}),
\eea

\bea
    &&G_{D,d=3}((-)^F,(-)^\chi,\mathcal{C},\mathcal{R}_1,\mathcal{R}_2,\mathcal{R}_3,\mathcal{T})\nn\\
    &\sim& G_{D,d=3}((-)^F,(-)^\chi,\mathcal{C},\mathcal{R}'_1=(-)^\chi\mathcal{R}_1,\mathcal{R}'_2=(-)^\chi\mathcal{R}_2,\mathcal{R}_3,\mathcal{T}'=(-)^\chi\mathcal{T})\nn\\
    &\xrightarrow{m_1}&G_{D,d=2}((-)^F,0,\mathcal{C},(-)^F\mathcal{R}_1,(-)^F\mathcal{R}_2,1,(-)^F\mathcal{T})\sim G_{D,d=2}((-)^F,\mathcal{C},\mathcal{R}_1,\mathcal{R}_2,\mathcal{T}),\nn\\
    &&G_{D,d=3}((-)^F,(-)^\chi,\mathcal{C},\mathcal{R}_1,\mathcal{R}_2,\mathcal{R}_3,\mathcal{T})\sim G_{D,d=3}((-)^F,(-)^\chi,\mathcal{C},\mathcal{R}_1,\mathcal{R}_2,\mathcal{R}_3'=(-)^\chi\mathcal{R}_3,\mathcal{T})\nn\\
    &\xrightarrow{m_2}&G_{D,d=2}((-)^F,0,\mathcal{C},\mathcal{R}_1,\mathcal{R}_2,(-)^F,\mathcal{T})\sim G_{D,d=2}((-)^F,\mathcal{C},\mathcal{R}_1,\mathcal{R}_2,\mathcal{T}),
\eea

\bea
    &&G_{D,d=4}((-)^F,\mathcal{C},\mathcal{R}_1,\mathcal{R}_2,\mathcal{R}_3,\mathcal{R}_4,\mathcal{T})\nn\\
    &\xrightarrow{m_1}&G_{DW,d=3}((-)^F,\mathcal{T},\mathcal{CR}_1\mathcal{T},\mathcal{CR}_2\mathcal{T},\mathcal{CR}_3\mathcal{T},1,\mathcal{T})\sim G_{DW,d=3}((-)^F,\mathcal{T},\mathcal{CR}_1\mathcal{T},\mathcal{CR}_2\mathcal{T},\mathcal{CR}_3\mathcal{T}),
\eea

\bea
    &&G_{D,d=5}((-)^F,(-)^\chi,\mathcal{C},\mathcal{R}_1,...,\mathcal{R}_4,\mathcal{R}_5,\mathcal{T})\nn\\
    &\sim& G_{D,d=5}((-)^F,(-)^\chi,\mathcal{C}'=(-)^\chi\mathcal{C},\mathcal{R}'_1=(-)^\chi\mathcal{R}_1,...,\mathcal{R}'_4=(-)^\chi\mathcal{R}_4,\mathcal{R}_5,\mathcal{T})\nn\\
    &\xrightarrow{m_1}&G_{D,d=4}((-)^F,0,(-)^F\mathcal{C},(-)^F\mathcal{R}_1,...,(-)^F\mathcal{R}_4,1,\mathcal{T})\sim G_{D,d=4}((-)^F,\mathcal{C},\mathcal{R}_1,...,\mathcal{R}_4,\mathcal{T}),\nn\\
    &&G_{D,d=5}((-)^F,(-)^\chi,\mathcal{C},\mathcal{R}_1,...,\mathcal{R}_4,\mathcal{R}_5,\mathcal{T})\nn\\
    &\sim& G_{D,d=5}((-)^F,(-)^\chi,\mathcal{C}'=(-)^\chi\mathcal{C},\mathcal{R}'_1=(-)^\chi\mathcal{R}_1,...,\mathcal{R}'_4=(-)^\chi\mathcal{R}_4,\mathcal{R}_5,\mathcal{T})\nn\\
    &\xrightarrow{m_2}&G_{D,d=4}((-)^F,0,\mathcal{C},\mathcal{R}_1,...,\mathcal{R}_4,(-)^F,\mathcal{T})\sim G_{D,d=4}((-)^F,\mathcal{C},\mathcal{R}_1,...,\mathcal{R}_4,\mathcal{T}),\nn\\
\eea

\bea
    &&G_{D,d=6}((-)^F,\mathcal{C},\mathcal{R}_1,...,\mathcal{R}_5,\mathcal{R}_6,\mathcal{T})\nn\\
    &\xrightarrow{m_1}&G_{DW,d=5}((-)^F,\mathcal{C},\mathcal{CR}_1\mathcal{T},...,\mathcal{CR}_5\mathcal{T},1,\mathcal{C})\sim G_{DW,d=5}((-)^F,\mathcal{C},\mathcal{CR}_1\mathcal{T},...,\mathcal{CR}_5\mathcal{T}),
\eea

\bea
    &&G_{D,d=7}((-)^F,(-)^\chi,\mathcal{C},\mathcal{R}_1,...,\mathcal{R}_6,\mathcal{R}_7,\mathcal{T})\nn\\
    &\sim& G_{D,d=7}((-)^F,(-)^\chi,\mathcal{C},\mathcal{R}'_1=(-)^\chi\mathcal{R}_1,...,\mathcal{R}'_6=(-)^\chi\mathcal{R}_6,\mathcal{R}_7,\mathcal{T}'=(-)^\chi\mathcal{T})\nn\\
    &\xrightarrow{m_1}&G_{D,d=6}((-)^F,0,\mathcal{C},(-)^F\mathcal{R}_1,...,(-)^F\mathcal{R}_6,1,(-)^F\mathcal{T})\sim G_{D,d=6}((-)^F,\mathcal{C},\mathcal{R}_1,...,\mathcal{R}_6,\mathcal{T}),\nn\\
    &&G_{D,d=7}((-)^F,(-)^\chi,\mathcal{C},\mathcal{R}_1,...,\mathcal{R}_6,\mathcal{R}_7,\mathcal{T})\nn\\
    &\sim& G_{D,d=7}((-)^F,(-)^\chi,\mathcal{C},\mathcal{R}_1,...,\mathcal{R}_6,\mathcal{R}'_7=(-)^\chi\mathcal{R}_7,\mathcal{T})\nn\\
    &\xrightarrow{m_2}&G_{D,d=6}((-)^F,0,\mathcal{C},\mathcal{R}_1,...,\mathcal{R}_6,(-)^F,\mathcal{T})\sim G_{D,d=6}((-)^F,\mathcal{C},\mathcal{R}_1,...,\mathcal{R}_6,\mathcal{T}),\nn\\
\eea

\bea\label{dom-reduc-Dirac-8}
    &&G_{D,d=8}((-)^F,\mathcal{C},\mathcal{R}_1,...,\mathcal{R}_7,\mathcal{R}_8,\mathcal{T})\nn\\
    &\xrightarrow{m_1}&G_{DW,d=7}((-)^F,\mathcal{T},\mathcal{CR}_1\mathcal{T},...,\mathcal{CR}_7\mathcal{T},1,\mathcal{T})\sim G_{DW,d=7}((-)^F,\mathcal{T},\mathcal{CR}_1\mathcal{T},...,\mathcal{CR}_7\mathcal{T}),
\eea

For higher spatial dimensions, the same procedure is applied to prove the symmetry reduction.

\end{document}